\documentclass [10pt,twocolumn]{IEEEtran}
\usepackage{algorithm,algorithmic,amsbsy,amsmath,amssymb,epsfig,bbm,mathrsfs,multirow,amsthm}
\usepackage[T1]{fontenc}
\usepackage[latin9]{inputenc}
\usepackage{amsmath}
\usepackage{fixltx2e}
\usepackage{algorithm}
\usepackage{array,multirow,graphicx}
\usepackage{color}
\usepackage{cite}
\usepackage{float}

\DeclareMathOperator{\E}{\mathbb{E}}

\begin{document}
\title{Data Collection and Wireless Communication in Internet of Things (IoT) Using Economic Analysis and Pricing Models: A Survey}

\author{Nguyen Cong Luong, Dinh Thai Hoang, \textit{Student Member, IEEE}, Ping Wang, \textit{Senior Member, IEEE}, Dusit Niyato, \textit{Senior Member, IEEE}, Dong In Kim, \textit{Senior Member, IEEE}, and Zhu Han, \textit{Fellow, IEEE}
\thanks{N.~C.~Luong, D.~T.~Hoang, P.~Wang, and D.~Niyato are with Nanyang Technological University, Singapore. E-mails: clnguyen@ntu.edu.sg, thdinh@ntu.edu.sg, wangping@ntu.edu.sg, dniyato@ntu.edu.sg.}
\thanks{D.~I.~Kim is with Sungkyunkwan University, Korea. E-mail: dikim@skku.ac.kr.} 
\thanks{Z.~Han is with University of Houston, Houston, TX, USA. E-mail: hanzhu22@gmail.com.} 
}
\maketitle
\begin{abstract}
This paper provides a state-of-the-art literature review on economic analysis and pricing models for data collection and wireless communication in Internet of Things (IoT). Wireless Sensor Networks (WSNs) are the main component of IoT which collect data from the environment and transmit the data to the sink nodes. For long service time and low maintenance cost, WSNs require adaptive and robust designs to address many issues, e.g., data collection, topology formation, packet forwarding, resource and power optimization, coverage optimization, efficient task allocation, and security. For these issues, sensors have to make optimal decisions from current capabilities and available strategies to achieve desirable goals. This paper reviews numerous applications of the economic and pricing models, known as intelligent rational decision-making methods, to develop adaptive algorithms and protocols for WSNs. Besides, we survey a variety of pricing strategies in providing incentives for phone users in crowdsensing applications to contribute their sensing data. Furthermore, we consider the use of some pricing models in Machine-to-Machine (M2M) communication. Finally, we highlight some important open research issues as well as future research directions of applying economic and pricing models to IoT.  

{\it Keywords}- Wireless sensor networks, crowdsensing network, M2M communication, IoT, pricing models, economic models.
\end{abstract}
\section{Introduction}
\label{sec:Intro}

Internet of Things (IoT) is a novel paradigm which allows billions of smart devices to be connected to the Internet. Such devices can be sensors/actuators which are able to operate and transmit data to other systems without or with minimal human intervention~\cite{Minerva2015towards}. The development of IoT has brought a great influence to many areas, and there have been many IoT applications implemented to improve the system performance as well as the quality of life such as healthcare, transportation, manufacturing, and so on~\cite{gubbi2013internet}. Surveys of technologies and applications of IoT were presented in \cite{al2015internet}, \cite{whitmore2015internet}, \cite{atzori2010internet}, \cite{kantarci2015sensing}.

In IoT systems, Wireless Sensor Networks (WSNs) are one of the most important components mainly used to collect data from the environment and convey such data to the central controllers for further processing. However, different from conventional wireless sensor networks, sensors in IoT are required to be ``smarter''~\cite{Minerva2015towards}. In particular, sensors in IoT can not only perform normal functions, e.g., sensing information from the surrounding environment, but also make optimal decisions without or with minimal human intervention given their constrained resources and the dynamic of the environment for the requested IoT services. In addition, with billion of devices connecting to the Internet, it leads to many challenges in efficiently controlling and managing IoT's sensors. Consequently, new approaches with more efficiency and more flexibility to adapt to dynamic IoT networks need to be developed.

Apart from classical approaches, e.g., optimization-based approaches, economic and pricing based approaches have been widely applied in IoT systems. Compared with the optimization-based approaches, the economic and pricing based approaches provide the following benefits:
\begin{itemize}
\item The primary and most important benefit of economic approach is the revenue generation. The profit of IoT systems must be maximized given the revenue and cost incurred. 
\item Components in IoT have different objectives and constraints since they may belong to different entities, e.g., sensor owners, spectrum providers, and data center operators. Pricing approaches are introduced to determine optimal interactions among these self-interested and rational entities.
\item IoT has employed new sensing paradigms such as participatory sensing and crowdsensing networks to gather data from portable smart devices. Thus, to attract the users to contribute their data, incentive mechanisms using pricing and payment strategies can be adopted in order to guarantee the stable scale of participants and to improve the accuracy, coverage, and timeliness of the sensing results.
\item Using economic and pricing models, e.g., auctions, allows selecting the sensors with the highest remaining resources to perform sensing tasks. This can guarantee a trade-off between maximizing the network lifetime and providing the required data quality for sensors.
Moreover, the pricing models can easily eliminate data redundancy without the complexity computation.
\end{itemize}

Although there are some research works related to data collection and communication in wireless sensor networks, they focused on traditional approaches~\cite{wang2011networked, ehsan2012survey}. Besides, there are surveys related to the pricing approaches, e.g., \cite{he2012internet}, \cite{dasilva2000pricing}, \cite{gizelis2011survey}, but they addressed the issues in Internet or wireless networks. Moreover, the surveys on applications of game theory for wireless sensor network focus on general problems not specifically from economic and pricing aspects~\cite{shi2012game}, \cite{machado2008survey}. To the best of our knowledge, there is no survey specifically discussing the use of economic models to deal with the data collection and communication in IoT systems. This motivates us to deliver the survey with the aim of providing the state-of-the-art literature review on the economic models in IoT networks. Thus, through this paper, the readers will understand how economic models and pricing theory can be applied to address data collection and communication issues in IoT systems. 

\begin{figure*}[ht]
 \centering
\includegraphics[width=\linewidth]{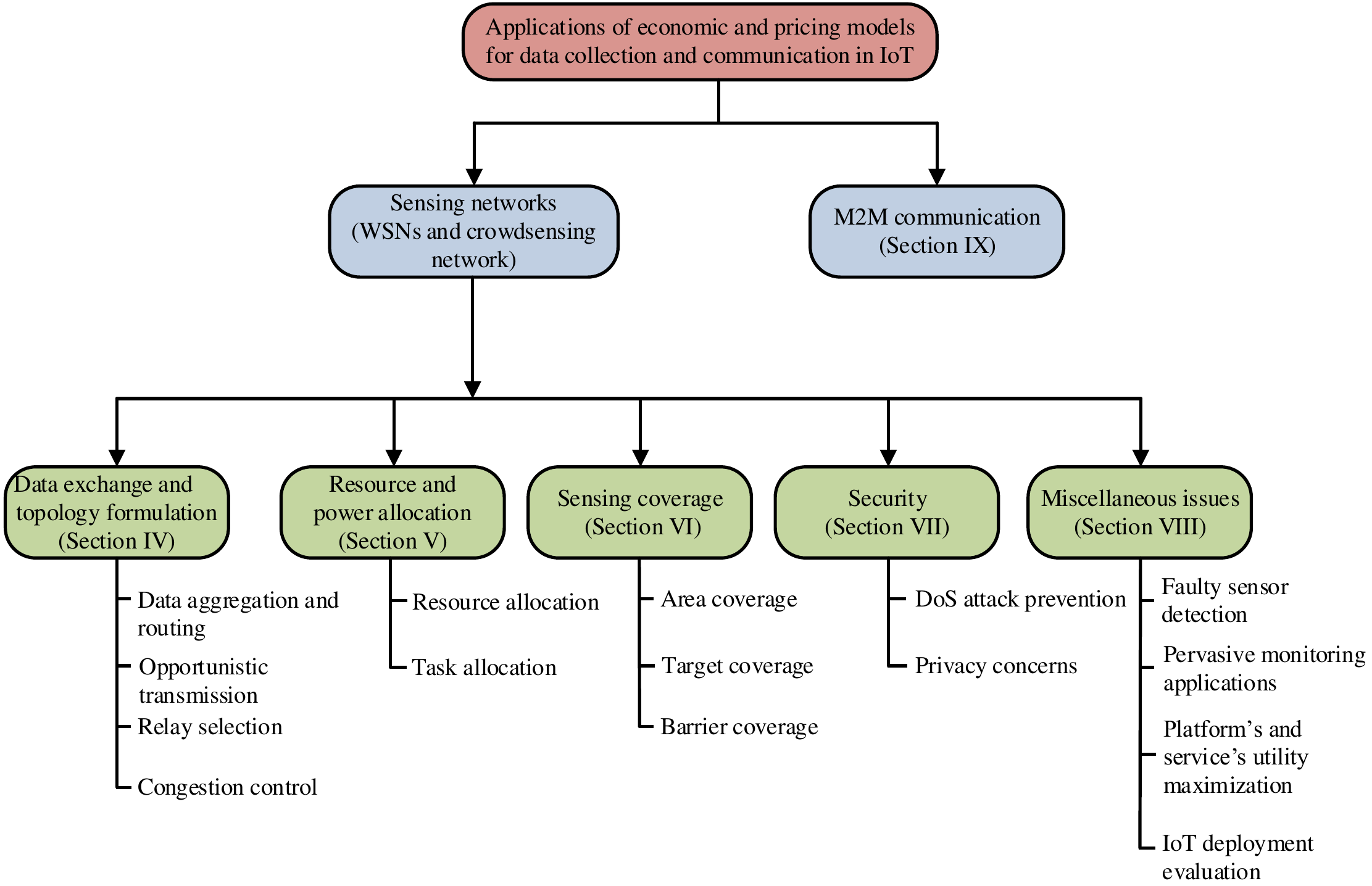}
 \caption{A taxonomy of the applications of economic and pricing models for data collection and communication in IoT.}
 \label{application_pricing_model}
\end{figure*}

In this paper, we mainly survey the economic models for solving data collection and communication in IoT. In particular, we classify the related works based on the WSN's issues as shown in Fig.~\ref{application_pricing_model}. The issues consist of four major sections, i.e., data exchange and topology formation, resource and power allocation, sensing coverage, and security. Furthermore, some related problems in IoT networks, e.g., faulty sensor detection, pervasive monitoring, service's utility maximization, and deployment evaluation, as well as the M2M's resource allocation are also discussed. We also review economic models together with efficient solutions for aforementioned problems. In addition, advantages and disadvantages of the solutions are also highlighted. 

The rest of this paper is organized as follows. Section~\ref{sec:General_IoT} describes a general architecture and services of IoT. Section~\ref{sec:Intro_Price} introduces the background and basic of economic and pricing models. In Sections~\ref{sec:Data_exchange}-\ref{sec:M2M_pricing}, we review the existing work of IoT from various perspectives. More specifically, Section IV discusses how to apply economic and pricing models for aggregating the sensing data in interested areas. Applications of economic and pricing models for resource allocation to maximize the resource efficiency in the data aggregation are given in Section V. Section VI employs pricing models for solving problems concerning sensing coverage and target tracking. Section VII considers economic approaches to address the DoS attack and the privacy issues in the sensing data collection. Section VIII reviews economic approaches for some miscellaneous issues in IoT. In addition, applications of pricing models for resource allocation in M2M communication are discussed in Section IX. We highlight trends, outline some open issues, and present some research directions in Section~\ref{sec:Open_issues}. Finally, we summarize and conclude the paper in Section~\ref{sec:Conclusion}.

\section{General architecture and services of IoT}
\label{sec:General_IoT}

\subsection{Definitions of IoT}
\label{sec:Def_IoT}
The concept of IoT is broad, and there are a few definitions available \cite{gusmeroli2010vision}, \cite{chui2010internet}, \cite{gubbi2013internet}, \cite{stankovic2014research}. In general, IoT refers to a self configuring, adaptive, complex network that allows a variety of things or objects, e.g., Radio Frequency IDentification (RFID) tags, sensors, actuators, and mobile phones, through unique addressing schemes, to interact and cooperate with each other to reach common goals \cite{atzori2010internet}. In addition to the definition, some key features of IoT are given as follows \cite{borgia2014internet}:
\begin{itemize}
\item \textit{Sensing capability:} Things in IoT context are able to perform sensing tasks.
\item \textit{Heterogeneity:} IoT may support for different underlying networks, e.g., wired, wireless, and cellular, and a variety of diverse communication devices, e.g., access point-based, and Peer-to-Peer (P2P) fashion.
\item \textit{Addressing modes:} IoT may support for anycast/unicast/multicast/broadcast transmissions. 
\item \textit{High reliability:} IoT guarantees connectivity and reliable transmissions based on different solutions.
\item \textit{Self-* capabilities:} Self-* capabilities of IoT are: (i) high configuration autonomy, (ii) self-organization and self-adaptation to dynamic scenarios, and (iii) self-processing of the huge amounts of exchanged data.
\item \textit{Secure environment:} IoT guarantees the robustness to secure issues such as network attacks (e.g., hacking and DoS), authentication, data transfer confidentiality, data/device integrity, privacy, and trusted secure environment.
\end{itemize}
\subsection{Architeture of IoT}
\label{sec:Arch_IoT}
\begin{figure}[ht]
 \centering
\includegraphics[width=6 cm, height=8.4cm]{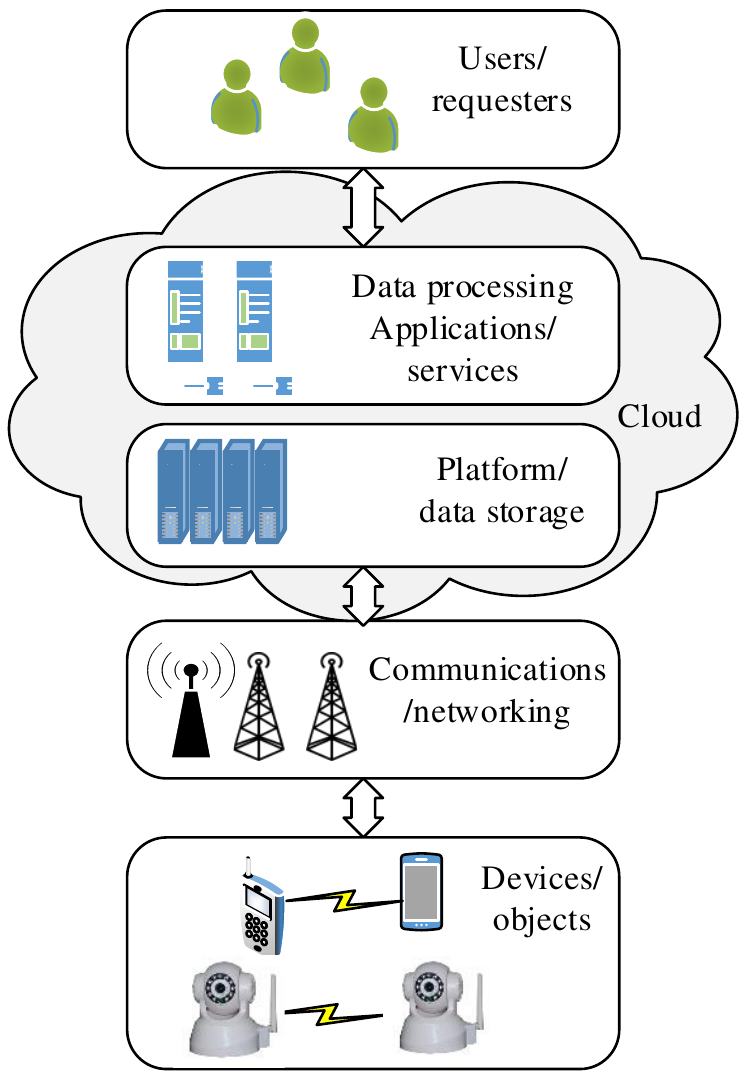}
 \caption{A general model of IoT.}
 \label{IoT_architechture}
\end{figure}

To meet the above features, several IoT architectures have been provided \cite{gubbi2013internet}, \cite{atzori2010internet}, \cite{uckelmann2011architectural}, \cite{gluhak2011survey}. An IoT architecture is shown in Fig. \ref{IoT_architechture} which consists of the following different tiers:
\begin{itemize}
\item \textit{Devices:} This physical layer is composed of low-level devices, e.g., RFID tags, sensors, and smart devices. Since they have limited resources in computing and storage, their functions are to perform only primitive tasks such as gathering data from the physical entities, e.g., environment conditions, or monitoring area of interests. Other functions of the devices are to connect with Internet gateways for data aggregation or to forward the data. 
\item \textit{Networking and communications:} This layer consists of data communication and networking infrastructures for delivering data gathered from devices at the physical layer to higher layers, e.g., cloud platform.
\item \textit{Platform and data storage:} This layer contains hardwares and platforms in data centers or services in the cloud with the aim of providing facility for the data access and storage.
\item \textit{Data management and processing:} This layer can be an application software which provides access services for IoT users. 
\end{itemize}

\subsection{Resources and Services of IoT}
\label{sec:Res_IoT}
Since IoT is a heterogeneous large-scale system with multiple different resources and services, the resource management is a key challenge to efficiently deliver IoT resources and services to users \cite{xu2014non}. In general, there are following resources and services in IoT which can be traded through adopting market and pricing models:
\begin{itemize}
\item \textit{Sensing data:} Sensing data is gathered from devices at the physical layer. The sensing information which is extracted from the sensing data can be traded and priced as to optimize the profits of owners \cite{nesse2013assessment}, \cite{sarvary1997marketing}, \cite{feng2014price}. 
\item \textit{Power:} IoT components, e.g., sensors, access points, base stations, and servers, need energy for their operations. In monopoly markets, energy providers can maximize their profits by optimizing the prices of energy supplied to the IoT components \cite{muller2002can}, \cite{ilic2012energy}. Recently, energy harvesting from renewable resources is a viable solution for providing "self-energy recycling" to IoT sensors through energy cooperation \cite{beeby2006energy}, in conjunction with the smart grid.

\item \textit{Cloud services:} They can be cloud data storage and computation services which are available and traded to IoT users \cite{buyya2008market}. 
\item \textit{Spectrum and network bandwidth:} In wireless network, spectrum and bandwidth are precious resources for data transmission. For example, in cognitive radio networks, the spectrum can be traded in a dynamic fashion among licensed users and unlicensed users with the aim of earning more revenue and improving the spectrum utilization. In practice, caching can be a viable solution for saving spectrum and network bandwidth in a large-scale IoT network by utilizing the sensing data and information stored in the cache, as long as the information value is not outdated and are temporally valid. 

\item \textit{Data and information services:} They can be offered and integrated to support IoT applications in proving services to IoT users, e.g., information searching, data mining \cite{bin2010research}, and data security. 

\item \textit{Location-based services:} By using real-time geographical information from user devices such as smartphones or tablets, IoT provides location services to interested individuals, organizations, and the government. The services
involve indoor and outdoor location services such as identifying a location of a person or an event, discovering the nearest places, e.g., restaurants, coffee shops, and stores \cite{chen2013localization}. Location-based services are also primary services of the IoT with an expected revenue of $£$34.8 billion in 2020 (http://iotbusinessnews.com).
\end{itemize}

\begin{figure}[ht]
 \centering
\includegraphics[width=\linewidth]{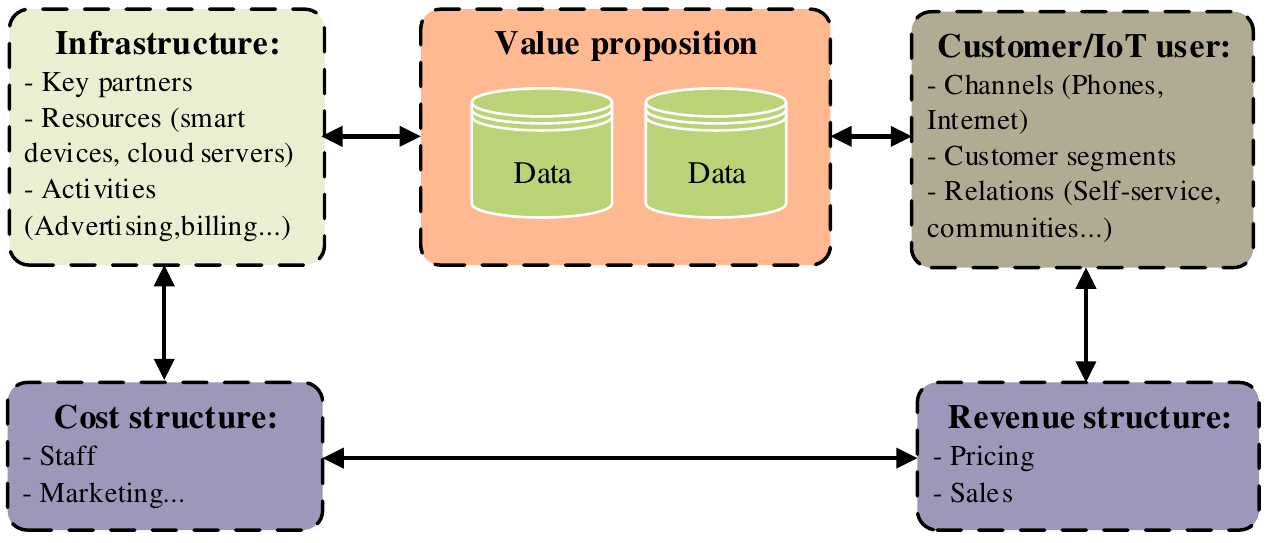}
\caption{A general business model of IoT systems.}
 \label{business_model}
\end{figure}

Among these resources, the sensing data and information are the most important ones in IoT since they can maximize the utility and profit of owners and providers \cite{niyatoeconomics}. To describe how IoT changes business models, an IoT business model (see Fig.~\ref{business_model}) was presented in \cite{niyato2015smart}. Basically, the IoT business model consists of four main components in which the value proposition of data resources is the core of the model. The value proposition involves setting the price of the sensing data resources and encouraging IoT users, i.e., customers' willingness to pay. Since the value proposition will create the major revenue for the businesses, choosing appropriate pricing strategies is important in IoT business models. In the next section, we will present the basics of pricing mechanisms and market models that can be applied to IoT.

\section{Overview and fundamentals of economic and pricing theory applied in IoT}
\label{sec:Intro_Price}
\begin{figure*}[ht]
 \centering
\includegraphics[width=\linewidth]{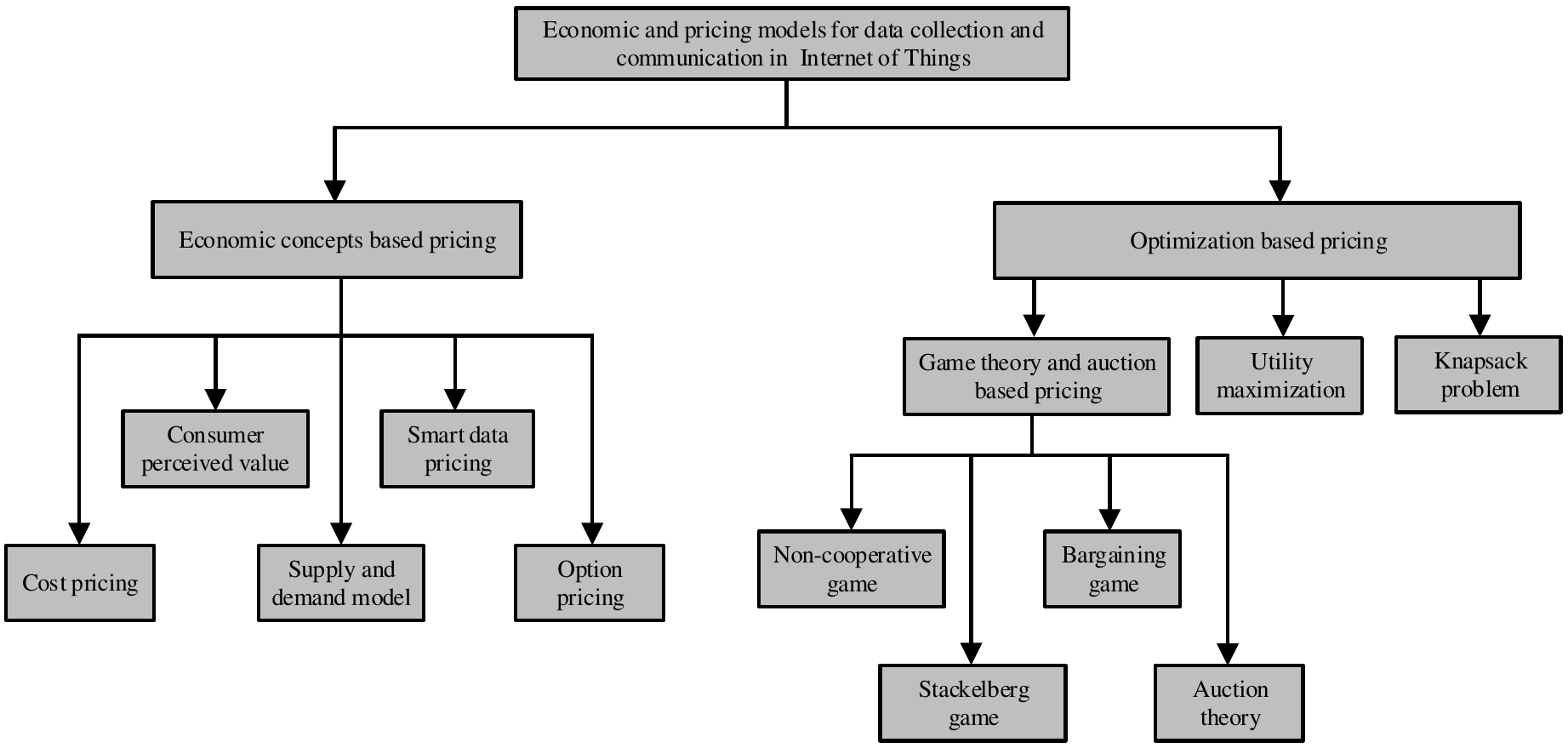}
 \caption{A taxonomy of economic and pricing models in IoT.}
 \label{taxonomy_pricing_model}
\end{figure*}

Economic and  pricing approaches in developing IoT systems have received a lot of attention due to their benefits as summarized in Section \ref{sec:Intro}. The classification of the commonly-used economic and pricing approaches used in the data collection and communication of IoT is shown in Fig.~\ref{taxonomy_pricing_model}. For convenience, they are categorized into three groups based on how to set the price: economic concepts based pricing, game theory and auction based pricing, and optimization based pricing.    

\subsection{Economic Concepts Based Pricing}
\label{subsec:Economic_concept}
This section discusses well-known pricing strategies based on the classical economic concepts such as cost, profit, demand and supply functions. 

\subsubsection{Cost-based pricing}
\label{subsec:Cost_theory}
Cost-based pricing is a pricing strategy which involves determining total cost of a commodity and adding a percentage of the cost as a desired income or profit. Mathematically, the selling price of a commodity is defined by $p=C \times (1+m)$, where $C$ is the total cost, $m$ is the markup which is the profit percentage added to the total cost. The total cost of the commodity typically consists of the fixed costs and variable costs. For example, to set the selling price for the sensing data, the following costs are considered: the fixed costs which are infrastructure costs, resource costs, e.g., energy and computational costs, and the variable costs which are the data delivery costs and the labor cost. In the cost-based pricing models, the profit maximization of the data seller was proposed in \cite{yang2013selfish} or of the information service provider as shown in \cite{pasura2005pricing}. Moreover, the model was employed to select data communication mode at low costs, i.e., opportunistic networks, as shown in \cite{conti2010opportunities}. The main advantage of the cost-based pricing is its simplicity since it requires only the internal cost information to set and adjust the selling price. However, this strategy does not consider external market factors, e.g., the competitors, the demand and the response of buyers. More information about the cost based-pricing model can be found in \cite{pindyck2005microeconomics}, \cite{nagle2008strategy}. 

\subsubsection{Consumer perceived value pricing}
\label{subsec:Perceived_value}
Pricing strategies based on covering costs are easily copied by competitors. Therefore, to maximize long-term profits, a seller needs to set the price by considering the buyer's perceived value from the commodity or the service rather than using traditional costs. The buyer's perceived value is the overall benefit derived at the price that the buyer is willing to pay. To set the price based on the perceived value, it is necessary to identify the set of value drivers that (1) present value perceptions about the commodity and the seller, (2) create positive attitudes and feelings, (3) provide the basis for differentiation, and (4) provide the reason to buy the commodity. In particular for the sensing data market in which the buyers may be service providers or data collectors, there are five major value drivers which impact the data pricing, that is $p_v=f(v_e,v_p,v_s,v_m, v_c)$ \cite{harmon2009pricing}, where:
\begin{itemize}
\item $v_e$ is the economic value which is based on buyer's perceptions about the costs of generating the sensing data. 
\item $v_p$ is the performance value which is based on buyer's perceptions of the utility of the data. A utility is an economic term referring to the total satisfaction received from consuming a commodity or service \cite{berry1996capture}. Therefore, the data utility is the most important factor for the pricing decision which may involve associated features such as the accuracy, quality, and timeliness. 
\item $v_s$ is the supplier value which is based on buyer's perceptions about the credibility, i.e., reputation, of the data seller.
\item $v_m$ is the buyer's psychological motivations for a particular purchase.
\item $v_c$ is the situation context that may have effect on purchasing behavior of the buyer. 
\end{itemize}
Generally, since the perceived value is the tradeoff between the perceived utility to be received and the perceived price for acquiring the information, sellers should understand these tradeoffs to maximize the buyer's value and seller's outcomes. In IoT market, consumer perceived value pricing model is efficient to estimate the consumers' potential demand for the data through determining their willingness and affordability \cite{gao2012analysis}. It is also used to maximize the profit of the information service provider in the market based on the variation of consumers' perception value of the information \cite{li2013research}.

\subsubsection{Supply and demand model}
\label{subsec:Supply_Demand_theory}
\begin{figure}[h]
 \centering
\includegraphics[width=7.3cm, height = 6cm]{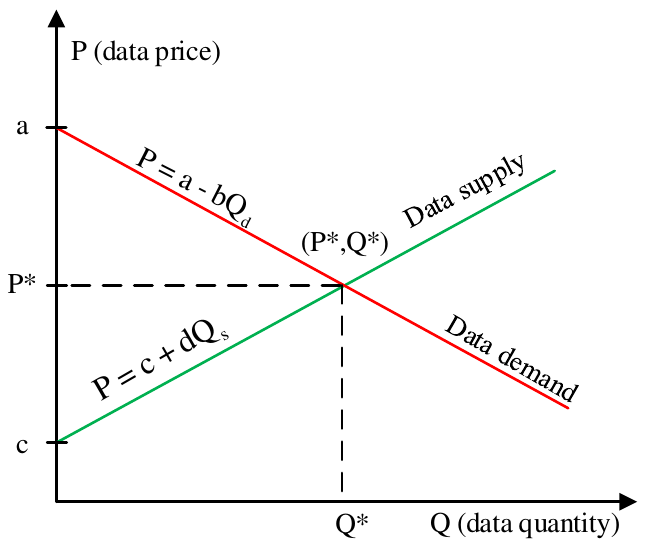}
 \caption{Supply and demand model for sensing data market.}
 \label{demand_supply_theory}
\end{figure}

Although the consumer perceived value pricing considers more about the demand of buyers, it ignores the incomplete information conditions and market competition. Supply and demand model can tackle the problem. The supply and demand are parts of the economic model in which the relations between these two factors can be exploited to determine the price of a commodity in a market \cite{pindyck2005microeconomics}. Consider a competitive market in which multiple sellers and buyers compete for sensing data. Assume that the relationship between the price $P$ for a data unit and the quantity demanded $Q_d$ is expressed by a linear demand function as $P=a-bQ_d$, with $a$ and $b$ are coefficients of the function. Similarly, a linear supply function which presents the relationship between $P$ and the quantity supplied $Q_s$ is given by $P=c+dQ_s$. Accordingly, the sellers and the buyers tend to change the price and the quantity to maximize their objective functions, e.g., profit and utility, subject to some constraints, e.g., budget and technology. As shown in Fig. \ref{demand_supply_theory}, the behaviors of the buyers and the sellers can be used to determine a unique point called a market equilibrium, i.e., an economic equilibrium, at which the quantity demanded equals the quantity supplied, i.e., $Q_d=Q_s=Q^*$, and $P=P^*= (a-c)/(b+d)Q^*$. At the market equilibrium, the price and the quantity are respectively called the competitive price, i.e., market clearing price, and the competitive quantity or the market clearing quantity. The market equilibrium has some basic properties as indicated in \cite{dixon1990equilibrium}: (i) the behavior of sellers and buyers is consistent since the demand equals supply, (ii) no seller or buyer has an incentive to change its behavior since their actual trades equal their desired trades, (iii) and the equilibrium is the outcome of the tatonnement process, meaning that at any non-equilibrium price, there will be excess supply or demand which leads to a movement in price towards the equilibrium. Since the buyers are highly sensitive to the price change, this model can guarantee fairness among the buyers \cite{xu2013internet} and provides an optimal allocation. 

The market equilibrium balances the supply and the demand of the network resources and the sensing data in multi-modal sensor networks \cite{chavali2012managing}. It also helps consumers achieve a desired information quality \cite{wang2012truth}, \cite{tham2015quality}. Also, the model was used for the flexible spectrum sharing as proposed in \cite{jain2012network}, \cite{berry2012network}. 
\subsubsection{Smart data pricing}
\label{subsec:Congest_theory}
To avoid large peak demands created by users' simultaneous consumption of scarce resources, the Smart Data Pricing (SDP) schemes are proposed. The SDP refers to various techniques such as charging buyers depending on the resource usage time, setting location based tariffs, and imposing prices based on buyer activity levels. We consider two common approaches used in IoT which are the time-dependent pricing and the usage-based pricing. The objectives are to reduce the network congestion by stimulating buyers to consume less resources or shift their usage to less congested times. 
 \begin{itemize}
 \item{\textit{Time-dependent pricing:}} In time-dependent pricing, prices of the resource have a temporal component, i.e., the prices vary
over different time to discourage large peak demands \cite{sen2015smart}. Only then will buyers be incentivized
to diffuse their demands over time, thus improving network resource efficiency by reducing the peaks and filling up the valley periods. 
 \item{\textit{Usage-based pricing:}} The classical flat pricing models, i.e., price is set statically, fail to adapt to the resource demand and to guarantee fairness. A more effective resource pricing model called usage-based pricing can be used. Accordingly, users are charged according to access
rate or resource usage \cite{edell1995billing}. It is stated that if the
price is set based on the usage, a fair and efficient use of resources would be moderately
promoted to some extent. 
 \end{itemize}
The two above approaches to managing the network congestion can help create a "win-win" solution
for both service providers and their users. The service providers benefit by reducing peak congestion while users are offered
more choices and technologies to save on their costs. Compared to the demand-and-supply model, the SDP has no need of the iteration to determine the market-clearing price, meaning that the market is cleared as soon as buyers have submitted their bids for access \cite{mackie1995pricing}. In IoT systems, the SDP has been introduced as an alternative to address resource management issues, e.g., power management \cite{chui2010internet}, \cite{liang2013udp}, resource demand management \cite{sen2013smart}, and price setting for resources in cloud computing applications \cite{fox2009above}.
\subsubsection{Option pricing}
\label{subsec:Option_theory}
An option concept in finance is a contract which gives the buyer, i.e., the owner of the option, the right to buy or sell an underlying asset or instrument at a fixed price, i.e., a specified strike price, at any time on
or before a specified date \cite{cox1979option}. The buyer then pays a price, also known as the premium, to the seller in exchange for the right granted by the option. To determine the price or value of an option, the Black-Scholes model in the option pricing theory is typically used. Accordingly, the price of a real option depends on various factors such as the current value and the uncertainty of expected cash flows, the value of fixed costs, the risk free rate of return, the time to maturity of the option, and any value lost over the duration of the option. Mathematically, the price, i.e., the value, of a real option based on the Black-Scholes model is given by \cite{macbeth1979empirical}
\begin{equation}
P=SN(d_1) -N(d_2)Ke^{-rt},
\label{Option_equ}
\end{equation}
where $P$ is the value of the option, $S$ is the current price of expected cash flows, $t$ is the expiration time of the option, $K$ is the exercise price, and $r$ is the risk-free interest rate. $N(d_i)$ denotes the cumulative standard density function evaluated at $d_i$, where $d_1$ and $d_2$ are defined as follows:
\begin{equation}
d_1= \frac{ln(S/K) + (r+\sigma^2/2)t}{t^{1/\sigma}} 
 \text{ and }  d_2= d_1-t^{1/\sigma},
\label{Option_equa_ex}
\end{equation}
with $\sigma$ is the variance of expected cash flows. Since the expected cash flows are uncertain, they must be estimated via the use of fuzzy logic. 

In IoT, the Black-Scholes model can be applied for the IoT investment evaluation \cite{lee2015internet}, the resource reservation in M2M communication \cite{lee2012optimal}, and the task schedule for sensors \cite{nguyen2012system}. For example, for evaluating IoT investment projects, the relation between variables in equation (\ref{Option_equ}) and those of the project are given as follows: the current value of project is current value of the real option, the investment cost of the project is the exercise price, the riskiness/uncertainty of the project is the price uncertainty, and the time value of money is the risk-free rate. The applications of the real option theory for resource reservation and the task schedule will be given in the survey section.

\subsection{Game Theory and Auction Based Pricing}
\label{subsec:Game_theory}
Game theory and auction are the formal study of decision-making where several players must make
choices that potentially affect the interests of the other players. The players in this context may be sensing data buyers, sellers or service providers. This section discusses how to set the price for sensing data in IoT using the auction and the game theoretic frameworks.

\subsubsection{Non-cooperative game}
\label{subsec:Noncooperative_game}
A game is known as non-cooperative if it is not possible for the players to form coalitions or make agreements. To interpret the definition of the game, some terminologies are given below.

\begin{itemize}
\item{\textit{Payoff}:} A payoff is a real number, i.e., a utility or an interest, which reflects the desirability of an outcome to a player. 
\item{\textit{Player}:} A player is a participant which makes decisions in a game. 
\item{\textit{Rationality}:} A player is rational if it always plays in a way which maximizes its own payoff. 
\item{\textit{Strategy}:} Player's strategy is a complete plan of actions/instructions that the player
can choose \cite{shi2012game}, and the payoff depends on not only its own
actions but also the action of others.
\end{itemize}

Consider a sensing data market in which sellers compete for selling their data. The market can be modeled as a non-cooperative game in which the players are the sellers with their pricing strategies. Let $(P, \pi)$ be a game with $n$ sellers and $P_i$ be a set of pricing strategies of seller $i$, where $P=P_1 \times \dots \times P_n$ is the Cartesian product of the individual strategy sets, and $\pi=(\pi_1(p),\dots,\pi_n(p)) \in R^n$ is a vector of payoffs of the sellers. Let $p_i \in P_i$ be the pricing strategy of seller $i$, then we have a vector of strategies $p=(p_1,...,p_n)$ of $n$ sellers, and a vector $\overline{p}_i$ which consists of the strategy choices of all sellers except the $i$th seller, i.e., $\overline{p}_i=(p_1,...,p_{i-1}, p_{i+1},..., p_n)$. Given a vector of strategies $p$, seller $i$ achieves payoff $\pi_i(p)$ which is a real value function of the seller's chosen strategy and strategies of the others. A set of strategies $p^*=(p_1^*,...,p_n^*) \in P$ is the Nash equilibrium if none of the sellers can improve its payoff by deviating its own strategy when other sellers keep their strategies fixed, that is \cite{friedman1971non}
\begin{equation}
\label{Nash_equilibrium}
\forall i, p_i \in P_i : \pi_i(p_i^*, \overline{p}_i^*) \geq \pi_i(p_i, \overline{p}_i^*).
\end{equation}

 Inequality (\ref{Nash_equilibrium}) implies that sellers at the Nash equilibrium have no incentive to change their own strategies since
 the payoffs are worse off. This makes the Nash equilibrium a consistent solution concept for games. However, sometimes there is no Nash equilibrium at all. Moreover, there may exist multiple Nash equilibria in a game, thus players may not be clear about which one to focus on. Therefore, it is important to check the existence and uniqueness of the Nash equilibrium when setting price using the non-cooperative game. 

The non-cooperative game theory covers a broad range of areas of IoT. For example, it can be used for resource allocation in wireless networks \cite{teng2010resource} or in sensor networks \cite{kim2011cross}. Also, it can be used for the spectrum sharing among a primary user and multiple secondary users \cite{niyato2008competitive}. In the cloud computing area, the non-cooperative game theory is used to maximize the profits of the competitive cloud providers \cite{pal2013economic}. Moreover, it can be used for the sensing coverage optimization \cite{nisan2007algorithmic}.

\subsubsection{Stackelberg game}
\label{subsec:Stackelberg_game}

The Nash equilibrium solution obtained from non-cooperative games assumes that the sellers know each other's strategies and they are announced at the same time. However, in real markets, these assumptions may not always hold, and thus the sellers cannot calculate its Nash strategy. Therefore, a seller can wait until the other seller's strategy is announced and then solves an optimization problem for its corresponding strategy, i.e., the strategies are given sequentially. Such a game is known as the Stackelberg game \cite{simaan1973stackelberg}. It was stated that in a Stackelberg game, if a seller has to announce its strategy first, its payoff is not lower than the corresponding Nash solution. Definition and properties of the Stackelberg game are given as follows. 

Assume that the game consists of two sensing data sellers and $P_1$ and $P_2$ are the sets of pricing strategies for sellers 1 and 2, respectively. The strategy of seller 1 is to maximize its payoff function $\pi_1(p_1,p_2)$ while seller 2 maximizes its payoff function $\pi_2(p_1,p_2)$, with $p_1$ and $p_2$ are the chosen strategies of sellers 1 and 2, respectively. Assume that seller 2 selects its strategy first, and thus it is called the leader. Seller 1 selects its strategy second, and it is called the follower. 

\textbf{Definition 1.} \textit{If there exists a mapping $T: P_2 \rightarrow P_1 $ such that, for any fixed $p_2 \in P_2 $, $\pi_1(Tp_2,p_2) \geq \pi_1(p_1,p_2) $ for all $p_1 \in P_1 $, and if there exists a $p_{2s2} \in P_2 $ such that $\pi_2(Tp_{2s2},p_{2s2}) \geq \pi_2(Tp_2,p_2) $, then the pair $(p_{1s2}, p_{2s2}) \in P_1 \times P_2$, where $p_{1s2}= Tp_{2s2} $, is called a Stackelberg strategy pair with seller 2 as leader and seller 1 as follower.}

In other words, the Stackelberg strategy is optimal for the leader when the follower reacts with the follower's optimal strategy. Let the graph $D_1= \{(p_1,p_2) \in P_1 \times P_2: p_1 = Tp_2\}$ of the mapping $T$ be called the rational reaction set of seller 1 when seller 2 chooses strategy $ p_2 \in P_2$. By playing according to the set $D_1$, seller 1 is a rational player. Similarly, $D_2$ is the rational reaction set of seller 2 if seller 1 is the leader. From $D_1$ and $D_2$, there are the two following propositions.

\textbf{Proposition 1.} \textit{A strategy pair $(p_{1s2}, p_{2s2})$ is a Stackelberg strategy with seller 2 as leader iff $(p_{1s2}, p_{2s2}) \in D_1 $ and $\pi_2(p_{1s2}, p_{2s2}) \geq \pi_2(p_1,p_2) $, $\forall (p_1,p_2)\in D_1$. }

\textbf{Proposition 2.} \textit{A strategy pair $(p_{1N}, p_{2N})$ is a Nash strategy pair iff $(p_{1N}, p_{2N}) \in D_1 \cap D_2 $.}  

These propositions show that in the Stackelberg game, the leader obtains at least
as good performance in terms of payoff function as the corresponding Nash solution. The reason is that when a leader chooses a Stackelberg strategy, the leader is actually imposing a solution which will facilitate itself. 

In sensing information market, the Stackelberg game can be applied to analyze the strategic options and payoff function between information provider and intermediaries \cite{li2013research}, or to maximize the profits of different participants of IoT industry value chain \cite{lv2012competition}. Besides, it was used for the distributed relay selection and power control for multiuser cooperative communication \cite{wang2009distributed}, or for enhancing both the QoS and the network's robustness in sensor networks \cite{danak2006inner}.  

\subsubsection{Bargaining game}
\label{subsec:Bargaining_game}
Bargaining games refer to the situations in which two or more players must reach an agreement regarding how to distribute an object. Consider a simple sensing data market including a single data seller which faces a single potential buyer. A bargain is successful if and only if the sensing data is transferred at a mutually acceptable price. Let $r_s$ be the seller's reservation price which is the smallest monetary sum that it will accept, and $r_b$ be the buyer's reservation price that it is willing to pay for the data. The seller and the buyer respectively submit price offers, $p_s$ and $p_s$, which are considered as their strategies. The strategy of the seller is to determine $p_s^*$ at which its expected profit $\pi_s (p_s,r_s)$ is maximized, i.e., $\pi_s (p_s^*,r_s)  \geq  \pi_s (p_s,r_s) $, $\forall p_s$. Similarly, the buyer determines its offer $p_b^*$ which makes its expected profit $\pi_b (p_b,r_b)$ maximum, i.e., $\pi_b (p_b^*,r_b)  \geq  \pi_b (p_b,r_b) $, $\forall p_b$. If $p_b^* \geq  p_s^*$, a bargain is enacted and the data is sold at the sale price, $p = kp_b^* + (1-k) p_s^*$, with $0 \leq  k \leq 1$. A pair of the best response offer strategies $(p_b^*, p_s^*)$ constitute the Nash bargaining equilibrium at which no player can increase its expected profit by unilaterally altering its chosen strategy. 

Since the Nash bargaining equilibrium is a solution of a cooperative game, it is used to enhance the users' own transmission opportunities and achieve high spectrum efficiency \cite{yang2010optimal} or an efficient energy management \cite{niyato2007wireless}. It also guarantees a full utilization of the spectrum resources while improving the utility of selfish users \cite{liu2009bargaining} or addresses the resource sharing problem between the secondary users and the primary user \cite{zhang2012selective}. 

\subsubsection{Auction}
\label{subsec:Auction_theory}
An auction is a public sale of buying and selling items, e.g., sensing data and bandwidth. The auction consists of an explicit set of rules which determine resource allocation and prices of the basis from market participants \cite{mcafee1987auctions}. Basic terminologies used in auction theory can be defined as follows:

\begin{itemize}
\item{\textit{Bidder}:} A bidder, i.e, buyer, wants to buy items in auction. In data aggregation, a bidder may be a data collector or a service provider who wants to buy data for their own purposes.  
\item{\textit{Seller}:} A seller, e.g., a sensor and a mobile device, wants to sell items.
\item{\textit{Auctioneer}:} An auctioneer controls an auction and announces the winner.
\item{\textit{Price}:} A price in auction, i.e., a hammer price, is the price at which the buyer and the seller agree to make a deal. There are two types of auction price, i.e., asking price and bidding price. A seller submits an ask which indicates the asking price on the item to be sold while a buyer can submit a bid indicating the bidding price for the requested item. 
\end{itemize}

Theoretically, there are several types of auctions. A taxonomy of auctions for resource allocation in wireless networks was given in \cite{zhang2013auction}. For the rest of this section, we introduce the details of auction types often used in data aggregation and communication in IoT.

\textit{(a) Sealed-bid auctions:}
\label{subsubsec:Auction_seal}
In sealed-bid auction, buyers submit simultaneously their sealed bids to the auctioneer. Different from open-cry auctions (e.g., English auction and Dutch auction) which allows buyers to observe their rival's bids and then to choose or revise their own bids, in the sealed-bid auction, each buyer does not know the other buyers' bidding strategies and thus cannot change its bid. There are two kinds of sealed-bid auctions. 
\begin{itemize}
\item{\textit{k-th-price sealed-bid auction:}} First-price and second-price sealed-bid auctions are the two most important \textit{k-th}-price sealed-bid auction. In the first-price sealed-bid auction, a buyer with highest price wins the item and pays the seller the highest price. The price in the first-price sealed-bid auction is defined by $p_F=  \max \limits_{p \in P }p$, with $P$ is the set of bidding prices. In the second-price sealed-bid auction, i.e., the Vickrey auction \cite{vickrey1961counterspeculation}, the winner pays the second-highest price rather than the price that it submitted, i.e., $p_V=  \max \limits_{p\in P \backslash\{p_i\}}p$, with $p_i$ is the highest price of the winner. With this rule, the winner can never affect the price that it pays, so there is no incentive for any bidder to misrepresent its value. The bidding true valuation is called the dominant strategy. Therefore, the outcome of the second-price auction meets a strong criterion of dominant strategy equilibrium, meaning that each bidder has a well-defined best bid value regardless of how high the bidder believes its rivals will bid \cite{krishna2009auction}. It is noted that any dominant strategy equilibrium is always a Nash equilibrium. Differently, the first-price sealed-bid auction has no dominant strategy equilibrium but it satisfies the weaker criterion of Nash equilibrium.

\item{\textit{Vickrey-Clarke-Groves (VCG) auction:}} A VCG auction is actually a generalized Vickrey auction with multiple items for sale \cite{ausubel2006lovely}. The basic idea behind the VCG auction is that the winner has to pay for the loss of the social value due to its getting item. 
Assume that there is a set of items for sale $T= \{t_1, t_2,\dots, t_M\} $, with $M$ is number of items and $t_i$ implies $i$th item, and a set of $N$ bidders $B= \{b_1, b_2,\dots, b_N\} $. Let a bid of bidder $b_i$ for item $t_j$ be $v_i(t_j)$ and also $V_N^M$ be social value created by $M$ items. According to the VCG auction rule, if $v_i(t_j)$ is the highest, i.e., the bidder $b_i$ wins to obtain item $t_j$, the price that the bidder has to pay for the winning is given by
\begin{equation}
V_{N\backslash\{b_i\}}^M - V_{N\backslash\{b_i\}}^{M\backslash\{t_j\}}.
\end{equation}

The outcome of a VCG auction is a Bayes-Nash equilibrium \cite{lucier2012revenue} and will be further described in Section \ref{sec: Aggregation_Participatory}. 

\end{itemize}

\begin{figure}[ht]
 \centering
\includegraphics[width=\linewidth, height = 10.3cm]{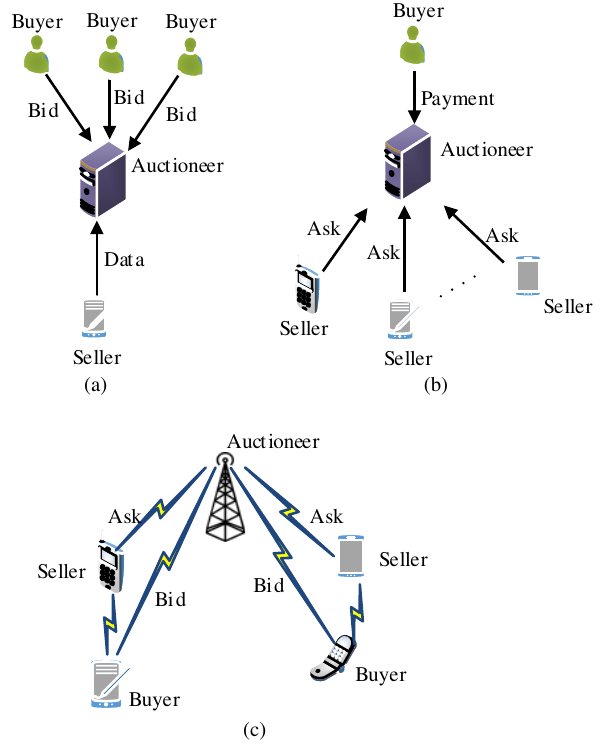}
 \caption{Different kinds of auctions: (a) forward auction for data aggregation with a single seller, (b) reverse auction for data aggregation with a single buyer, and (c) double auction for cooperative communication \cite{guo2014optimal}.}
 \label{single_double_auction_theory}
\end{figure}

\textit{(b) Forward, reverse, and double auctions:}
\label{subsubsec:Auction_reverse}
The auctions mentioned above are classified based on the seller's side and they are the forward auctions. Considering the buyer's side, there are reserve and double auctions. More specifically, we have the following definitions:

\begin{itemize}
\item{\textit{Forward auction:}} In a forward auction, buyers bid for items by offering increasingly higher prices, as illustrated in Fig.~\ref{single_double_auction_theory}(a). 
\item{\textit{Reverse auction:}} In a reverse auction, sellers compete for buyers' attraction by submitting their asks to the auctioneer as shown in Fig.~\ref{single_double_auction_theory}(b), and thus the price typically decreases to the lowest price that the buyer can accept. The reverse auction can be combined with other auction mechanisms, e.g., sealed-bid reverse auctions.  
\item{\textit{Double auction:}} A double auction is reminiscent of Walrasian auction \cite{mullen1996market} which perfectly matches the supply and demand model. In a double auction, buyers and sellers simultaneously submit their bids and asks to an auctioneer, respectively \cite{friedman1993double}. The auctioneer defines a price $p$, i.e., a clearing price, at which the asks from sellers are less than $p$ while the bids from buyers are more than $p$. Typically, the clearing price is set as $p=(p_i + a_j)/2$, where $p_i$ is price of the $i$th bidder and $a_j$ is price of the $j$th seller. The clearing price and the corresponding allocated commodities determine a competitive equilibrium, i.e., a market equilibrium, where the supply in the market is equal to the demand in the market.
An example of using the double auction for cooperative communications is given in Fig.~\ref{single_double_auction_theory}(c). Accordingly, relay nodes (i.e., sellers) submit their asking prices to sell their relay services while source nodes (i.e., buyers) bid these services for cooperative communication. Once receiving the asks and the bids, the base station (i.e., auctioneer) determines winners and clearing prices. The detail algorithms of the double auction process will be described in Section \ref{sec:Allocation_Double_auction}. 
\end{itemize}

\textit{(c) Combinatorial auction:}
\label{subsubsec:Combi_Cooper}
 In a combinatorial auction, a buyer submits its bid along with the need of a whole bundle of multiple items \cite{cramton2006combinatorial}. Based on the constraints included in the bid as well as the capacity of item allocation from sellers, the auctioneer finds an optimal allocation which is also considered as the winner of the auction. Compared to standard auctions, e.g., the sealed-bid auction, combinatorial auctions have more advantages, e.g., little global information requirement, economic efficiency, utility maximization for buyers, and revenue maximization for sellers. However, a computational burden is a challenge of the combinatorial auction since the winner determination for this auction is an NP-complete problem, meaning that there is no known polynomial-time algorithm to find the optimal allocation. Typically, only algorithms which find approximated solutions for the combinatorial auction have been proposed, e.g., the Lagrangian relaxation approach as shown in \cite{hsieh2010combinatorial}. 

The auctions mentioned above and their corresponding applications in IoT are summarized in Table \ref{table_auction_sum}. We can observe that several applications in IoT can employ auction mechanisms. Moreover, the sealed-bid auctions, e.g., the first-price and second-price sealed-bid auctions, are more commonly used than the others due to their simplicity and privacy guarantee \cite{suzuki2001efficient}.

\begin{table*}[h]
\caption{Summary of common auctions and key features.}
\label{table_auction_sum}
\scriptsize 
\begin{centering}
\begin{tabular}{|>{\centering\arraybackslash}m{2cm}|>{\centering\arraybackslash}m{2cm}|>{\centering\arraybackslash}m{5.5cm}|>{\centering\arraybackslash}m{4cm}|>{\centering\arraybackslash}m{2cm}|}
\hline
\multirow{2}{*} {\textbf{Auction type}}&\multirow{2}{*} {\textbf{Market structure}} &\multirow{2}{*} {\textbf{Key descriptions}} &\multirow{2}{*} {\textbf{Suitable scenarios}}  &\multirow{2}{*} {\textbf{Solution}} \tabularnewline 
& & & & \tabularnewline 
\hline
\hline
 First-price sealed-bid auction \cite{krishna2009auction} & A seller, multiple buyers, and an auctioneer & Private bidding process, and winning buyer pays the highest price &\begin{itemize} \item {Economics: buyer's payoff maximization}  \item{Technical: data aggregation, resource allocation, routing protocol, and privacy concerns} \end{itemize} &Nash equilibrium\tabularnewline \cline{2-5} 
\hline
Second-price sealed-bid auction/ Vickrey auction \cite{vickrey1961counterspeculation}& A seller, multiple buyers, and an auctioneer & Private bidding process, and winning buyer pays the second highest price &\begin{itemize} \item {Economics: buyer's payoff maximization}  \item{Technical: data aggregation, congestion management, resource allocation, and privacy concerns} \end{itemize} &Nash equilibrium\tabularnewline \cline{2-5} 
\hline
VCG  \cite{ausubel2006lovely}& A seller; multiple buyers; and an auctioneer & Generalized Vickrey auction with multiple items; private bidding process;  winner pays for the loss of the social value due to its winning the item &\begin{itemize} \item {Economics: social welfare maximization}  \item{Technical: data aggregation; and resource allocation} \end{itemize} &Bayesian Nash
equilibrium\tabularnewline \cline{2-5} 
\hline
 Double \cite{friedman1993double}& Multiple sellers, multiple buyers, and an auctioneer & Both buyers and sellers submit their bids and asks, respectively &\begin{itemize} \item {Economics: ordinary markets}  \item{Technical: resource allocation} \end{itemize} &Market
equilibrium\tabularnewline \cline{2-5} 
\hline
Combinatorial \cite{cramton2006combinatorial}& A seller, multiple buyers, and an auctioneer &Multiple heterogeneous auction items, and certain bundles of items have the highest value &\begin{itemize} \item {Economics: payoff maximization for buyers, and resource efficiency maximization}  \item{Technical: data aggregation, and resource allocation} \end{itemize} &Optional solution\tabularnewline \cline{2-5} 
\hline
\end{tabular}
\par\end{centering}
\end{table*}

Apart from the above-mentioned pricing strategies, we close this section with a posted price mechanism which is typically used in online procurement markets. 
\subsubsection{Posted price mechanism}
\label{subsec:Posted_price}
In an online procurement market, i.e., a digital market, sellers arrive in a sequential order to offer their
commodities, and a buyer decides to accept or reject the offer without full knowledge of the actual costs of the commodities \cite{hartline2001dynamic}. Assume that sellers are rational, they may exaggerate their costs to earn more profits. To learn the sellers' actual costs, bidding mechanisms (e.g., sealed-bid auctions) can be used through soliciting an ask from each seller upon its arrival. However, the bidding mechanisms are unnecessarily complex \cite{badanidiyuru2012learning}, and the posted price mechanism which gives sellers "take-it-or-leave-it" offers is a preferred alternative. The posted price mechanism posts a specific price for each seller. If a seller's actual cost is below the price offered by the mechanism, the seller accepts the price, and the buyer purchases from the seller at the price. Otherwise, the seller rejects the offer. Based on the current buyer's responses, the mechanism will set prices for subsequent sellers' commodities. Generally, compared with the bidding models, the posted price mechanism is less complex since it decides an offer price without observing the sellers' asks and pays the seller only if its ask does not exceed the offer. Therefore, in IoT, the posted price mechanism is widely used for the data aggregation in crowdsensing networks in which the data sellers, i.e., phone users, arrive in a sequential order. 

\subsection{Pricing Schemes Based on Utility Maximization and Knapsack Problem}
\label{subsec:Optimization_theory}
This section discusses two approaches using the mathematical optimization to determine the price. 

\subsubsection{Utility maximization}
\label{subsec:Utility_theory}

Utility is a term to represent the level of preference, i.e., the satisfaction, that a buyer receives from consuming goods or services. The different satisfaction of different
buyers to various price scenarios can be modeled analytically by adopting the utility function from microeconomics \cite{mas1995microeconomic}. A good approach to understand how to set the price based on utility functions is the introduction of the Network Utility Maximization (NUM) problem \cite{kelly1998rate} to allocate the transmission rate to sensors, i.e., buyers, in WSNs. Accordingly, the rate allocation can be characterized as the solution of the following NUM problem:
\begin{align}
\label{NUM_SYS}
 SYSTEM(U,A,C): \text{        maximize  } \sum_{s}U_s(r_s) \\
\text{        subject to } Ar \leq C \notag \\ 
\text{        over  }  r \geq 0, 
 \notag 
\end{align}
where $U_s(r_s)$ is the utility function of buyer $s$ depending on the allocated rate $r_s$, $A$ is a 0-1 matrix, each of which $A_{js}$ allows to determine whether the resource $j$ belongs to the route of buyer $s$ or not. The constraint $C$ contains elements $C_j$ which is the finite capacity of resource $j$. 
Typically, the utility functions of buyers are not known to the seller, e.g., a resource allocator. Therefore, one can divide (\ref{NUM_SYS}) equally into two sub-optimization problems. The first sub-optimization is the utility maximization problem for buyer $s$ 
\begin{align}
\label{NUM_USER}
 USER_s(U_s ; \lambda_s):  \text{        maximize  } U_s(\frac{p_s}{\lambda_s}) - p_s \notag \\ 
\text{        over  }  p_s  \geq 0, 
\end{align}
where $p_s = r_s \lambda_s$ is the price that buyer $s$ must pay for allocated rate $r_s$, and $\lambda_s$ is the charge per unit for the buyer. The second sub-optimization problem is the network's optimization problem given by
\begin{align}
\label{NUM_NET}
NETWORK(A, C  ;p):  \text{        maximize  } \sum_{s \in S}  p_s \lg r_s \notag \\ 
\text{        subject to } Ar \leq C \notag \\ 
\text{        over  }  r \geq 0,
\end{align}
Assume that the utility function $U_s(r_s)$ is an increasing, strictly concave and continuously
differentiable function of $r_s$. The authors in \cite{kelly1997charging} indicated that there exists a unique optimal solution ($r^*=(r_1^*,\dots, r_S^*),\lambda^*=(\lambda_1^*,\dots, \lambda_S^*), p^*=(p_1^*,\dots, p_S^*)$) which can solve the three above optimization problems. The vector $r^*$ is the unique optimal allocating vector rate and $\lambda^*$ is the current optimal resource price vector. 

In addition to the rate allocation for sensors, the utility maximization scheme is used for the bandwidth supply and demand balance \cite{chiang2004balancing}, or the optimal congestion and contention control \cite{lee2006jointly}.  

In practice, the above optimal solution is obtained only when the user utilities are assumed as concave functions. However, for hybrid service
systems, e.g., video and voice applications, such utilities vary with different types of applications with inelastic flows. Therefore, the pricing and resource allocation problem may be a non-convex optimization problem \cite{lee2005non}, and the NUM framework is no longer appropriate. 
%
\subsubsection{Multi-objective knapsack problem}
\label{subsec:Knapsack_theory}

The knapsack problem is a problem in combinatorial optimization which has been widely applied in both industry and financial management. The knapsack problem can be described as follows: given a non-negative real number $M$ and a set of $N$ objects where each object $i \in \{1,\dots,N\}$ is assigned a pair $(v_i,w_i)$ of non-negative real numbers. The values of $v_i$ and $w_i$ can be interpreted as the utility and the weight of an object $i$ while $M$ can be interpreted as the capacity of a knapsack, meaning that the maximum weight that the knapsack can hold. The goal is to find subset $S$ of objectives that maximizes $v(S)$ subject to the constraint $w(S) \leq M$, with $v(S)= \sum_{i \in S} v_i$ and $w(S)= \sum_{i \in S} w_i$ \cite{kellererknapsack}. In other words, the goal is finding a collection of the most valuable objects with the knapsack capacity constraint.

For more clarity, we take an example of applying the knapsack problem to selecting the active sensors, i.e., objectives, for performing a sensing task. The goal is to select a subset of the sensors such that the sum of their utilities is optimized under the constraint of a certain energy budget $M$. The value of $M$ is defined as the amount of energy necessary to accomplish a specific task, and weight $w_i$ of sensor $i$ is its energy cost when participating in the task. The utility of sensor $i$ may involve its potential relevance $R_i$ (e.g., sensing data quality), and its residual energy $e_i$, as shown in \cite{delicato2006efficient}. Then, the objective function of the knapsack problem is given by 
\begin{align}
\label{Knapsack_problem}
\text{        maximize } \sum_{i=1}^N x_i (\alpha R_i + \beta(e_i-w_i)) \notag \\
\text{        subject to }  \sum_{i=1}^N x_i w_i \leq M,   \notag \\
\end{align}
where $x_i=1$ if sensor $i$ is selected to participate in the task, and 0 otherwise. Coefficients $\alpha$ and $\beta$ are used to balance the priorities depending on specific applications. Although the knapsack problem is NP-hard, it can be solved optimally in pseudo-polynomial time through dynamic programming algorithm \cite{leiserson2001introduction}. The term \textit{pseudo-polynomial} means polynomial if the input (e.g., the value $M$ in equation (8)) is given in unary encoding. A pseudo-polynomial time algorithm is the algorithm that runs in time which is a polynomial in the input size, not the value the input represents.

Apart from the above example, we can also find other applications of the knapsack problem in IoT, e.g., sensing data collection \cite{pham2011novel}, \cite{davis2012heuristic}, or building new IoT-aware placement algorithms \cite{spinnewyn2015towards}.

\section{Applications of economic and pricing models for data exchange and topology formation}
\label{sec:Data_exchange}

As stated earlier, a WSN normally consists of a large number of autonomous and resource-limited nodes which cooperate to perform one or more sensing tasks within an area of interest. The coordination among them forms a communication network such as a homogeneous multi-hop network, or a hierarchical network. The main sensors' functions are to sense, process, and transmit data to a base station which is also known as a sink node. A sensor may send its data directly to a sink node or through other sensors. However, the topology and transmission routes of sensor networks may considerably fluctuate over time, e.g., a new sensor joins or leaves the network, or congestion appears at nodes. This section reviews applications of economic and pricing models of the data exchange and topology formation which can be summarized as follows:

\begin{itemize}
\item{\textit{Data aggregation and routing}:} Data aggregation is defined as the process where the data is aggregated from
multiple sensors. It is crucial in removing the data redundancy as well as saving the energy required for information transmission to the sink node \cite{maraiya2011wireless}, \cite{vaidyanathan2004data}. Towards this, the use of sophisticated "compressed sensing" algorithms \cite{donoho2006compressed} is of great importance, which leads to more energy-efficient data aggregation. In conjunction with this compressed sensing algorithm, the "sparsity level" can be controlled adaptive to the time-varying channel conditions, and hence the data aggregation may be jointly designed and optimized with the opportunistic  transmission strategy below. With such joint design and optimization, the economic and pricing models for data exchange and topology formation can be more adaptive to the energy state and channel state of mobile sensors, subject to the QoS of the requested services offered to the IoT users. 

\item{\textit{Opportunistic transmission strategy}:} Opportunistic transmission strategy is a technique which exploits available channels for data transmission. The use of pricing strategies helps sensor nodes achieve the highest usage efficiency of bandwidth as well as energy. 
\item{\textit{Relay selection}:} Due to the energy and bandwidth constraints, the data transmission between a sensor node and its destination may be accomplished by relay nodes. Pricing models have been proved as efficient approaches to select relay nodes for forwarding the data.  
\item{\textit{Congestion control}:} When there are more than one packet arriving at a node at the same time, the congestion occurs. Pricing mechanisms can be used to manage the congestion efficiently. 
\end{itemize}

\subsection{Data Aggregation and Routing}
\label{sec:Aggregation_Routing}
This section considers the applications of economic and pricing models for the data aggregation process in WSNs and in crowdsensing networks. 
\subsubsection{Data aggregation and routing for typical WSNs}
\label{sec:Sensor_Network}

\begin{figure}[ht]
 \centering
\includegraphics[width=\linewidth, height = 7.3cm]{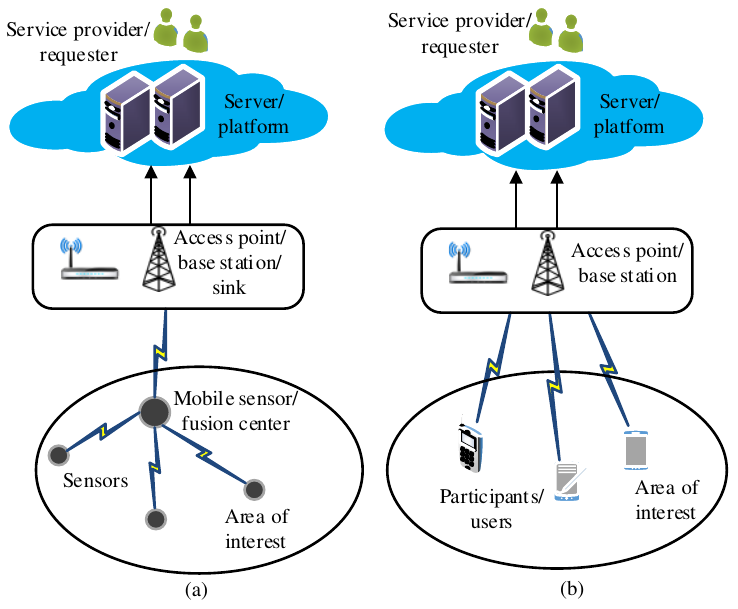}
 \caption{Data aggregation using in (a) WSNs, and (b) crowdsensing networks.}
 \label{data_aggregation_sensor_crowdsensing_networks}
\end{figure}

A common model of data aggregation in WSNs is shown in Fig.~\ref{data_aggregation_sensor_crowdsensing_networks}(a) which is composed of sensors, a mobile collector and a sink. The mobile collector is to gather data from sensors and then transmit it to the sink node via wireless channels. Due to the energy and capacity constraints of sensors and wireless links, improving energy efficiency and network lifetime is one of the most important issues \cite{alskaif2015game}. However, they also have to guarantee the required Quality of Service (QoS), e.g., high data quality and low latency. This conflict makes economic and pricing models become very useful in WSNs. In this section, we review the most recent approaches using economic and pricing models to achieve a trade-off between the network lifetime maximization and the required service quality. 

\textbf{Sealed-bid reverse auction:} The reverse auction selects the sensors with the highest remaining energy as winners for contributing sensing data. This approach allows to minimize the redundant information while maintaining high energy efficiency. Such an approach can be found in \cite{cao2013incentive} with the aim of prolonging the network lifetime which is defined as the time span from the deployment to the time at which the network becomes non-functional. The model consists of selfish sensors which act as sellers to provide the target localization data to a fusion center, i.e., a buyer (an auctioneer). The fusion center conducts the first-price sealed-bid reverse auction by soliciting asks from the sellers. The sensors submit simultaneously and confidentially to the fusion center their asks including estimated values of data and the asking prices that they are willing to be paid. In particular, the asking price of a sensor reflects the remaining energy of the sensors, e.g., a high price means that the remaining energy of the sensor depletes. Based on the estimated values and the asking prices, the fusion center defines its own utility function and selects an optimal subset of sensors which maximize the utility function as the winners. Since only the sensors with high energy, i.e., low asking prices, are selected, the energy consumption balance and long network lifetime are achieved. The simulation results showed that the lifetime in terms of the time step at which the network becomes non-functional of the proposed approach can achieve up 25 while that of the popular strategy for sensor selection based on the FIM (Fisher Information Matrix) \cite{zhao2002information} is around 15. However, the authors in this proposed approach did not specify the unit of the time step. 

The approach in \cite{cao2013incentive} did not consider the QoS of sensing data transfer. Thus, the authors in \cite{shah2013price} adopted the sealed-bid reverse auction to address the routing problem in WSNs with the aim of minimizing the latency and maximizing the network throughput for all applications. In fact, the latency is also closely related to the energy efficiency since a data packet with a large delay due to traveling more hops in the network consumes more energy \cite{al2013priced}. The model consists of a router, i.e., a buyer, who selects one of its neighbors, i.e., sellers, as the next hop for the packet delivery. Accordingly, each neighbor submits its ask including a path price which reflects the estimated time delay of a route from the neighbor to the destination. Generally, the path price of a neighbor is a function of the number of hops from the neighbor to the destination, the winning ask on the neighbor, and the application priority. The router selects a neighbor with the lowest path price as its next hop, i.e., the winner. The process is repeated on every hop and the minimal cost route is selected on a hop-by-hop basis which leads to a path with an overall minimal delay. Since each node always checks the path price of its neighbors whether it is smaller than the current node, the packet always moves towards the destination without routing loops. The simulation results showed that compared with the destination-sequenced distance vector algorithm in \cite{he2002destination}, the proposed approach increases delivered packets while reducing more the delay for damaged networks (up to 30\% of damage). However, for more highly damaged networks, the performance difference between the two approaches is slight. 

\textbf{Value-based pricing:} The reverse auction mentioned above aims at maximizing the utility of the requesters. However, due to the energy and capacity constraints of the sensors and the wireless links, it is essential to determine the payment expectation of the requesters and then adapt efficiently the corresponding resource usage. Hence, value-based pricing is suitable for this situation. Different from the reverse auction which defines the lowest price among sellers, the value-based pricing sets price primarily based on the perceived value to the buyer rather than the cost of the product, the market price, or even the historical prices. The authors in \cite{al2013priced} employed the value-based pricing to set the data selling price for the sensors, i.e., sellers, according to the requirements of the requester, i.e., the buyer. The general model includes entities as shown in Fig.~\ref{data_aggregation_sensor_crowdsensing_networks}(a), but there is an additional access point for pricing at the requester's side. The data selling price is set in two stages at the sink and the access point. The sink sets the price based on the expectations of the requesters about the data quality, packet delay, and packet lifetime. The access point then uses the price set by the sink along with the other requirements of the requester to determine the maximum price that the requester is willing to pay. The simulation results showed that the proposed scheme improved significantly compared with mobile ad-hoc schemes with respect to scalability, lifetime, delay, delivery ratio, and price.

\subsubsection{Data aggregation and routing in participatory sensing and crowdsensing networks} 
\label{sec: Aggregation_Participatory}

Participatory sensing allows a large number of users using mobile devices, e.g., smartphones and tablets to gather sensing data from interest areas without requiring the fixed and expensive infrastructure of WSNs \cite{burke2006participatory}. Similar to the participatory sensing, a crowdsensing network stimulates ordinary citizens to contribute data which is sensed or generated from their mobile devices \cite{guo2014participatory}. However, the crowdsensing network further explores the sensed data from mobile Internet services, e.g., mobile social network services, and heterogeneous crossspace. In other words, the crowdsensing network is an extension of the participatory sensing by aggregating the sensing data from mobile devices and the user-contributed data from other data sources. Since this paper considers the data collection from the mobile devices, a common model can be used for both the two paradigms. The common model is shown in Fig. \ref{data_aggregation_sensor_crowdsensing_networks}(b) in which the participants, i.e, users, using mobile phones to provide their sensing data to the service providers, i.e., buyers, via the servers or platforms. Since the sensing data contribution requires the users to consume resources, e.g., energy and bandwidth, a key challenge in the data aggregation is to provide incentive to the users. There were some surveys discussing incentive mechanisms in the participatory sensing, e.g., \cite{gaosurvey}, \cite{jaimessurvey}, \cite{zhangincentives}, but they did not focus on economic issues. Therefore, this section reviews economic-based approaches
used in the participatory sensing and the crowdsensing networks. 

\textbf{Sealed-bid reverse auction:} In participatory sensing and crowdsensing networks, the reverse auction mechanisms are efficiently used in stimulating users to provide their sensing data. Using the data aggregation model as shown in Fig.~\ref{data_aggregation_sensor_crowdsensing_networks}(b), the server which acts as an auctioneer first broadcasts the sensing task description from the requesters to all users, i.e., sellers. Users which are interested in the sensing task will accept and perform the sensing task. Once completing the sensing task, the users submit their asks including the sensed data and the corresponding prices to the server. The server selects a subset of them who own the lowest asking prices and makes the payments to them. 

\begin{figure}[ht]
 \centering
\includegraphics[width=\linewidth, height = 8.4cm]{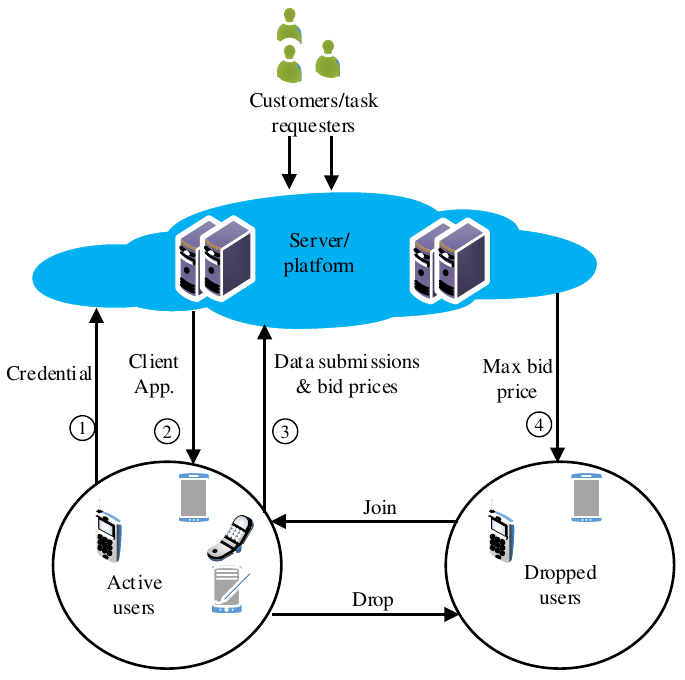}
 \caption{Data aggregation using RADP-VPC.}
 \label{data_aggregation_sealed_bid_reverse_auction}
\end{figure}

To minimize the payment, i.e., incentive cost (asking price), a high price competition between the users is always remained, meaning that the number of users participating in the auction needs to be large enough. The authors in \cite{lee2010sell} proposed a Reverse Auction based Dynamic Price incentive mechanism with Virtual Participation Credit (RADP-VPC) to achieve this goal. The model is shown in Fig. \ref{data_aggregation_sealed_bid_reverse_auction} in which the users perform the sensing task assigned by the requesters and provide the sensing data to the requesters via the server. The proposed approach consists of two functions: RADP and VPC. The RADP employs the first-price sealed-bid reverse auction to select the users with lowest asking prices as the winners. The server will purchase the sensing data from the winners and pay them rewards. Naturally, the losers in the current round have no incentive to participate in the next round and price competition will be decreased. As a result, in the next round, the winners increase the asking price to gain their utilities which may cause an incentive cost explosion. To tackle this problem, the VPC mechanism is introduced. Accordingly, loser $i$ in the previous auction round $r-1$ will receive an amount of virtual credit $\alpha_i$ when participating in the current auction round $r$, as follows:
\[
v_i^r= 
\begin{cases}
    v_i^{r-1} + \alpha_i,& \text{if user $i$ is lost in round $r-1$},\\ 
    0,              & \text{otherwise},
\end{cases}
\]
where $v_i^r$ is the cumulative VPC. It will be used for decreasing the asking price in current round as $a_i^{rc}=a_i^{ra} - v_i^r$, where $a_i^{ra}$ is the actual ask which is claimed by user $i$ and $a_i^{rc}$ is his competition ask. The server uses $a_i^{rc}$ to select the winners rather than $a_i^{ra}$, and thus the loser has more opportunities to win in the current round. 
The simulation results showed that using the VPC, the proposed strategy can reduce the incentive cost up to 63\% compared with the random selection based fixed pricing mechanism in \cite{mobile2010real} while stabilizing this cost over auction rounds.

However, in \cite{lee2010sell}, the winners' rewards may not be their expectations and they drop out of the next round. This also makes the asking price increase in the next round. To overcome this problem, the authors in \cite{lee2010dynamic} introduced a participant ReCruiting (RC) mechanism, in addition to the RADP-VPC. The RC allows the server to broadcast  the maximum price paid to the winner in the previous auction rounds to
the dropped users. Like an invitation, the price information helps the dropped users to re-evaluate their return-on-investment and potentially rejoin in the next round. The simulation results illustrated that the RADP-VPC with the RC dramatically suppressed the auction price, thereby alleviating the pressure on choosing the right value of VPC in \cite{lee2010sell}. 

The results in \cite{lee2010dynamic} can be improved if we determine the satisfactory levels of the credit expectations for the dropped out users. However, determining the expectation of the users in advance is a challenge to the server since they may only vote for satisfaction/dissatisfaction after getting the credit. To deal with this problem, the authors in \cite{liu2011efficient} employed the exponential smoothing method which takes the most recent payment history to predict the next expected credit from the users. The simulation results showed that the total allocated credit is close enough to the total required for the users with a small tolerance of $\pm 0.05$. Nevertheless, this method cannot be applied to users which have no previous payments. 

Similar to \cite{lee2010sell}, the authors in \cite{luo2014profit} argued that since all participants contributed their data, all of them should be paid rewards as incentives to execute the future
tasks. However, in this approach, monetary is used as rewards instead of the VPC. In fact, the incentive cost can be reduced further if we eliminate the winners which have too-closed locations since their coverages can overlap with others. To select the subset of the winners according to their locations (in additions to the data price), the authors in \cite{jaimes2012location} and \cite{jaimes2014spread} combined the RADP-VPC scheme \cite{lee2010sell} with the Greedy Budgeted Maximum Coverage (GBMC) algorithm in \cite{khuller1999budgeted} which guarantee that the number of selected winners (thus the incentive cost) is minimum while still achieving the same sensing coverage as \cite{lee2010sell}.

\textbf{Sealed-bid multi-attribute reverse auction:}
By means of the combination of RADP,
VPC, and RC, the proposed mechanism in \cite{lee2010sell} tackled two typical issues in reverse auctions which are the cost explosion and the user dropping, and thus it is more economically feasible. However, other problems such as data quality (i.e., QoS) are not explicitly addressed while not all sensing data have the same quality \cite{boutsis2012dynamic}. Therefore, negotiating only on the data price is not enough to meet data quality requirements of service providers. Hence, the authors in \cite{krontiris2012monetary} proposed a sealed-bid Multi-Attribute reverse Auction (MAA) scheme to aggregate sensing data with the aim of maximizing the servicer provider's utility. The MMA is a more complex form of the RADP by integrating additional attributions of the sensing data, e.g., the data price, the location accuracy, and the sensing time, into submitted asks. These attributes have different weights and typically, the price attribute has the largest weight since it is the most important. Based on the attributes and the corresponding weights, the service provider evaluates the utility score of an ask submitted from a user. The users with highest utility scores are selected as winners and get monetary for their contributions. Compared with a typical reverse auction, e.g., RADP, the simulation results showed that the utility of the MMA for the service provider is much higher, meaning that the service provider gets more beneficial. The reason is that the negotiation on more attributes brings more choices for the service providers to maximize the utility for application objectives. However, there exist some problems which were not identified in this approach, e.g., how to weigh the attributes and the privacy concerns of users when they submit their data.

Given the framework of MMA, the authors in \cite{albers2013coupons} investigated the payment for users using coupons. Accordingly, the coupons which allow the users to get discounts when buying something in stores, restaurants, and so on, are used as rewards instead of the real-money. A coupon itself contains two monetary values: the internal value which is the amount of money by redeeming the coupon, and the external value which is determined by the valuation of each user for the coupon. A user who wants to redeem its coupons can feed back the service provider so that the service provider evaluates the preference of the user and will suggest the coupons to a more interested user in the next auction round. This guarantees that the external value of coupon is always maintained at a high level which reduces the incentive costs for the service provider. Similarly, the authors in \cite{li2015reputation} used the coupons as rewards for the user incentive. However, different from \cite{albers2013coupons}, the number of coupons for a user is calculated according to its reputation score using the trimmed-mean method \cite{marazzi1999truncated} on the reliability of disseminating data. Users with high reputation scores mean a good participation history and get more coupons. However, there are still some limitations when the coupons are used. For example, from users' perspective, the coupons are usually less worth than money since money can be used universally. In addition, it is difficult to evaluate the coupons with two monetary values as defined above.

\textbf{VCG reverse auction:} 
Although the MMA scheme allows a service provider to evaluate the asks based on different attributes of data, the price is still the most important attribute. However, the price suggested by users is private information and the service provider does not know the actual price which may contain the energy and bandwidth costs. They often declare higher costs than the actual one to earn more money. To prevent such a misreport, the authors in \cite{koutsopoulos2013optimal} proposed a VCG reverse auction scheme which formulates the Bayesian game among the users, i.e., sellers. The VCG reverse auction is a type of the sealed-bid auction of multiple items in which the roles of the buyers and the sellers are reversed. The auction scheme works in a socially optimal manner by charging each seller who causes the harm of the social value to other sellers. Accordingly, each user submits an ask including the sensing costs and the participation level. In particular, the participation level of a user here is the amount of data that the user contributes. Assume that user $i$ knows only its actual cost $c_i$ and the probabilistic information about costs of the others which can be denoted as $\textbf{c}_{-i}=(c_j: j \in N, j \neq i)$, where $N$ is the number of participating users. The declared cost of user $i$ is determined to maximize his expected utility given by
\begin{equation}
\E_{\textbf{c}_{-i}} [u_i(\textbf{c})] = \E_{\textbf{c}_{-i}} [p_i(\textbf{c})-c_i x_i(\textbf{c})],
\end{equation}
where $u_i()$ is the utility function of user $i$ with participation level $x_i$ and payment $p_i$, $c_i$ is the cost realization which is determined according to the Bayesian game as follows. A declared cost vector $\textbf{y}^*$ is the Bayesian Nash equilibrium if for each user $i \in N$, we have 
\begin{equation}
\E_{\textbf{y}^*_{-i}} [u_i(y^*_{-i}, \textbf{y}^*_{-i})] \geq \E_{\textbf{y}_{-i}} [u_i(y_{-i}, \textbf{y}_{-i})], \text{ for all } y_i  \neq y^*_{-i}. 
\label{bayesian_equil}
\end{equation}
Equation (\ref{bayesian_equil}) means that no user wants to change his cost declaration strategy since this change will lead to a lower utility. 

Assume that $c_i$, with $i \in \{1,\dots, N\}$, is defined. The service provider selects a user $\widetilde{i} =  \arg \min \limits_{i}c_iq_i$, where $q_i$ is the quality indicator for user $i$ which is computed based on the data quality provided by the user in the past. The service provider finally pays the winner the compensation defined by the difference between the total cost of sensors $j \neq \widetilde{i}$ when the winner $\widetilde{i}$ does not participate in the auction and the total cost of sensors $j \neq \widetilde{i}$ when the winner $\widetilde{i}$ participates in the auction. Although the proposed mechanism is optimal in the sense of
minimizing the compensation cost to participants, it did not consider the similarity of device measurements which leads to the potential data redundancies.  

\textbf{Posted price mechanism:} As stated earlier, the posted price mechanism is used in online procurement markets in which sellers arrive in a sequential manner to offer their goods. Therefore, this mechanism is feasible in continuous crowdsensing applications where the phone users, i.e., sellers, arrive in a sequential order, and asks cannot be solicited. 

\begin{figure}[ht]
 \centering
\includegraphics[width=\linewidth, height = 4.5cm]{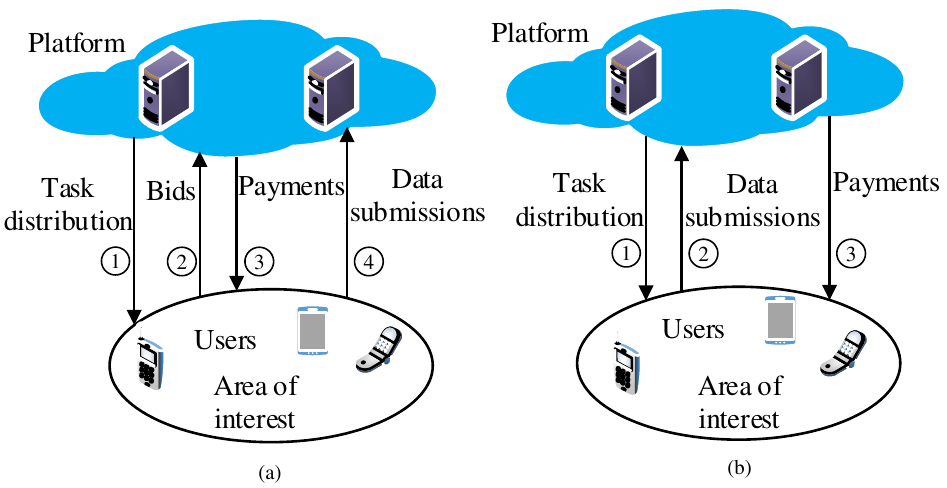}
 \caption{(a) Sealed-bid reverse auction based data aggregation, and (b) posted price mechanism based data aggregation.}
 \label{data_aggregation_posted_pricing}
\end{figure}

To make the posted price mechanism applicable, the authors in \cite{sun2014collection} proposed a new framework based on the combination of the posted price mechanism and an all-pay auction to provide higher quality sensing data and guarantee the extensive user participating under given budget constraints. The all-pay auction is similar to the standard (winner pay) first-price auction except that losers must also pay their bids \cite{baye1996all}. In the proposed approach, the roles of the buyers and the sellers of the all-pay auction and the posted price mechanism are reversed, i.e., the buyer sets the offer price instead of the seller. The process of the proposed data aggregation model is shown in Fig.~\ref{data_aggregation_posted_pricing}(b) which is similar to that of a general sealed-bid reverse auction as shown in  Fig.~\ref{data_aggregation_posted_pricing}(a). However, the platform, i.e., the buyer, divides the task deadline and the budget into several smaller periods and corresponding budgets. The phone users, i.e., sellers, find their own suitable periods and submit their sensing data to the platform. Based on the sellers' contributions and the allocated budget in each period, the buyer determines the number of optimal winners in the period using the all-pay auction and a threshold price for the next period. Subsequently, when there is an arrival seller, the seller compares its cost (for the task execution) with the threshold price. If the cost is smaller than the threshold and the budget in the period has not been exhausted, the seller chooses to accept the data sensing task. The simulation results implemented in \cite{sun2013behavior} showed that when more budgets are provided, the number of participants of the proposed solution increases more quickly compared to the winner-take-all, in which a single winner obtains all prizes. This result showed the extensive participating of the proposed strategy.

\textbf{Value-based pricing:} To motivate the users, i.e., sellers, to provide their sensing data and share their bandwidth, the authors in \cite{lan2012incentive} employed the value-based pricing to determine the payment for the users based on the utilities, e.g., the data quality and the timeliness, of the data that they contribute. The model consists of two types of seller: helpers who share their 3G connectivity or relay the data to the service provider and source users who own data to upload. Once the data is generated, an initial utility value is assigned to the data which will decrease over time due to its freshness reduction. The data can be uploaded to the service provider, i.e., the buyer, by either the helpers or the source if the source has the 3G connectivity. The service provider calculates an amount of credits for the helpers and the source according to the freshness and the quality of the data. To measure the motivation level of the users, a delivery rate which is defined  as the amount of data uploaded to the server over the amount of data generated was introduced. The simulation results showed that the success rate of the proposed solution can achieve up to 70\% while that of the First-Come-First-Serve (FCFS) policy \cite{schwiegelshohn1998analysis} is around 55\%. However, it is difficult to evaluate the rate in the real world since it may need several months of data. The same approach can be found in \cite{tsujimori2013senseutil}, but the number of credits for the sellers is calculated according to more attributes of the sensing data including the location information, starting time, sensing frequency, expiry time and kind of data. 

\textbf{Demand and supply model:} Demand and supply model is used to allow sellers to adjust price of goods according to buyers' demands to reach the market equilibrium. The authors in \cite{masazade2013market} modeled the sensing data bit allocation problem as the demand and supply model in which a fusion center is buyer, and sensors are sellers. However, different from a traditional demand and supply model, the buyer adjusts the price instead of the seller. The fusion center first broadcasts a price vector to purchase sensing data bits from sensors. Given the price vector, sensors find the demand vector to determine their production plans which maximize their profits. If the demand amount is less than the supply one, the price is decreased, and vice versa. The price update process continues until the market equilibrium is reached where the data bits supplied by the sensors meet the fusion center's demand. Since the sensing data produced by sensors is always close to that of the fusion center's demand, the resources are used efficiently. However, as indicated in \cite{scarf1960some}, the equilibrium may not be stable and the computational cost for converging to the equilibrium can be high.

\textbf{Bargaining:} Before being uploaded to the server, the data can be filtered by participants to guarantee that the data sent to the server is not redundant. Thus, the participants may employ the bargaining model to exchange their data before uploading as proposed in \cite{xie2009bargain}. The model consists of two selfish participants, i.e., sellers, who provide their data to the server, i.e., the buyer. Based on the probability of successful data delivery to the server, the sellers weigh all of their data and exchange their data lists to learn data that the participant $1$ does not have but the participant $2$ has, and vice versa. From each other's list, the participants 1 and 2 exchange the equal weight data from high to low by using a greedy algorithm. After exchanging, the expected credit reward $R_i^m(r)$ is used by participant $1$ to anticipate the reward of trading message $m$ of type $r$ with participant $2$, and is defined by
\begin{equation}
 R_1^m(r)=A_1^m(r)p_1(r), \notag
\end{equation}
where $A_1^m(r)$ denotes the appraisal of the message $m$. $A_1^m(r)$ indicates the probability that only the participant 1 delivers this message to the server, i.e., the probability that all other participants have not delivered the message $m$. $p_1(r)$ is the probability of participant 1 to meet the server at the first time. The simulation results showed that the bargaining-based scheme achieves more fairness while incurring less communication overhead compared with the direct transmission schemes. However, as the number of participants increases, the scheme may require a large number of exchange messages among them. 

Bargaining was also adopted in traffic prediction systems providing the data negotiation among participants and the server, e.g.,  \cite{lan2013providing} and \cite{luo2012fairness}. In this model, the roles of the participants and the server can be buyers or sellers. In \cite{lan2013providing}, drivers who participate on a route desire to get the road traffic information from others via the server. The server will only provide this information to the participants who upload their sensing data. The participants will get more information if they upload higher quality data which allows the server to predict the traffic situation more accurately. To distinguish the quality of the data, two attributes of data are considered which are the data update speed and the route where the data update comes from. For example, if the data is updated more frequently from a participant, then the data value from the participant would be greater than others'. The experiment results showed that the proposed mechanism could improve the prediction accuracy of traffic prediction up to 80\%, and thus reducing significantly traffic congestions. However, the obtained result also requires a large number of participants. 

The same approach was found in \cite{luo2012fairness} which aims at maximizing
the total amount of collection data and their quality. However, this approach studied more about the fairness of incentive distributions among participants. The fairness reflects the relationship between each participant's data contribution and its received service quota for future data consumptions. In other words, extra rewards are given to participants with more contributions. An optimization problem was formulated to achieve the maximum fairness which is measured by the Jain's index \cite{jain1998quantitative} for all
participants. However, this work did not provide specific attributes to evaluate the data updated from the participants which can prevent some participants from executing future tasks due to the insufficient payments. A stricter data exchange protocol was developed in \cite{chou2012using} in which a participant cannot download data directly from the server or indirectly from other participants if it did not perform data uploads to the server. Therefore, the participants are motivated to get information from the server by uploading sensing data or sharing their bandwidth with other participants. Nevertheless, this approach was not verified by experiments.

Apart from the common schemes described above, others economic models have also been applied to the data collection in the participatory sensing including:

\textbf{Multi-objective Knapsack problem:} The previous data collection approaches like RADP-VPC or MMA allow the service provider to select users with the lowest payment. However, when the allocated budget is constrained, the number of selected users is strict. To guarantee the QoS of the service provider with a limited number of users, the multi-objective Knapsack problem can be used. Based on this theory, the authors in \cite{pham2011novel} modeled the selection of participants as a multi-objective optimization problem which aims at achieving the high quality data with the low payment. The model consists of a service provider, i.e., a buyer, who needs to select strictly a limited number of participants, i.e., sellers, to provide the sensing data. The participants are selected according to the knapsack problem formulation as presented in Section \ref{subsec:Knapsack_theory} where the weight constraint is the limited number of participants while the high data quality and the low payment are the objectives of the optimization problem. In the case that the number of selected participants is not equal to the weight constraint, a criterion of crowding distance in \cite{deb2002fast} which measures the distance from a participant to its neighbors can be taken to add or eliminate the participants. For example, if the number of selected participants is more than the weight constraint, some participants with the smallest crowding distance can be eliminated. The experiment results showed that the proposed scheme outperforms the auction-based strategies, e.g., RADP-VPC \cite{lee2010sell}, in terms of the ratio of the total data quality to the total payment. It means that although the RADP-VPC scheme has the smallest payment, it provides a very low quality data, which makes it difficult for the service provider to ensure the QoS.

\begin{table*}
\caption{Applications of economic and pricing models for data aggregation and routing}
\label{table_data_aggregation_sum}
\scriptsize 
\begin{centering}
\begin{tabular}{|>{\centering\arraybackslash}m{0.2cm}|>{\centering\arraybackslash}m{0.4cm}|>{\centering\arraybackslash}m{1.6cm}|>{\centering\arraybackslash}m{1cm}|>{\centering\arraybackslash}m{0.8cm}|>{\centering\arraybackslash}m{1.2cm}|>{\centering\arraybackslash}m{5.1cm}|>{\centering\arraybackslash}m{2.4cm}|>{\centering\arraybackslash}m{1.3cm}|}
\hline
\multirow{2}{*}  {\textbf{}} & \multirow{2}{*}  {\textbf{Ref.}} &  \multirow{2}{*}  {\textbf{Pricing model}}  & \multicolumn{3}{c|} {\textbf{Market structure}} & \multirow{2}{*}  {\textbf{Mechanism}} & \multirow{2}{*}  {\textbf{Objective}} & \multirow{2}{*} {\textbf{Solution}} \tabularnewline 
\cline{4-6}
 & & & \textbf{Seller} & \textbf{Buyer} & \textbf{Item}  & & &\tabularnewline
\hline
\hline
\parbox[t]{2mm}{\multirow{9}{*}{\rotatebox[origin=c]{90}{ \hspace{-18cm} Data aggregation and routing}}}
&\cite{cao2013incentive} & Sealed-bid reverse auction & Selfish sensors& Fusion center &Target localization data & Sellers submit to the auctioneer their ask values inversely proportional to the remaining energy, the auctioneer selects an optimal subset of sellers with the lowest ask values & Energy consumption balance, and buyer's utility maximization&Nash equilibrium\tabularnewline \cline{2-9}

&\cite{shah2013price} & Sealed-bid reverse auction & Neighboring sensors & Router &Packet forwarding service & Sellers submit to the buyer their asks including the path price, the buyer selects a seller with the lowest path price& Overall minimal delay, and network throughput maximization&Nash equilibrium\tabularnewline \cline{2-9}

&\cite{al2013priced} & Value-based pricing & The sink node & Requester &Packet forwarding service & Seller sets prices according to the requester's requirements to maximize the utility of the requester& Buyer's utility maximization&Value optimization\tabularnewline \cline{2-9}

&\cite{lee2010sell} & Sealed-bid reverse auction & Phone users & Server &Sensing data & Sellers submit their asking prices including the task execution costs to the buyer, the buyer selects a subset of sellers with the lowest asking prices and gives them rewards. The losers also get virtual credit for incentive& Service quality guarantee, incentive cost minimization, improved fairness, and social welfare improvement&Nash equilibrium\tabularnewline \cline{2-9}

&\cite{lee2010dynamic}& Sealed-bid reverse auction & Phone users & Server &Sensing data &Same as \cite{lee2010sell} but the buyer adds a recruiting mechanism to stimulate the dropped users to join in the future&Service quality guarantee, incentive cost minimization, improved fairness, and social welfare improvement&Nash equilibrium\tabularnewline \cline{2-9}

&\cite{luo2014profit}& Sealed-bid reverse auction & Phone users & Server &Sensing data &Same as \cite{lee2010sell} but the buyer pays the users by monetary instead of the virtual credit&Incentive improvement&Nash equilibrium\tabularnewline \cline{2-9}

&\cite{jaimes2012location} \cite{jaimes2014spread} & Sealed-bid reverse auction & Phone users & Server &Sensing data &Sellers submit their asks including costs and their locations. The buyer selects a subset of the winners based on the GBMC algorithm&High data quality, area coverage improvement, and incentive cost minimization&Nash equilibrium\tabularnewline \cline{2-9}

& \cite{krontiris2012monetary} \cite{albers2013coupons} & Sealed-bid
multi-attribute reverse auction & Phone users & Service provider&Sensing data &Sellers submit their asks including multiple attributes. The buyer calculates the utility scores based on the attributes and selects a subset of the users with highest utility scores&Utility maximization&Bayes-Nash equilibrium\tabularnewline \cline{2-9}

& \cite{koutsopoulos2013optimal} & Vickrey-Clarke-Groves reverse auction & Phone users & Service provider&Sensing data &A Bayesian game among sellers is introduced. Sellers submit their asks including the costs and data contributions to maximize their utilities. The buyer selects a user as the winner depending on the cost and a data quality indicator &Individually rational, and incentive-compatible&Bayesian Nash
equilibrium\tabularnewline \cline{2-9}

&\cite{sun2014collection} \cite{sun2013behavior} & Posted price mechanism & Phone users  & Platform&Sensing data &Sellers compare their costs and a threshold price set by the buyer to decide to perform the tasks. They submit the sensing data to the buyer and get rewards &High data quality, and extensive participation improvement&Nash
equilibrium\tabularnewline \cline{2-9}

&\cite{lan2012incentive} \cite{tsujimori2013senseutil} & Value-based pricing & Phone users & Service provider&Sensing data and relay service &Seller sets prices according to the service provider's requirements to maximize the utility of the requester &High transmission success rate, and utility maximization&Value optimization\tabularnewline \cline{2-9}

& \cite{masazade2013market}  & Demand and supply model & Sensors & Fusion center&Sensing data &Buyer adjusts iteratively the price of data so that its data demand matches with the data supply of sellers &Resources efficiency&Market equilibrium\tabularnewline \cline{2-9}

&\cite{xie2009bargain}  & Bargaining game & Two phone users & Server &Sensing data &Two sellers exchange their data with each other by using a greedy algorithm to provide the non-overlapped data to the buyer &Overhead and delay reduction, and fairness&Pareto efficiency\tabularnewline \cline{2-9}

&\cite{lan2013providing}  \cite{luo2012fairness}  & Bargaining game & Phone users & Server &Sensing data and road traffic information &Sellers provide their sensing data to the buyer. The buyer pays the sellers for the road traffic information &Fairness, utility maximization, and prediction accuracy improvement of traffic &Pareto efficiency\tabularnewline \cline{2-9}

&\cite{pham2011novel} & Multi-objective Knapsack problem & Phone users & Service provider &Sensing data&Buyer employs a pairwise tournament selection approach to select a set of optimal sellers which maximize high data quality and minimize the payment  &Fairness, utility maximization, and prediction accuracy improvement of traffic &Pareto efficiency\tabularnewline \cline{2-9}

\hline
\end{tabular}
\par\end{centering}
\end{table*}

\subsection{Opportunistic Transmission and Neighbor Discovery}
\label{sec:Opportunistic}
In the previous section, we discussed the applications of economic and pricing models to attract phone users in aggregating sensing data with high quality at low cost. The majority of sensing data is then sent to the server through single-hop 3G/4G cellular networks. However, due to limitations such as 3G/4G costs and cellular capacity, the cellular networks would not be economically feasible. Therefore, the opportunistic networking can be adopted to achieve the minimum transmission costs, maximum energy efficiency, and minimum traffic.

\begin{figure}[ht]
 \centering
\includegraphics[width=\linewidth, height=7.3cm]{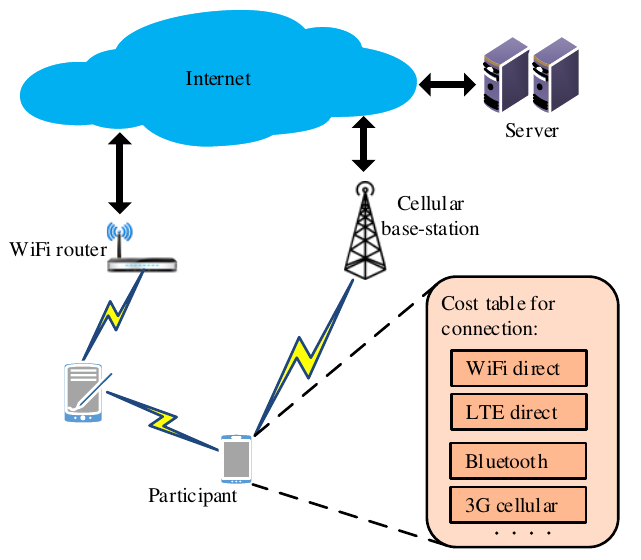}
 \caption{Opportunistic transmission based on cost pricing.}
 \label{opportunistic_transmission_cost_minimization}
\end{figure}

Such a scheme can be found in \cite{adeel2014self} and \cite{al2013online} in combining with the cost-based pricing to minimize the global system cost. Cost-based pricing is a strategy to set the price of a product based on costs to produce the product. The model in \cite{adeel2014self} is shown in Fig.~\ref{opportunistic_transmission_cost_minimization} which consists of a source phone user, i.e., a seller, selling its sensed data to a server, i.e., a buyer, through the 3G cellular radio or the WiFi routers nearby or its neighboring users. First, the source builds a one-hop neighbor table including the available neighbors and corresponding relay costs. The source then considers the costs as its input to calculate its profit and select a neighbor that has the minimal input cost for the data relay. Since the cost using the short-range communication (via neighbors or WiFi routers) is much lower than the 3G/4G connectivity, the proposed approach encourages the seller to transmit its data to neighbors using the short-range communication. The simulation results illustrated that the global system cost can be arbitrarily close to the minimal according to the Lyapunov optimization theory. However, these results are only reachable if the system parameter which is multiplied by the total monetary value of the data to determine the data selling price is sufficiently small.

Similarly, the authors in \cite{al2013online} adopted the cost-based pricing to compute the packet delivery cost from the source sensor to the sink via mobile devices. In this model, there are two types of relay nodes: integrated relays which act as traditional wireless relays and courier nodes. To save resources, critical packets are relayed by the integrated relays while less critical packets can be forwarded by a courier depending on its forwarding charge with respect to the packet's threshold price. Each courier node decides on its forwarding price based on the delivery time, buffer capacity, and the remaining energy. Depending on the criticalness of data, the source selects a courier node with low forwarding price so that its profit is maximized. Since the forwarding price is inversely proportional to the availability of resources, the proposed delivery scheme achieves high efficiency in terms of energy, cost and delivery rate. For example, at pause time of 500 seconds, the average consumption energy can be saved up to 40\% compared with the traditional MANET delivery protocols, e.g., Ad-hoc On-demand Distance Vector (AODV) protocol \cite{broch1998performance}. However, in the proposed approach, the authors did not define the threshold price which allows the source to decide whether to forward the packet to the courier or to directly transmit it.

In practice, an adaptive control of the "sparsity level" in compressed sensing \cite{donoho2006compressed} for the data aggregation can be combined with this opportunistic transmission strategy, which will achieve the minimum transmission costs, maximum energy efficiency, and minimum traffic. Therefore, the use of appropriate economic and pricing model in jointly designing the data aggregation with the opportunistic transmission is of great importance, which seems to not be studied before. This is a kind of future research direction to come up with such a nice economic and pricing model for joint design and optimization towards adaptive control of "sparsity factor" in the compressed sensing based data fusion, considering the channel/path conditions, subject to the QoS of the IoT services. 

\subsection{Relay Selection}
\label{sec:Relay_Selection}

To meet the QoS requirements, the source node needs  to choose optimal routes (e.g., the
shortest route and the least-energy consumption route) for its data forwarding. However, nodes on the different routes are typically rational and selfish, thus pricing strategies can be employed as incentive mechanisms for the nodes to forward the sensing data.

\subsubsection{First-price sealed-bid reverse auction} 
\label{sec:relay_first_reverse_auction}
\begin{figure}[ht]
 \centering
\includegraphics[width=\linewidth, height = 6.8cm]{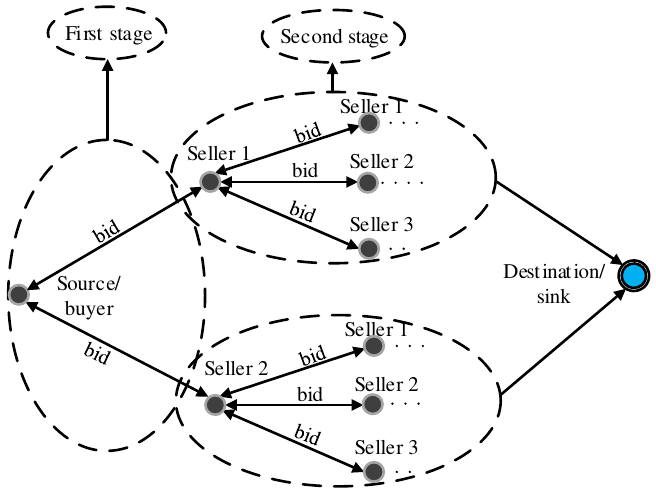}
 \caption{Auction-based relay selection mechanism.}
 \label{routing_Protocol}
\end{figure}

Similar to the data aggregation, the sealed-bid reverse auction schemes are also applied efficiently to solve the relay selection problem. As described in \cite{liu2011game}, the scheme was introduced for the relay node selection with the aim of decreasing the energy consumption and prolonging the network lifetime. The model is illustrated in Fig.~\ref{routing_Protocol} in which the whole forwarding process can be implemented by multiple stages. Each stage consists of one buyer, i.e., a source node, who buys the data forwarding service from several sellers, i.e., neighbor nodes. Since the stages have the same relay selection process, we only consider the relay selection in the first stage. The source node builds its own table containing information of the link quality and the residual energy of neighbors. The link quality is a function of the residual energy and the hop count of the neighboring node (to the sink node). Different from the typical sealed-bid reverse auction where only the neighbors (sellers) submit their asks, in this approach, both the source node and the neighbors submit their asking prices at the same time. The asking price of a neighbor is a function of the hop count, the link quality and the service price that is willing to get from the source. The asking price of the source is an average price calculated based on the information in its table. The source will buy the relay service from a neighbor at the deal price of the neighbor if its asking price is smaller the asking price of the source. In case of multiple selected neighbors, the source chooses the one with the lowest asking price. The simulation experiment showed that the proposed algorithm outperforms the Low Energy Adaptive Clustering Hierarchy (LEACH) protocol \cite{heinzelman2000energy} in terms of energy consumption. This is because of the single-hop communication strategy of the LEACH protocol. 

\subsubsection{Dutch reverse auction} 
\label{sec:relay_Dutch_reverse_auction}
Another auction scheme which can be used for the relay selection strategy is the Dutch reverse auction. In this scheme, the seller initially asks a high price and then lowers the price at each iteration until the buyer accepts the price. Similar to the first-price sealed-bid reverse auction, a seller in the Dutch reverse auction has no bidding information of the others and the winner gets a payment that is equal to his asking price. Such approach for the relay selection was proposed in \cite{lima2008game} where the model is used similarly to \cite{liu2011game}. Accordingly, the source acts as a buyer who purchases the relay service from one of the relay nodes, i.e., sellers. However, instead of asking the relay nodes to suggest their asking prices as in the first-price sealed-bid reverse auction, the source provides forwarding regions with different priorities as the asking-price, i.e., the initial hint, via the request-to-send frame of the request-to-send/clear-to-send handshake. The priority of a region is defined according to the advance that the node forwards to the destination. It means that the regions which are closest to the destination have the highest priority. Each relay node determines independently and autonomously its priority to build its own ask. They then respond to the source sequentially according to the priority via the clear-to-send frame. The relay node with the first arrival clear-to-send frame is selected as the winner for the data forwarding. The simulation results showed that the proposed strategy achieved the stable packet delivery success ratio up to 90\% while this ratio of the scheduled relay selection is around 70\%.
\subsubsection{Stackelberg game}
\label{sec:Relay_Stackelberg} 
Auction approaches allow the source to select the best neighbor for its data forwarding. However, there may be rational interactions between the nodes. For example, the selected relay node just accepts to transmit part of the packet instead of the whole packet from the source, and thus the whole packet should be divided into different parts. However, the splitting strategy depends on the independent decisions of the relay nodes rather than the source. Therefore, the authors in \cite{karapistoli2014routing} modeled the relay selection problem as a hierarchical game, i.e., a two-stage Stackelberg game between the source, i.e., the buyer, and the relay nodes, i.e., sellers. Accordingly, the relay nodes act as leaders and they first set their own relay prices. Then, the source, i.e., the follower, selects the corresponding splitting strategy, i.e., the flow configuration, to maximize its utility. Given the source's splitting strategy, the leaders then may re-determine the prices with the aim of maximizing their own utilities.  It was proved that the utility functions of the follower and the leaders are convex, and thus there exists a unique solution for the flow configuration and the relay pricing which is the Stackelberg equilibrium outcome of the game \cite{gintis2000game}. The simulation results verified that the outcome of the Stackelberg game achieves the best utilities among all Nash equilibrium strategies which are obtained by assuming a simultaneous choice for leaders and the follower. Moreover, compared with the Nash equilibrium strategies, the Stackelberg strategy achieved a more balanced energy dissipation among the relay nodes since the relays make price announcements taking into account not only the competition, but also the battery reserve capacity. 

\subsection{Congestion Management}
\label{sec:Congestion_mana}
When multiple sources transmit simultaneously the data to the sink node, the congestion may occur at the relay nodes. This congestion level increases network delays and decreases the value of information. Traditional congestion control techniques which often employed the destination-to-source feedback scheme, e.g., \cite{ee2004congestion}, are not applicable in WSNs since the flow paths can change before the feedback loop can be formed. Therefore, dynamic pricing strategies are considered as appropriate solutions for the congestion management in WSNs. 

\subsubsection{Second-price sealed-bid auction}
\label{sec:Congest_second_auction} 
An auction may be conducted at congested nodes to prioritize the sensing data forwarding. The authors in \cite{chen2009auction} employed the second-price sealed-bid auction for the congestion management in which the congested node acts as an auctioneer, i.e., a seller, and the packets act as bidders, i.e., buyers. The bidders compete with each other for the current transmission slot and the auction may be repeated until the congestion is removed. The reason for using the second-price sealed-bid auction is that the winner only gets the current transmission slot instead of  the entire resources as the first-price sealed-bid auction. Consider a target tracking application, each packet contains the target information, e.g., the target velocity, the application priority, and a delay from the source to the sink node. However, since the auction is conducted at the congested node, not the sink, an approximate value of the delay is used instead. Specifically, the value is taken from the previous auction round. The value of the packet is evaluated in terms of loss of information utility. The auctioneer selects a packet with the highest loss of information utility as the winner of the current transmission slot. The losers also get an additional fund for the future auction rounds which is proportional to their information loss incurred due to its transmission delay. This enables equalizing the information utility loss according to the different target tracking
applications and also minimizing the total utility loss in the network. The simulation results indicated that the difference in utility losses between the proposed solution and the analytical congestion management approach is small on average across all the tracking applications. The auction-based congestion management, by extension, was used in directional binary sensor networks to provide the additionally directional information of the target \cite{chen2009dynamic} or to provide the congestion information for the source nodes in the path selection \cite{shah2012dynamic}. In particular for the path selection application, the winning bid values in the auctions to resolve the congestion on the next hops can be included in the path price as discussed in \cite{shah2013price}. Therefore, this helps the source node to dynamically obtain the path with the smallest predicted utility loss. 

The auction used in \cite{chen2009auction}, \cite{chen2009dynamic}, and \cite{shah2012dynamic} can be considered as a static auction since each congested node executes the auction based only on its local information and does not take the auction information about other nodes into account. This can make the congested node evaluate the loss of information utility of the packet (i.e., bid) incorrectly. For example, if the packet has not passed the congested node yet, the node does not approximate the time delay, meaning that the loss of information utility cannot be evaluated exactly. To allow the congested node to make more efficient decisions, the authors in \cite{geyik2012market} introduced a \textit{traveling auction} scheme in which a node can share its local auction information with its neighbors. The traveling auction is similar to the auction scheme in \cite{chen2009auction} with the addition of the information sharing mechanism among nodes in the network. The shared information is added to the packet header which includes the application \textit{ID}, the time at which the packet is generated as well as the delay. However, increasing the overhead messages also affects the traffic in the network. 

Apart from the above congestion management, the second-price sealed-bid auction was also adopted in the data queue management for priority deviation problem in delay-tolerant WSNs \cite{li2013price}. Accordingly, data packets are scheduled for transmitting based on their priorities. To distinguish the priority of data and avoid the collision due to data priority deviations, each packet is associated with the price which is a function of three parameters, i.e., priority dimension which identifies the network class, priority class which shows the priority order of each data group, and priority measure to quantify the priority value of each data group. The data packets in the queues will be prioritized based on their prices and the higher-price data packets will be sent first. Compared with the direct data transmission schemes, the simulation results indicated that the proposed approach achieves a better performance in terms of the data delivery ratio. However, high complexity of computing the priority for each price can result in increasing average delay of the network.

\subsubsection{Value-based pricing}
\label{sec:Congest_value_congestion} 
The value-based pricing can be also employed for the packet dropping strategy at the congested node as proposed in \cite{qiu2009efficient}. However, in this approach, the buyer sets the price instead of the seller. Specifically, the sink, i.e., the buyer, sets prices for packets receiving from sensors, i.e., sellers, depending on which the sink is willing to pay and then notifies the sensors these prices. The packets with higher prices are expected to contain more important information, and the sink prefers to buy such important information. Therefore, when the congestion arises at the sink, the packets with higher prices are selected. However, to guarantee the fairness for all packets and the coverage fidelity of the whole network, the \textit{Jain's fairness index} \cite{jain2010art} was introduced which allows to determine the accepting probability of each packet with the aim of maximizing the relatively fair opportunity to all packets. Since the accepting probability, i.e., the accumulated survival probability, contains information of the price and the coverage fidelity, it is considered as a decision variable which the sink uses to select or drop the packets. The simulation results showed that the throughput of the proposed scheme is much higher than that of the First-In First-Out (FIFO) policy, especially when the network congestion occurs.

Apart from the pricing models used above, the authors in \cite{zhou2005port} adopted a term called \textit{node price} to address the congestion control and the energy consumption minimization. Specifically, each packet sent from the source to the sink is associated with a node price which is defined as the total number of transmission attempts across the network before a
successful packet is delivered to the sink. The node price reflects the communication cost, i.e., the consumed energy to successfully deliver a packet from each source to the sink. When the congestion arises at the sink node, it directs individual sensors to increase or decrease their reporting rates depending on their communication costs so that the energy
consumption of the WSN is minimized while alleviating the congestion. However, in
dense networks, it is difficult to send the control information to every
node since the sink can be at distance from the sensors. 

\textbf{Summary:} The reviewed papers in this section have shown clearly the efficiency of the economic and pricing models for the problems related to the data exchange and
the topology formation. We summarize the issues along with references in Table~\ref{table_data_aggregation_sum} for the data aggregation and Table~\ref{table_topo_form_sum} for the topology formation. From these tables, we observe that many works investigated the data aggregation problem while the topology formation problem was less studied. For the data aggregation, auctions are considered as the most efficient pricing models in stimulating users to contribute high quality data at low incentive cost. The next section discusses the use of economic and pricing models for sensing resource and task allocation. 

\begin{table*}
\caption{Applications of economic and pricing models for data aggregation and routing}
\label{table_topo_form_sum}
\scriptsize 
\begin{centering}
\begin{tabular}{|>{\centering\arraybackslash}m{0.55cm}|>{\centering\arraybackslash}m{0.4cm}|>{\centering\arraybackslash}m{1.6cm}|>{\centering\arraybackslash}m{1cm}|>{\centering\arraybackslash}m{0.8cm}|>{\centering\arraybackslash}m{1.4cm}|>{\centering\arraybackslash}m{4.8cm}|>{\centering\arraybackslash}m{2.4cm}|>{\centering\arraybackslash}m{1.3cm}|}
\hline
\multirow{2}{*}  {\textbf{}} & \multirow{2}{*}  {\textbf{Ref.}} &  \multirow{2}{*}  {\textbf{Pricing model}}  & \multicolumn{3}{c|} {\textbf{Market structure}} & \multirow{2}{*}  {\textbf{Mechanism}} & \multirow{2}{*}  {\textbf{Objective}} & \multirow{2}{*} {\textbf{Solution}} \tabularnewline 
\cline{4-6}
 & & & \textbf{Seller} & \textbf{Buyer} & \textbf{Item}  & & &\tabularnewline
\hline
\hline
\parbox[t]{2mm}{\multirow{9}{*}{\rotatebox[origin=c]{90}{ \hspace{1.2cm} Opportunistic}}}
\parbox[t]{2mm}{\multirow{9}{*}{\rotatebox[origin=c]{90}{ \hspace{1.2cm} transmission}}}

& \cite{adeel2014self} & Cost-based pricing & Source phone user& Server &Sensing data&Seller updates costs from their neighbors for forwarding the sensing data to the buyer. The best neighbor is chosen to maximize the seller's profit & Throughput optimality, global costs minimization, and self-organization&Pareto optimum\tabularnewline \cline{2-9}

& \cite{al2013online} & Cost-based pricing& Relay nodes&Source sensor &Packet delivery&Sellers set their forwarding prices which are inversely proportional to the availability of resources& Energy consumption minimization, and high reliability &Pareto optimum\tabularnewline \cline{2-9}
\hline
\parbox[t]{2mm}{\multirow{9}{*}{\rotatebox[origin=c]{90}{Relay selection}}}

&\cite{liu2011game} & First-price sealed-bid reverse auction &  Neighboring sensors&Source node &Data relay service& Sellers submit their asking prices which is a function of hop count, link quality. Buyer selects a seller with lowest price as the winner& Energy consumption minimization, and network lifetime enhancement&Nash equilibrium\tabularnewline \cline{2-9}

&\cite{lima2008game} & Dutch reverse auction & Relay nodes&Source sensor &Data relay service&Buyer sends the different prices corresponding to their forwarding regions to sellers. A seller will relay the data if it accepts the offer price& Energy consumption reduction, and high reliability &Nash equilibrium\tabularnewline \cline{2-9}

&\cite{karapistoli2014routing} & Stackelberg game & Relay nodes&Source sensor &Data relay service&Sellers set their relay prices. The buyer selects corresponding data splitting strategy so that its utility is maximized& Utility maximization of sellers, and energy consumption reduction&Stackelberg equilibrium\tabularnewline \cline{2-9}
\hline
\parbox[t]{2mm}{\multirow{9}{*}{\rotatebox[origin=c]{90}{ \hspace{-4cm} Congestion management}}}

& \cite{chen2009auction}  \cite{chen2009dynamic}  \cite{shah2012dynamic}&Second-price sealed-bid auction & Congested sensor&Packets  &Current transmission slot& Buyers submit their bids including the loss of information utility. The seller selects a buyer with the highest loss of information utility as the winner& Information utility loss balance, and total utility loss minimization&Nash equilibrium\tabularnewline \cline{2-9}

& \cite{geyik2012market} &Second-price sealed-bid auction & Congested sensor&Packets  &Current transmission slot& Same as \cite{chen2009auction} but the sharing information scheme between the congested sensors is introduced & Information utility loss balance, and total utility loss minimization&Nash equilibrium\tabularnewline \cline{2-9}

&\cite{li2013price} &Second-price sealed-bid auction & Congested sensor&Packets  &Transmission slots& Buyers submit their bids including priority parameters. The seller assigns the transmission slots to higher-price packets&High packet delivery ratio, and average delay reduction &Nash equilibrium\tabularnewline \cline{2-9}

&\cite{qiu2009efficient} &Value-based pricing & Sensors &Sink node  &Sensing information& The buyer sets price for each packet depending on which it is willing to pay. When the congestion arises at the sink, the packets with higher prices are selected & Throughput maximization, information utility maximization, and fairness &Value optimization\tabularnewline \cline{2-9}   

& \cite{zhou2005port} &Value-based pricing & Sensors &Sink node  &Sensing information& Each seller set a node price on each its packet which is the function of the previous unsuccessful attempts. The buyer adjusts the reporting rate of the sellers depending on the price & Network lifetime maximization, and reliability maintenance &Value optimization\tabularnewline \cline{2-9}   

\hline
\end{tabular}
\par\end{centering}
\end{table*}

\section{Applications of economic and pricing models for resource and task allocation}
\label{sec:Resource_alloc}
As aforementioned, WSNs are very limited by resources. The resources include the available energy, wireless bandwidth, and computational resources (e.g., processing capability, storage capacity, and battery power). Therefore, the resource usage optimization in WSNs has become a critical issue. In this section, we review the use of economic and pricing models for the following issues:
\begin{itemize}

\item{\textit{Resource allocation}:} Traditional resource allocation algorithms often assume that the available resources in a system such as network bandwidth do not change. However, the amount of available resources on the devices and the communication channel in WSNs is not constant. Pricing mechanisms are used as solutions in which all of WSN's scarce resources, e.g., CPU, storage, bandwidth, and battery, can be best utilized with the variability of resource budget.

\item{\textit{Energy Control}:} The energy consumption of the sources are reserved for sensing, processing and communication. Pricing strategies based on the utility are employed to stabilize and optimize the transmit power level for sensors. 

\item{\textit{Task allocation}:} Task allocation is to assign
sensing tasks to specific sensors in the network for
execution. Given the constrained resource and distributed
structure of WSNs, the goal of the task allocation scheme is to achieve a fair energy balance among the sensors while minimizing the delay. To address this problem, price formulations can be used as they can continuously adapt to changes of the resource availabilities. 
\end{itemize}

\subsection{Resource Allocation}
\label{sec:Resource_All}

Resource allocation in WSNs is to assign the network resources, e.g., time slot, communication bandwidth, and energy, to sensors for performing their tasks. In traditional resource management, resources are assigned to the sensors in a static manner. Such system resource management is inefficient since the demand and supply of the resources do not usually match \cite{zhang2013auction}. Market-enabled pricing schemes can alleviate the inefficiency by creating an artificial market for exchanging dynamically the resources between the sensors. 

\subsubsection{Combinatorial auction}
\label{sec:Allocation_combinatorial_auction}
The aforementioned auction schemes are based on a single-item auction mechanism in which individual items are auctioned independently. However, in many cases, using multiple items may yield a higher utility than using each of them separately, i.e., the items are complementary. For example, in moving target tracking applications, using measurement data from multiple sensors increases the tracking accuracy. Therefore, instead of using the single-item auctions, we can use combinatorial auction schemes which allow consumers, i.e., buyers, to request a complete bundle of auction items. 

\begin{figure}[ht]
 \centering
\includegraphics[width=\linewidth]{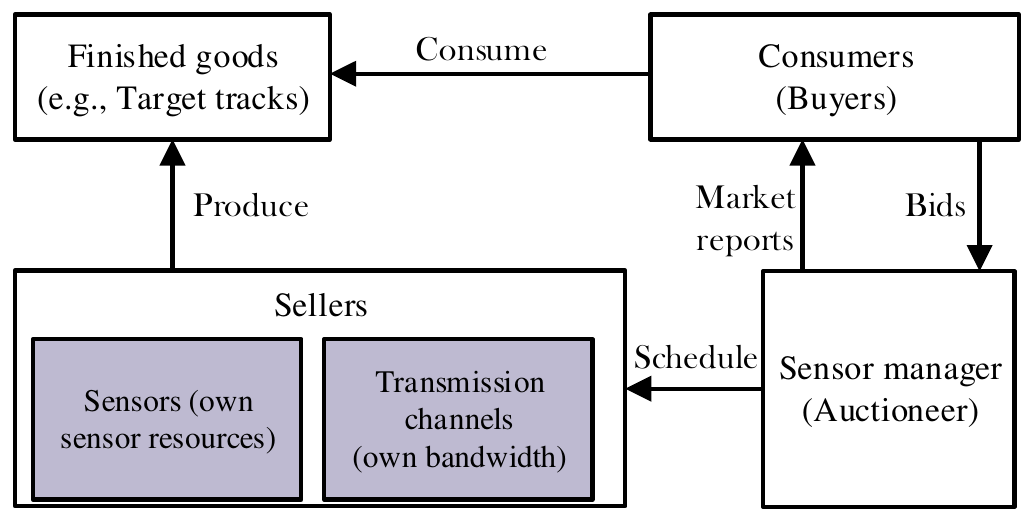}
 \caption{Market based architecture for multiple resource allocation.}
 \label{resource_combinatorial_auction}
\end{figure}

\cite{viswanath2005masm} is an example in which the combinatorial auction was adopted for the resource allocation problem in WSNs through proposing a market architecture. The model is shown in Fig.~\ref{resource_combinatorial_auction} where the sensor manager acts as an auctioneer. Individual sensors and transmission channels are represented as sellers to provide goods, e.g., area scans, and target tracks, consumed by goods requesters, i.e., buyers. First, the buyers submit their bids to the auctioneer involving a task description, e.g., the minimum task quality. Completing this task requires the combination of resources from the sensors and communication bandwidth. For each task in the consumers' bids, the sensor manager enumerates the possible allocations and uses the price quotes included in these bids to evaluate the clearing price for each allocation. Evaluating the allocation can be performed via an iterative procedure called \textit{tatonnement} process depending on the current usage rate and the availability of the resource, as described in \cite{avasarala2009market}. For finding the winner of the auction, unlike the auction schemes mentioned earlier, e.g., sealed-bid auctions, the winner determination of the combinatorial auction in \cite{avasarala2009market} is actually to find an optimal resource allocation. Due to computational complexity, different approaches were proposed for the winner determination. 

One of the winner determination approaches can be found in \cite{sandholm2003bob} in which the authors employed an anytime combinatorial search algorithm called CABOB. The approach finds a resource allocation so as to maximize the revenue of the auctioneer under the constraint in which each item, i.e., a resource, can be allocated to at most one bidder. Generally, by using a modified depth-first search on the space of possible bids, the CABOB is considered as one of the fastest exact winner determination algorithms. However, its complexity grows exponentially with the number of items and may not be applicable with a large number of bids (i.e., combinations of resources) to be allocated. Therefore, the authors in \cite{avasarala2006approximate} proposed
a new algorithm called Seeded Genetic Algorithm (SGA) which has polynomial time complexity in the number of bids. The SGA is a ranking-based 
representation scheme which is a combination of the genetic algorithm and a \textit{bid-ranking scheme} with the purpose of
producing only feasible allocations and finding high quality
solutions quickly. The simulation results indicated that compared with the information theoretic sensor manager algorithm in \cite{mcintyre1996information}, the proposed solution in \cite{avasarala2006approximate} achieves better resource allocation in terms of more number of destroyed targets. Although these results are promising, they only use a naive implementation of seeding. Investigating more sophisticated mechanisms that adds incentive compatible measures to approximate winner determination algorithms needs to be considered.


\subsubsection{Double auction}
\label{sec:Allocation_Double_auction}
\begin{figure*}[ht]
 \centering
\includegraphics[width=\linewidth]{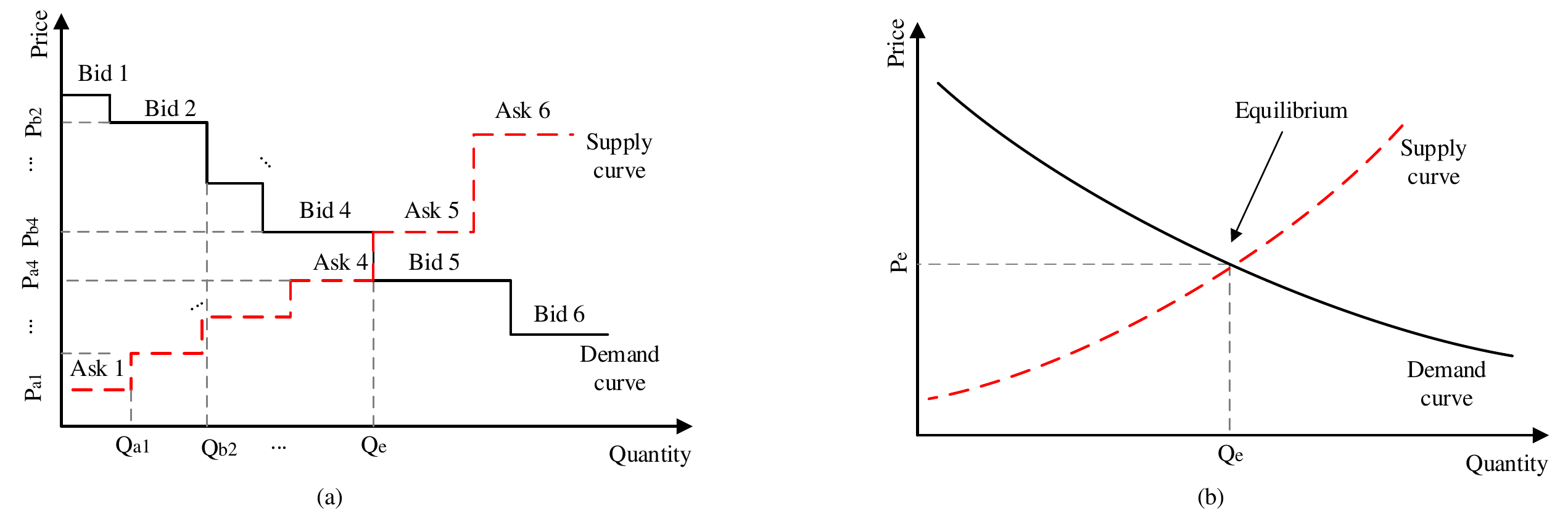}
 \caption{(a) Discrete supply and demand curves of double auction, and (b) continuous supply and demand curves from economics.}
 \label{Double_auction_resourc}
\end{figure*}

The scheme discussed in Section \ref{sec:Allocation_combinatorial_auction} aims at allocating the resources to different sensors, each of which executes an individual task. For multifunction sensor (e.g., multifunction MEMS (Micro-Electro-Mechanical Systems) sensors) which can perform multiple tasks, executing a task is assigned a
resource. However, since the demanded resources and the supplied resources for a task may not be equal, it is essential to exchange the resources between the tasks with the aim of maximizing the global performance. Accordingly, each task must determine how much the amount of allocated resources will be used. If there are some residual resources, the task can resell them to other tasks such that its benefit is maximized. Since a task may act as a resource buyer and a resource seller simultaneously, the double-sided auction can be used to allow multiple sellers and buyers to exchange their resources by submitting their asks and their bids to an auctioneer. The asks from sellers and the bids from buyers form the supply and demand curves, respectively as shown in Fig.~\ref{Double_auction_resourc}(a). The x-axis denotes the amount of supplied resources and y-axis denotes the ask or the bid prices. For example, on the supply curve, a seller asks to sell $Q_{a1}$ units of resources at price $P_{a1}$ (i.e., Bid 2) while on the demand curve, a buyer bids to buy $Q_{b2}$ units of resources at price $P_{b2}$ (i.e., Ask 1), and so on. It can be seen that Fig.~\ref{Double_auction_resourc}(a) is actually a discretized form of the supply and demand model which is shown in Fig.~\ref{Double_auction_resourc}(b). The intersection point of the supply and demand curves is the supply-demand equilibrium, i.e., the market equilibrium \cite{krishna2002auction}.    

Following this model, the authors in \cite{charlish2012multi} proposed the Continuous Double Auction Parameter Selection (CDAPS) scheme in which sensor tasks represented by agents exchange their resources via the double auction scheme. At any time, each agent submits a bid to buy and an ask to sell resources. The auctioneer decides a valid transaction for the resource exchange at which there exists the largest ask price $p_a$ and the lowest bid price $p_b$ with $p_a>p_b$, the transaction price, i.e., a clearing price, $p$ is set as $p=(p_b + p_a)/2$. From Fig.~\ref{Double_auction_resourc}(a), $p_a$ and $p_b$ are $P_{a4}$ and $P_{b4}$, respectively. The buyer obtains resources while the sellers receive payments according to the auction rules. This process is repeated to match each remaining pair of a buyer and a seller and define corresponding clearing prices. Theoretically, there exist multiple clearing prices. However, the auctioneer often selects one of them as the clearing price to avoid the complexity of the auction process. The competitive market equilibrium obtained by this auction scheme satisfies the Karush-Kuhn-Tucker conditions which guarantees that the algorithm converges to the optimal solution. With the target tracking resource allocation model described by Van Keuk \cite{van1993phased}, the simulation results demonstrated that the performance in terms of mean track utility of the CDAPS has a significant improvement over conventional rule based methods since it converges to the global optimum allocation. However, similar to the data allocation approach based on the supply and demand model in \cite{masazade2013market}, the global optimum allocation may not be stable.

\subsubsection{Non-cooperative game for auctions}
\label{sec:Allocation_non_cooperative_game}
The auction mechanisms are often used when the bidding strategies are unknown to the bidders. When the bidders have knowledge of bidding information of the others, a non-cooperative game can be used to determine a resource allocation equilibrium among the bidders \cite{an2007market}. The authors in \cite{yujie2011mobile} adopted the game model for the resource allocation between the players which are sub-applications or agents with different priorities. The function of each sub-application is to manage a task sequence, purchase resources from a resource sensor and assign the resource to its current task. When multiple agents simultaneously access the resources, a bidding process is conducted on the resource sensor to decide how to partition the resources. Accordingly, each agent submits its bid including the price that the agent is willing to pay. To determine how to bid for an agent, it is assumed that the payments of all other agents are constant, and each agent has knowledge of bidding strategy of other agents. The objective of the agent is to decide an optimal bidding strategy, i.e., a payment, to minimize its current task's execution time, constrained by its expenditure budget. The set of optimal bidding strategies yields a unique Nash equilibrium for resource allocation among interested agents at which no agent can gain an advantage by deviating its own payment strategy unilaterally \cite{bredin2003computational}. However, in a real market environment, one agent
cannot know how much others would like to pay for the resource since the agents have private
information and do not reveal their strategies. Therefore, a more efficient allocation algorithm named Bayesian Nash equilibrium allocation can be studied. 

\subsubsection{Demand and supply model}
\label{sec:Allocation_demand_supply}
 Demand and supply model is used to determine a market equilibrium which can also be obtained via the double auction scheme as discussed in Section \ref{sec:Allocation_Double_auction}. The authors in \cite{tiwari2011bit} adopted the demand and supply model for a rate allocation algorithm in wireless multimedia sensor networks. The model consists of a rate allocator which acts as a seller to assign the bit rate to the data sending sensors, i.e., buyers. Accordingly, in a time slot, the allocator broadcasts an initial resource price which can be taken from the previous time slot. Based on the price, each sending sensor calculates a resource demand with the aim of maximizing its average utility \cite{tiwari2009pricing}. If the total resource demand is not equal to the total resource supply, the allocator adjusts the price depending on the difference between them. The new price is announced to the sensors to recalculate their resource demand. This process is iterated until the total resource demand equals the total resource supply, i.e., a resource allocation equilibrium. Typically, several iterations of the price and demand messages would be exchanged between the allocator and the sensors, and thus the convergence speed is affected. However, as indicated in \cite{jian2015rate}, the allocation equilibrium can be obtained with only a single iteration by introducing a delay buffer with the size corresponding to the actual time delay. Accordingly, the delay buffer will store any extra bid rate from the sensor which is drained during those periods until the total resource demand is less than the total resource supply. The performance of the rate allocation algorithm was measured by the data quality in terms of peak-to-noise ratio. The simulation results showed that the peak signal-to-noise ratio of the proposed approach is much higher than that of the constant rate allocation \cite{lim2005text} which lacks any knowledge about the future demand. 

\subsubsection{Smart data pricing}
\label{sec:Congestion_resource}
The smart data pricing is employed to make users aware of high cost when consuming bandwidth during the peak demand periods. Adopting this policy to the WSNs, the authors in \cite{hui2009study} investigated a congestion price function at each relay sensor for the flow control, i.e., to manage the data transmission rate. Similar to the relay selection problem, the model here consists of a source, i.e., a buyer, which buys the data relay service from its neighboring sensors, i.e., sellers. There is a price regulation scheme at neighbors depending on the size of the forwarding data required by the source. When the source sends the data with a large size to a neighbor, the service price for the data relay increases. With the strategy of minimizing the cost function, the source will re-split the data traffic by adjusting the traffic at each neighbor which leads to a load balance among the neighbors. In addition to the load balance, a resource balance among the sensors is also achieved if the available resource, e.g., energy, is taken into account in the price function.

\subsubsection{Utility function}
\label{sec:Utility_resource}
This scheme aims at optimizing the resource allocation for each sensor through maximizing utility functions. Such approach often uses the NUM framework, i.e., the NUM problem, associated with the method of Lagrange multipliers with their interpretation
of resource prices as described in Section \ref{subsec:Utility_theory}. The authors in \cite{eswaran2007distributed} proposed a rate allocation algorithm in WSNs by using the NUM framework. Each source sensor is associated with a concave utility function depending on the source's transmission rate. Such concavity of the utility functions is to guarantee that the NUM problem has a unique optimal solution. Since the utility functions are known only to the sources, the NUM problem is decomposed into two optimization problems as defined similar to those in (\ref{NUM_USER}) and (\ref{NUM_NET}). The Lagrange multipliers are adopted with their interpretations as rates and resource prices that the source sensors are willing to pay. It was then proved that there always exists a unique optimal solution for the rates and the resource prices. The proposed algorithm was then implemented under a dynamic network where different sensors turn on and off at various time. The results showed that for each new state, the algorithm can immediately adapt to the dynamic of the system. 

The rate allocation algorithm in \cite{eswaran2007distributed} is only for unicast flows, meaning that each source transmits its data to a single sink. When multiple sinks consume data from the same source, i.e., multicast flows, the NUM framework needs to be modified. The authors in \cite{eswaran2008utility} extended the typical NUM framework to a more general WSN environment where individual sinks derive their utilities from a set of sources and intermediate nodes which use the link-layer multicast transmit scheme to forward data to multiple sinks. Based on the concavity property of utility function, each source sensor iteratively adapts its rate to reach an optimal solution.

Both \cite{eswaran2007distributed} and \cite{eswaran2008utility} use the gradient ascent method to find the optimal solution. Due to a small and constant step size, the convergence is slow. To achieve the optimal rates more quickly, the authors in \cite{eswaran2012utility} analyzed factors which have influences on the convergence speed. One of the critical factors is the step size which determines the magnitude of increase or decrease in rate at each iteration. If the value of step size is larger, the convergence of an algorithm is faster. However this also gives a rise to oscillations in system. Therefore, the step size needs to be set to guarantee the fast convergence and the stability. For this, the authors considered two methods to calculate the step size. The first method is Newton's method based on first-and second-order derivatives to compute the optimality of the objective function. The Newton's method is considered as a popular alternative to the gradient-approach because of its fast convergence. However, the convergences are not guaranteed if selecting an initial guess is too far from the exact root. The second method employs an Additive Increase Multiplicative Decrease (AIMD) \cite{rangwala2006interference} which takes into account the congestion level along the source's flow to sink nodes. Through evaluating the performance in terms of the utility versus time, the AIMD approach improves significantly the convergence speed compared with the Newton's method, especially in the dynamic environment (e.g., with frequent parameter changes in sources).

\begin{figure}[ht]
 \centering
\includegraphics[width=\linewidth, height = 4cm]{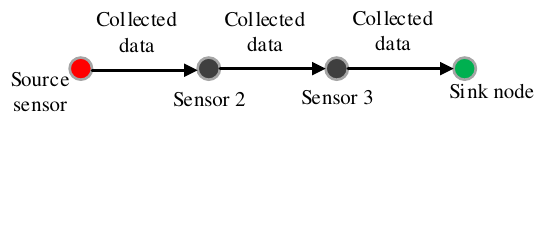}
\vspace{-1.5cm}
 \caption{An example of sensing data relay.}
 \label{resource_utility_function}
\end{figure}

In fact, there is always an inherent trade-off between the transmit rate and the consumed energy, e.g., a higher data rate requires greater sensing and transmit power. While the WSNs are energy-constrained, the energy allocation and the data transmission rate on each sensor are always considered together. Therefore, the authors in \cite{cheng2007price} and \cite{yang2009joint} proposed a joint power control and rate adaptation scheme for WSNs through using the NUM framework. The model can be illustrated by a simple example as shown in Fig.~\ref{resource_utility_function} in which the source sensor $s_1$ transmits its collected data to the sink node via relay sensors $s_2$ and $s_3$. Accordingly, the sensors, i.e., the source and the relay sensors, should allocate their relay power ratios properly so that the transmission of relaying traffic is guaranteed while still adjusting their data transmit rates so as to prolong their lifetime. Similar to \cite{eswaran2008utility}, the optimal solution for the power ratios and the transmit rates can be obtained via the NUM framework under the constraints of channel capacity and total energy consumed for relaying the collected data. By associating the NUM problem with the Lagrangian in which Lagrangian multipliers are the price per unit rate and the price per unit power/energy that a sensor needs to pay to the network, it was proved that there exists a unique solution for the rate and the power allocation. The simulation results illustrated that compared with the MaxUtility algorithm in \cite{nama2006optimal} which purely seeks for the utility maximization, the proposed algorithm requires less power for each link while achieving the better data rates. 

The same approach could be found in \cite{chen2009optimal}. However, the authors in \cite{chen2009optimal} considered more about the utility-lifetime tradeoff in WSNs by introducing of a weight between the rate and the consumed energy into the utility function of each sensor. The simulation results indicated that as the weight increases from 0 to 1, the total utility of the network also increases while the network lifetime decreases. The mechanism thus allows adjusting the weight according to the actual requirements and makes the network behave at a desired performance, i.e., utility maximization or lifetime maximization.

Similar to \cite{cheng2007price}, the authors in \cite{jin2007utility} studied the source rate optimization scheme taking into account the saturation of channel usage and energy consumption. However, the scheme is applied in heterogeneous sensor networks with several available routes or paths between the source and the sink node, i.e., the multipath routing. The goal is to adapt the path rates (and hence source rate) to maximize the utility achieved by each sensor. The authors introduced two price vectors associated with links (which is part of one path or route) and sensors. At each link, a link algorithm is executed to update the link price depending on the saturation of channel usage: if the aggregate path rate at a link exceeds the channel capacity, the link price will be increased. Similarly, each sensor updates its node price depending on the energy depletion. In particular, if energy consumption at a sensor exceeds the maximum energy allowed, the node price will be increased, and decreased otherwise. Both the link algorithm and node algorithm are iterative. Given these two prices, each source adopts the first-order Lagrangian algorithm to adapt its path rates and then calculate the source rate which is the sum of all path rates. It was also indicated that at the steady state, the source rate allocation of sensors is utility max-min fair and the global fairness is achieved. 

\subsection{Task Allocation}
\label{sec:Task_All}
In WSNs, task allocation schemes aim at assigning
tasks (e.g., target track and area scan) to specific sensors for
execution. Given limited resources of WSNs, static task assignment schemes may not meet an objective due to the lack of interactions between the elements of the network. Therefore, the dynamic task allocation schemes with negotiation mechanisms among elements in the network were developed to optimize the task allocation and resource utilization. This section reviews different approaches based on market models. 
\subsubsection{First-price sealed-bid auction}
\label{sec:Task_first_sealed_auction}
The authors in \cite{schrage2006market} designed a task allocation scheme by employing the first-price sealed-bid auction, called the \textit{Rust-In-Time (JIT) market}. The task allocation problem is modeled as a multi-agent negotiation including: sensor agents, i.e., sellers, which perform tasks, external systems known as buyers who may require data from the sensor network, a market entity, i.e., an auctioneer, to assign tasks to sensors, and resource agents to provide resources for sensors. First, buyers submit their bids including task requests and corresponding prices to the auctioneer. Then, these bids are broadcast to the sellers. Given the offered bid, each seller calculates its own utility based on its current capabilities to determine whether it can perform the task. If the sensor can perform the task, the task will be assigned accordingly. The auctioneer may terminate the auction to avoid the delay for real-time applications. This negotiation protocol provides the dynamic adaptation to network changes. However, the function of the auctioneer is only to broadcast the bids to the sellers, and thus this entity can be removed to reduce the system complexity. 

As indicated in \cite{gangadharaiahsoft}, instead of having a dedicated entity for an auctioneer, the seller (also the auctioneer) is a sensor selling its task that it cannot accomplish (e.g., due to the limited energy), to its neighboring sensors, i.e., buyers, who can perform the task. The seller first broadcasts the assigned task to the buyers and then based on the remaining energy, the buyers determine and submit the bid prices to the seller. The seller will select a winning buyer if the buyer's bidding price is higher than a predefined threshold. Similar to \cite{schrage2006market}, to get the highest bidding price, the seller waits for a deadline which is set according to the application requirement to decide the winner. Since the price is proportional to the remaining energy, the proposed strategy maintains the energy balance among the sensors in the network. Performance analysis showed a good balance obtained at the threshold of 0.6, but the authors did not indicate how to obtain this value. 

In fact, the sensing data is only useful to the auctioneer when the sensors are far enough from each other. Similar to  \cite{jaimes2012location}, the authors in \cite{zhu2013stamp} considered the distance among the sellers in a participatory sensing when assigning a task. The task allocation is modeled as the first-price sealed-bid auction in which the service aggregator, i.e., auctioneer, assigns the sensing task to more than one user, i.e., buyer. The buyers submit their sealed bids simultaneously to the auctioneer. To represent the conflict among buyers, the auctioneer uses graph composed of a set of vertices and a set of edges. Each vertex represents a buyer in the graph, and there is an edge between a pair of buyers, if their geographic distance is smaller than a threshold value. The auctioneer then allocates the task to the buyers
using the best known maximum independent set approximation
algorithm for general bounded degree graphs \cite{demange1997improved} so that
the buyers who are connected in graph are not allocated the task concurrently. Therefore, the proposed solution achieves an allocation efficiency as high as the best ever known maximum independent set algorithm in addition to the strategy-proofness property. In some cases, the users arrive sequentially in the geographic area, online auctions are efficient alternatives for the the sealed-bid auction.

\begin{figure}[ht]
 \centering
\includegraphics[width=\linewidth]{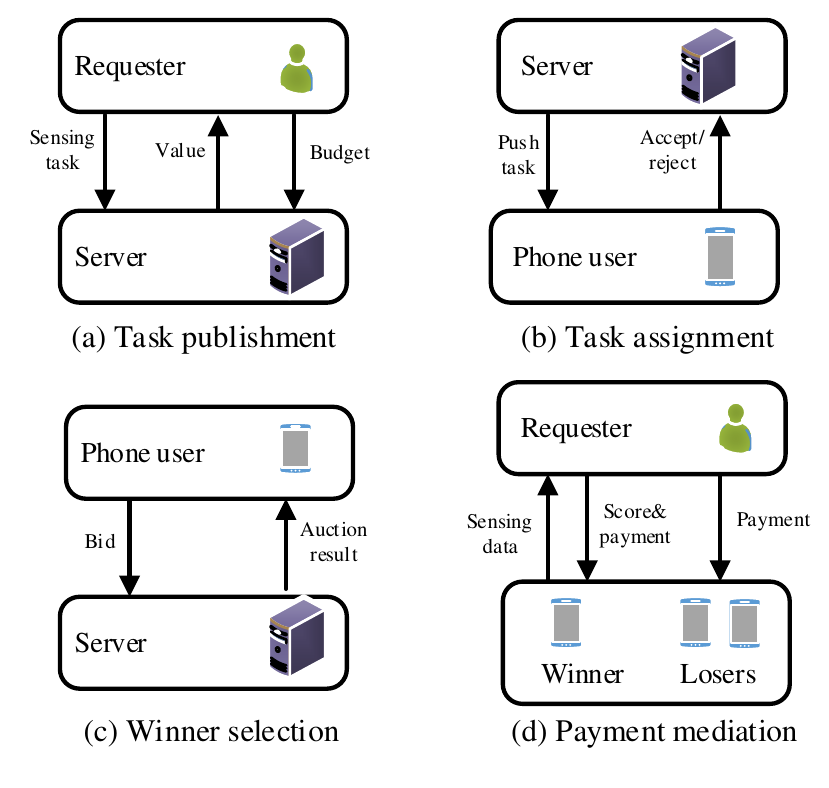}
 \caption{The multiple interactions among the three elements in the stages.}
 \label{task_allocation_reverse_auction}
\end{figure}
\subsubsection{First-price sealed-bid reverse auction}
\label{sec:Task_seal_reverse_auction}
This section discusses task allocation schemes based on the sealed-bid auction mechanism as described in Section \ref{sec:Task_first_sealed_auction}, but the roles of the buyers and the sellers are reversed. The objective of the reverse auction is to achieve a fair energy balance among sensors \cite{edalat2009price}, \cite{edalat2012auction} or high data quality \cite{nancross}. The model in \cite{edalat2009price} consists of an 
allocator acting as a buyer (or an auctioneer) and sensors (sellers) performing the required tasks. Accordingly, once receiving a required task from a buyer, the auctioneer broadcasts the task description (e.g., size and deadline of the task) to all sellers. Each seller calculates its own asking price for completing the task based on its
current status of available energy, communication cost, task deadline and resource release time. In particular, the asking price is inversely
proportional to the energy level. In particular, the seller who has higher remaining energy has the lowest asking price, and thus it may have more opportunities to be selected. To determine a seller with lowest asking price, i.e., the winner, centralized or distributed schemes can be executed. The former requires all sellers to submit their asking prices simultaneously, and the seller with the lowest price is selected for executing the task. In the distributed scheme, each seller sets a \textit{waiting time} value in proportion to its calculated price. The seller with the lowest price, i.e., the shortest waiting time, submits its asking price to the buyer before any other sellers, and it is selected as the winner. Other sellers receiving a winner determining message from the buyer will leave the auction without sending their asks to the buyer. The objective is to reduce the communication overhead and energy consumption for sending the non-winning messages to the buyer. However, the non-winning sensors still consume energy to listen to the winner determining message. Therefore, the authors in \cite{edalat2012auction} added \textit{waiting time reduction} phase which allows the sellers to compare their asking prices with a broadcast budget value from the buyer. If  the bidder's asking price exceeds
the budget, that bidder leaves the competition and switches sleep mode without waiting for the winner determining message. The simulation results indicated that the reverse auction based scheme balances the remaining energy for all the sensors and lowers the energy consumption compared with the static task allocation method, e.g., Energy Balance-Critical Node Path Tree (EB-CNPT)~\cite{tian2006dynamic}. However, the task allocation approaches in \cite{edalat2009price} or \cite{edalat2012auction} aims to enhance the performance of independent applications.

In the context of mobile crowdsensing, the authors in \cite{nancross} employed the reverse auction scheme to assign tasks to users. This scenario has three entities: 1) the task requester who posts a sensing task and acts as a buyer, 2) users, i.e., sellers, and 3) the server, i.e., the auctioneer, which is responsible for participant selection, task price evaluation, and payment mediation. The interactions among three entities are shown in Fig.~\ref{task_allocation_reverse_auction} which involve four stages: task publishment, task
assignment, winner selection, and payment mediation. Briefly, when the requester publishes the sensing task description, the server evaluates the task price according to its sensing region and sensing period. For example, when a task within a region is easy to be executed, its value (i.e., price) is low and vice versa, if the task within a square is difficult to be performed, which implies a high task value. This value is then sent back to the requester for suggesting a budget. Based on the budget, the requester selects phone users who satisfy the budgetary constraints and best match the requested sensing context for the task. The selected phone users then submit their asks for the requested sensing task, and the server selects the winner based on their asking price and reputation. In particular, the reputation of a phone user is a reflection of the quality about the historical sensed data that the phone user submitted to the server.  Both winner and losers receive reward, and moreover reputation of the winner is updated based on its historical data submissions. Although experiments showed that the proposed mechanism obtains a dynamic budget, optimal task allocation and high motivation for phone users to participate, the factors that have impact on phone user participation, such as user preference and privacy protection have been not considered. To tackle this problem, the authors in \cite{shahprofit} investigated the payment scheme to provide satisfying reward to the phone users. Accordingly, each seller is paid an amount of reward which is the highest asking price that the seller can submit to win the auction while contributing to the profit of the server. 

However, submitting the asking price which reflects the true costs of performing a task to the requester may
be challenging in real world settings. Indeed, the users may typically not reveal their true costs to get more reward or the costs may be difficult to determine. Instead of soliciting the users' costs, the authors in \cite{singla2013truthful} proposed to use the posted price model based on the link between the reverse auctions and the multi-armed bandits \cite{blum2007learning} which allows participants to express their reservation prices plus a profit. The selected winners
are rewarded with an amount computed by a $k$-armed bandit
system instead of their original bid prices as mentioned in \cite{shahprofit}. The idea behind this mechanism is to learn the reservation price curve for each participant. The prices are discretized
by creating a set of $k$ price arms, and at each time step, the
mechanism picks an arm based on some optimization criteria and user feedbacks, i.e., accept or reject the offer. After several
time steps, the mechanism may use a policy that balances the
exploitation, e.g., greedy policy, and the exploration to
reach a maximum average reward. The mechanism continues until
the budget is exhausted. Theoretical analysis showed that the proposed approach is budget feasible and incentive compatible (truthful). Moreover, the requester achieves a near-optimal utility. However, running the proposed mechanism in real time on crowdsourcing platforms to confirm these properties is still open for future studies.  

\subsubsection{Double auction}
\label{sec:Task_double_auction}
Compared with the single-sided auction schemes, double auction is more suitable for resource allocation problems with fairness and allocation efficiency \cite{sun2014sprite}. Indeed, in traditional one-to-many single-sided auction style, e.g., forward or reverse auctions, the centralized control belongs to the seller or the buyer who has the rights to establish the rule of transactions. This results in the collusion or the market manipulation problem. In the double auction procedure, both groups of the buyers and the sellers lose their relative dominant positions, and their relationship follows the supply and demand theory as described in Section \ref{sec:Allocation_Double_auction}.

The authors in \cite{chen2013sparc} proposed a strategy-proof (i.e., truthfulness) double auction for the sensing task scheduling in mobile participatory sensing systems. In this model, multiple sensing tasks from users which act as buyers will compete for the sensing times (i.e., items) from a large number of smartphone users, i.e., sellers. Also, the smartphone users compete to sell the sensing times to the sensing tasks. A trustworthy server that resides in the cloud acts as an auctioneer in this model. Each sensing task submits its bid including sensing time price and the sensing time demand to the server. At the same time, each smartphone submits its ask involving the sensing time price per unit and the limited sensing time. The server then sorts the smartphone users in a non-decreasing order by the bid prices and the sensing tasks in a non-increasing order by the ask prices. Then, the server finds the largest indexes of bid and ask prices in these two sets at which the first smartphone has sufficient sensing time to satisfy the demands of the first sensing task without losing any profit in the auction. The winners will be the first smartphone user and the first sensing task. The authors also proved theoretically that the given mechanism is strategy-proof since it possesses the property of individual-rationality and incentive-compatibility. The evaluation results indicated that the social welfare obtained with this algorithm linearly increases with the number of smartphone users and the number of sensing tasks. However, to adapt to the dynamic change of the number of crowdsourcing tasks and smartphones, the use of online double auctions is open for further study.

\begin{figure}
 \centering
\includegraphics[width=7cm,height=5.2cm]{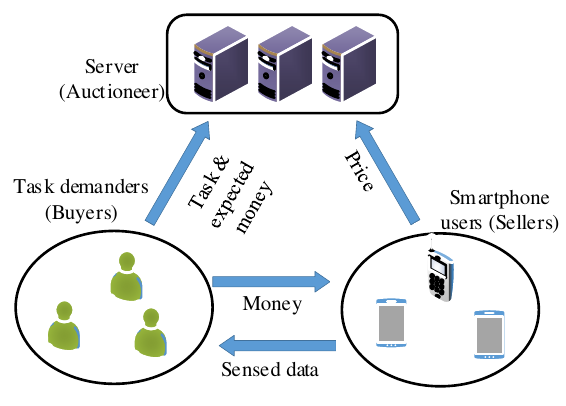}
 \caption{Double auction based task allocation in crowdsourcing system.}
 \label{task_allocation_double_auction}
\end{figure}

In \cite{xu2013data}, the authors also adopted double auction, but they considered a crowdsourcing system as illustrated in Fig.~\ref{task_allocation_double_auction}. The task requesters act as buyers who purchase the sensed data, i.e., items, from the users, i.e., sellers. In addition to the strategy-proof property, another economic property with ex-post budget balance is considered. A double auction is ex-post budget balance if the auctioneer's utility is no less than zero. Basically, the stages, i.e., the asks and bids sorting procedure, the task requesters and smartphone users matching procedure, are implemented similarly to those in \cite{chen2013sparc}. However, to maintain the ex-post budget balance, in the payment stage, the secondary price clearing rule of the second-price auction scheme is employed. Accordingly, the auctioneer, i.e., the server, pays each winning smartphone user by the bid of losing user with the highest bid and charges each winning task demander via the use of the uniform bidding price. The simulation results showed that for a given number of users, both the charge for task requesters and the payment to smartphone users are increased when the number of task requesters increases, thus improving the ex-post budget balance of the auction. 

\subsubsection{Combinatorial reverse auction}
\label{sec:Task_combinatorial_reverse_auction}
\begin{figure}[ht]
 \centering
\includegraphics[width=6cm,height=7.7cm]{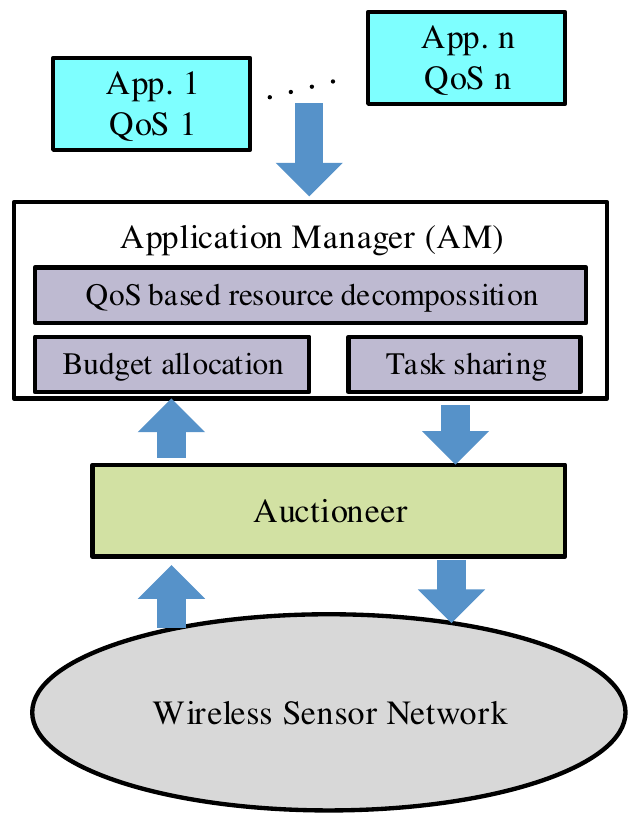}
 \caption{Market based architecture for multiple task allocation.}
 \label{Task_combinatorial_auction}
\end{figure}

For multiple concurrent applications in sensor networks (e.g., temperature monitoring, security
alarms, and light control), some common tasks across different applications can be shared to reduce the deployment and administrative costs, and increase the usability and efficiency of resources. However, the problem is how to identify the common tasks. If we consider a combination of the tasks is a bundle of items, a combinatorial auction is a suitable mechanism to identify the best bundle of items. The authors in \cite{edalat2011combinatorial} addressed the task allocation and sharing in multi-application WSNs via the reverse combinatorial auction. The model is illustrated in Fig.~\ref{Task_combinatorial_auction}, where the sensors are either bidders or resource sellers, and the auctioneer is an intermediate entity to distribute task messages from the application manager. Each application may consist of several tasks such as target surveillance, target positioning, and data acquisition. The objective is to maximize the network lifetime by sharing tasks and network resources between applications while improving the overall QoS of each application. Accordingly, after receiving a task message from the auctioneer, the sensors calculate and submit their asks including cost values for subsets of the applications' tasks in terms of available resources, e.g., energy and CPU cycle. 

Since the combinatorial reverse auction problem is a NP-complete, a heuristic two-phase Winner Determination Protocol (WDP) was proposed. The first phase employs a local decision making strategy to eliminate the number of asks with lower probability of winning. This reduces the message exchange overhead, the energy consumption, and the delay for winner determination. In the second phase, the suboptimal subsets are selected through an ordering heuristic algorithm as proposed in \cite{gonen2000optimal}. Simulations were implemented to evaluate the system efficiency and scalability when the number of concurrent applications and network size increase. The results illustrated that there is a significant difference in terms of energy consumption when the tasks are shared compared with the non-sharing case. Moreover, the proposed task allocation scheme outperforms the static task allocation method in~\cite{tian2006dynamic}, i.e., EB-CNPT, in balancing the energy in the network since the energy availability is considered in each epoch of the task allocation.

In the centralized WDP used in \cite{edalat2011combinatorial}, sensors are required to submit their asks simultaneously. This increases overhead of message exchanges with high energy consumption. Therefore, the authors in \cite{edalat2012auctionallocation} introduced the Energy and Delay Efficient Distributed Winner Determination Protocol (ED-WDP). Through the use of bid formation, ED-WDP reduces the message exchange overhead, energy consumption, and delay for the winner determination compared to the WDP. 

Different from \cite{edalat2011combinatorial}, the authors in \cite{khan2012resource} adopted a combinatorial auction for the task allocation with the aim of balancing the energy consumption among sensors. Specifically, when an object enters the field of view of a sensor, due to the energy depletion, the sensor, i.e., the seller, may sell the object tracking task to its neighboring sensors, i.e., buyers, through the combinatorial auction. Each neighboring sensor calculates its bid value which is a function of a bundle of attributes like the signal strength, the remaining resources, and the distance between the target and the neighboring sensor. In particular, the bid value of a neighboring sensor is high if its remaining resources are high and the distance is short. Among the neighbors, the seller selects a sensor with the highest bid value as the winner to perform the task. Through simulations, it was shown that the proposed approach achieves better energy balance among the sensors comparing with the static and random task schedules. However, the scenario used in the proposed approach is simple with a static object. In a dynamic environment with the movement of the object, more tracking tasks are required to be assigned to the sensors. 

\subsubsection{Option pricing}
\label{sec:Task_option_pricing}
For the task allocation problem, when an auctioneer accepts multiple task requests while there are limited sensors to perform, it needs to determine the values of the tasks for scheduling purpose. Evaluating the tasks can be implemented by auctions such as the first-price sealed-bid auction as mentioned in Section \ref{sec:Task_first_sealed_auction}. Accordingly, the value of a task is determined based on auction bids which is also known as a market price of the task \cite{nguyen2015system}. However, due to the variability of the value as well as the number of bids, the market price may not indicate the actual value of the task. Therefore, options pricing models based on the real option theory can be used to offer a robust method for
task valuation and scheduling by considering the time value of a dynamic task and the risk involved in the selection of an option for the requested task. Such options may include whether to abandon a
task (put), delay a task, and/or schedule a task (call). The most common model used to determine the price, i.e., the strike price, of an option, is the Black-Scholes model as shown in (\ref{Option_equ}). The authors in \cite{nguyen2012system} investigated a fuzzy real option approach for determining the market price of a sensor task for the task schedule. The approach called fuzzy cognitive map is a combination of the option theory and a fuzzy logic. Accordingly, the strike price is determined using the Black-Sholes pricing formula in (\ref{Option_equ}). In this case, the traditional financial variables, e.g., the price of the underlying security and the exercise price, are substituted by the mission variables, e.g., task priority and task duration, via the fuzzy logic. The obtained strike price for a
task is then used to decide whether a particular task is for the sensing task schedule. Nevertheless, experiment evaluations were not given in \cite{nguyen2012system}. 

\subsubsection{Discrimination pricing}
\label{sec:Task_discrimination_pricing}
For the price discrimination scheme, the price of  the same product is set differently at different time. Therefore, it can be used for managing resource demands through adjusting the resource price per unit at different time. In \cite{pletzer2010distributed}, the authors employed this scheme for allocating resources to execute tasks in visual sensor networks. The resource allocation problem is modeled as a multi-agent system in which the resource providers, i.e., sellers, sell resources and communication links to the tasks, i.e., buyers. Since the resource demand of the tasks strongly depends on the price of resource, two pricing models are considered. The first one, called "adjusted linear pricing", updates the resource price per unit based on the average of interest level (i.e., willingness to pay) of executed tasks and the number of allocated/requested resources. The second one, called "rate adaptive pricing", adjusts the resource price per unit in the next time slot based on the price set in the "adjusted linear pricing" phase and the current allocation rate (i.e., allocation level). The updated price is then used to control the behaviors of resource utilization of the tasks. Through adjusting the price for resources depending on their supply and demand relation, it was shown that the QoS level of the executed tasks is adapted efficiently according to the user-defined interest levels. However, simulation results for different scenarios with different interest level distributions did not show in this work. Moreover, non-linear price functions need to be considered in the first pricing model.

\textbf{Summary:} In this section, we have identified two major
issues in sensing networks and reviewed
applications of economic and pricing models for these issues (i.e., resource allocation and task allocation). We
summarize the issues along with references in Table \ref{table_sec4_resource_sum} for the resource allocation and Table \ref{table_sec4_task_sum} for the task allocation. From the two tables, we observe that the resource allocation are mainly solved by the utility function-based pricing model while for the task allocation problem, the first-price sealed-bid auction is widely used. Sensing coverage and target tracking are other
important issues in the development of sensing networks. In the following
section, we review the existing literature on pricing models-based
sensing coverage and target tracking strategies in WSNs.  

\begin{table*}
\caption{Applications of economic and pricing models for resource and power allocation}
\label{table_sec4_resource_sum}
\scriptsize 
\begin{centering}
\begin{tabular}{|>{\centering\arraybackslash}m{0.2cm}|>{\centering\arraybackslash}m{0.4cm}|>{\centering\arraybackslash}m{1.4cm}|>{\centering\arraybackslash}m{1.3cm}|>{\centering\arraybackslash}m{1.1cm}|>{\centering\arraybackslash}m{1.2cm}|>{\centering\arraybackslash}m{5cm}|>{\centering\arraybackslash}m{2.2cm}|>{\centering\arraybackslash}m{1.3cm}|}
\hline
\multirow{2}{*}  {\textbf{}} & \multirow{2}{*}  {\textbf{Ref.}} &  \multirow{2}{*}  {\textbf{Pricing model}}  & \multicolumn{3}{c|} {\textbf{Market structure}} & \multirow{2}{*}  {\textbf{Mechanism}} & \multirow{2}{*}  {\textbf{Objective}} & \multirow{2}{*} {\textbf{Solution}} \tabularnewline 
\cline{4-6}
 & & & \textbf{Seller} & \textbf{Buyer} & \textbf{Item}  & & &\tabularnewline
\hline
\hline
\parbox[t]{2mm}{\multirow{9}{*}{\rotatebox[origin=c]{90}{ \hspace{-9cm} Resource allocation and energy control}}}
&\cite{viswanath2005masm} \cite{avasarala2009market} \cite{avasarala2006approximate}& Combinatorial auction & Individual  sensors and transmission channels&Consumers & Area scans and target tracks & Buyers submit their bids to the auctioneer including task requirements. The auctioneer enumerates all possible resource allocations and find an optimal solution as the winner & QoS guarantee, and resource usage optimization&Optional solution\tabularnewline \cline{2-9}
& \cite{charlish2012multi} & Double auction& Sensor tasks & Sensor tasks & Resources &Buyers and sellers submit respectively their bids and asks to buy and sell the resources. The auctioneer decides the valid transaction for the resource exchange according to the supply-demand theory&Global optimum allocation& Market equilibrium\tabularnewline \cline{2-9}
& \cite{yujie2011mobile} \cite{bredin2003computational} &Non-cooperative game& Resource node  &Sub-applications &Resources  & Buyers submit their bid prices to the seller, and their strategies are to optimize theirs payments. The set of optimal bidding strategies yields a unique Nash equilibrium for resource allocation among interested buyers&Payment optimization of buyers, and efficient and balanced resource allocation & Nash equilibrium \tabularnewline \cline{2-9}
& \cite{tiwari2011bit} \cite{jian2015rate}& Demand and supply model &Rate allocator & Data sending sensors & Data rate & Buyers adjust their demands according to the prices from the seller. This process is iterated until total demand equals total supply which yields the marker clearing price & Data quality improvement & Market equilibrium\tabularnewline \cline{2-9}
& \cite{hui2009study} & Smart data pricing & Neighboring sensors& Source sensor & Transmission rate & Each seller builds own congestion price function of the data traffic size. When the source node sends the data with large size, the charge for the data forward is high & Load balancing& Market equilibrium\tabularnewline \cline{2-9}
& \cite{eswaran2007distributed} \cite{eswaran2008utility}& Utility function & Network &Source sensors &Transmission rate & Each buyer is associated with a concave utility function depending on its transmission rate and price per unit flow it must pay. Based on the NUM framework, a unique optimal solution for rate and price is determined &Optimal and proportionally-fair rate allocation & Optimal solution\tabularnewline \cline{2-9}
&\cite{eswaran2012utility} & Utility function  & Network& Source sensors & Transmission rate& Same as \cite{eswaran2007distributed} but the Newton's method and AIMD are introduced to improve the algorithm's convergence & Optimal and proportionally-fair rate allocation, convergence and stability improvement &Optimal solution\tabularnewline \cline{2-9}
&\cite{cheng2007price} \cite{yang2009joint} & Utility function & Network and relay sensors  & Source sensors& Transmission rate and energy & Same as \cite{eswaran2007distributed} but the price per unit energy is included in the utility function in addition to the price per unit flow & Optimal rate and energy allocation &Optimal solution\tabularnewline \cline{2-9}

& \cite{chen2009optimal}& Utility function& Network and relay sensors & Source sensors& Transmission rate and energy &Same as \cite{cheng2007price} but a weight between the allocated rate and the allocated energy is included in the utility function& Optimal utility-energy tradeoff & Optimal solution\tabularnewline \cline{2-9}

& \cite{jin2007utility} & Utility function&Network & Source sensors& Transmission rate&Same as \cite{chen2009optimal} but the saturation of channel usage and energy consumption of routes in networks are introduced & Rate optimization, energy consumption balance, and utility max-min fairness for resource allocation& Optimal solution\tabularnewline \cline{2-9}

\hline
\end{tabular}
\par\end{centering}
\end{table*}

\begin{table*}
\caption{Applications of economic and pricing models for task allocation}
\label{table_sec4_task_sum}
\scriptsize 
\begin{centering}
\begin{tabular}{|>{\centering\arraybackslash}m{0.2cm}|>{\centering\arraybackslash}m{0.4cm}|>{\centering\arraybackslash}m{1.5cm}|>{\centering\arraybackslash}m{0.9cm}|>{\centering\arraybackslash}m{1.1cm}|>{\centering\arraybackslash}m{1.4cm}|>{\centering\arraybackslash}m{5cm}|>{\centering\arraybackslash}m{2.2cm}|>{\centering\arraybackslash}m{1.2cm}|}
\hline
\multirow{2}{*}  {\textbf{}} & \multirow{2}{*}  {\textbf{Ref.}} &  \multirow{2}{*}  {\textbf{Pricing model}}  & \multicolumn{3}{c|} {\textbf{Market structure}} & \multirow{2}{*}  {\textbf{Mechanism}} & \multirow{2}{*}  {\textbf{Objective}} & \multirow{2}{*} {\textbf{Solution}} \tabularnewline 
\cline{4-6}
 & & & \textbf{Seller} & \textbf{Buyer} & \textbf{Item}  & & &\tabularnewline
\hline
\hline
\parbox[t]{2mm}{\multirow{9}{*}{\rotatebox[origin=c]{90}{\hspace{-9cm} Task allocation}}}
& \cite{schrage2006market} &First-price sealed-bid auction & Sensor & External systems & Task& Buyers submit bid prices, the seller accepts a task with the highest price&Utility maximization of seller& Competitive equilibrium\tabularnewline \cline{2-9}
&\cite{gangadharaiahsoft} & First-price sealed-bid auction &  Sensor & Neighboring nodes & Task &Same as \cite{schrage2006market} but the seller accepts a task from the first arrival bid& Energy balance& Competitive equilibrium \tabularnewline \cline{2-9} 
&\cite{zhu2013stamp}& First-price sealed-bid auction & Aggregator & Phone users & Sensing task &Buyers submit their bids. The auctioneer assigns the task to the buyers using the best known maximum independent set approximation
algorithm for general bounded degree graphs &Allocation efficiency and strategy-proof & Competitive equilibrium \tabularnewline \cline{2-9} 
&\cite{edalat2009price} \cite{edalat2012auction} & First-price sealed-bid reverse auction&Sensors & Allocator& Task&Sellers submit their asking prices. The task is assigned to a seller with the lowest asking price & Energy balance & Nash equilibrium\tabularnewline \cline{2-9}
&\cite{nancross} & First-price sealed-bid reverse auction& Phone users & Task requester& Sensing task&Same as \cite{edalat2009price} but the task is assigned to a seller with the low asking price and high reputation& High data quality & Nash equilibrium\tabularnewline \cline{2-9}
&\cite{shahprofit} & First-price sealed-bid reverse auction& Phone users & Server& Sensing data&Sellers submit their asks to the server to perform a subset of tasks, the buyer selects the sellers who are most likely to increase its profit by using the solution of a linear programming & High data quality, individual rationality, computational efficiency, and truthfulness & Optimal solution\tabularnewline \cline{2-9}
& \cite{chen2013sparc} & Double auction & Phone users & Sensing tasks&Multiple sensing times & Sellers and buyers, respectively, submit their asking prices and biding prices. Matching a buyer and a seller is based on the strategy-proof double rule&Good performance with social welfare and task satisfaction ratio &Market equilibrium \tabularnewline \cline{2-9}
& \cite{xu2013data} &Double auction & Requesters & Phone users& Sensed data&Same as \cite{chen2013sparc} but a payment stage based on the secondary price clearing is added& Maximized revenue of the winning buyers&Market equilibrium \tabularnewline \cline{2-9}
&\cite{edalat2011combinatorial} \cite{edalat2012auctionallocation} & Combinatorial reverse auction & Sensors& Sensor &A subset of applications' tasks &Sellers submit their asking prices. An ordering heuristic algorithm is employed to select the winner owning a subset of tasks which has the lowest price and covers the entire applications & Network lifetime maximization, and overall QoS enhancement& Optimal solution \tabularnewline \cline{2-9}
&\cite{khan2012resource} & Combinatorial auction &Sensor &Neighboring sensors &A target tracking task &Buyers submit their bundles including remaining resources, distances between them and the target. The seller selects a buyer with the highest bundle value as the winner& Network lifetime maximization& Optimal solution \tabularnewline \cline{2-9}
& \cite{nguyen2012system} & Options pricing & Task requesters &Sensor & Tasks & Black-Sholes pricing formula and a fuzzy logic are employed & Effective schedule & Real value of task\tabularnewline \cline{2-9}
&\cite{pletzer2010distributed}  & Price discrimination & Resource providers& Tasks  & Resources and communication links & Resource price is adjusted according to the willingness to pay of buyers through an exponential running average&  Resource utilization maximization, and QoS level adaptation & Pareto efficiency \tabularnewline \cline{2-9}
\hline
\end{tabular}
\par\end{centering}
\end{table*}

\section{Applications of pricing models for sensing coverage and target tracking}
\label{sec:Sensing_cover}
Sensing coverage is an important metric in many WSNs. It measures how well the sensors observe the physical phenomenon in an environment where they are deployed. A number of works studied and developed schemes to maximize the sensing coverage \cite{wang2009survey}. Classical approaches are based on centralized structure that can provide globally optimal solutions given resource constraints. However, they suffer from high computation cost and communication overhead, e.g., $O(n^3)$ for the Hungarian method \cite{kuhn1955hungarian}, where $n$ is the number of sensors. More importantly, the centralized schemes have a serious scalability issue. Thus, market based solutions, e.g., auction, were proposed, which are shown to achieve similar results to the optimal solutions. However, the computation cost is significantly reduced. Three types of sensing coverage are typically considered \cite{wang2009survey}.

\begin{itemize}
\item{\textit{Area coverage}:} Area coverage measures the area of sensing field that is covered, i.e., the collection of all space points within the sensing field. Static sensors can be deployed. Then, the uncovered area will be healed by selecting appropriate mobile sensors via pricing models. 
\item{\textit{Target coverage}:} Target coverage mainly deals with
how to cover a set of discrete targets (some space points) given
known locations. The requirement is to increase the sensing range of each sensor to cover interested targets. However, energy consumption is a critical constraint, and pricing models can be applied to address this problem. 
\item{\textit{Barrier coverage}:} Barrier coverage concerns finding a penetration path across a sensor field with the aim of detecting any intruders that attempt to cross the field. In this case, using pricing mechanisms may address key challenges of the barrier coverage, e.g, scalability and energy consumption. 
\end{itemize}

\subsection{Area Coverage}
\label{sec:Area_coverage}
\begin{figure}[ht]
 \centering
\includegraphics[width=7.4cm, height=6cm]{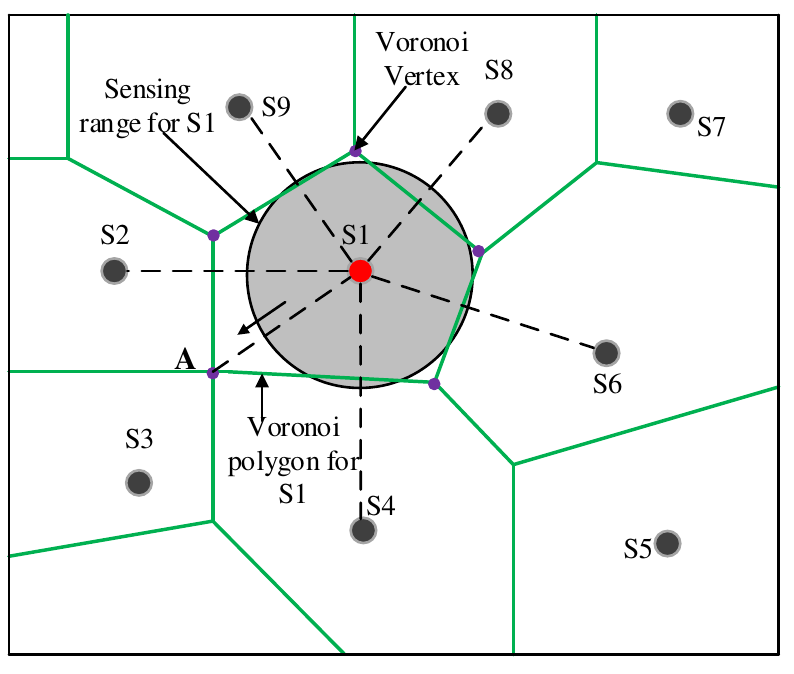}
 \caption{Illustration of using Voronoi diagram to detect a coverage hole and decide
the hole size.}
 \label{Voronoi_diagram}
\end{figure}

The authors in \cite{wang2003bidding} addressed the area coverage
problem by using the sealed-bid auction. The sensing area is covered partly by static sensors. To cover the rest of area, mobile sensors can be used. Specifically, the static sensor as a buyer buys the service from mobile sensor as a seller to move and perform sensing tasks in the uncovered area. The static sensors detect their local coverage holes by using Voronoi diagrams as in \cite{wang2006movement}. Accordingly, every static sensor forms a Voronoi polygon with respect to the position of its neighboring sensors as shown in Fig.~\ref{Voronoi_diagram}. The part of the polygon that lies outside its sensing range (i.e., disk coverage zone
centered at the sensor) is not covered by the sensor. Typically, a static sensor chooses the farthest vertex in its Voronoi polygon as the target location of the coming mobile sensor. For example in Fig.~\ref{Voronoi_diagram}, the location which needs to be healed for the sensor S1 is the vertex A. This is to avoid overlapping the coverage area between the static sensor and the mobile sensor and thus to guarantee that the achievable coverage is the highest. After that the static sensors calculate their bid prices based on the estimated sizes of the
holes they detect. The mobile sensors also calculate their own base prices based on the sizes of coverage holes formed at their current positions in the case that the mobile sensors decide to leave these positions. After receiving the bid prices from the static sensors, each mobile sensor decides to move to cover the hole if the highest price has a coverage hole size greater than the new hole size (i.e., base price) generated in its current location due to its movement. The accepted bid price will become the new base price of the mobile sensor. The bidding protocol repeats until no static sensor can give a bid price higher than the base price of any mobile sensor. Performance evaluation indicated that compared to the sensor random deployment algorithm, i.e., all sensors are static and VEC algorithm \cite{wang2006movement}, i.e., all sensors are mobile, the proposed algorithm can significantly reduce the number of sensors required to reach a certain coverage requirement. For example, when 10\% of the sensors are mobile, the required number of sensors is reduced by 30\%.   

Although the approach in \cite{wang2003bidding} achieves a high coverage, mobile sensors may consume much energy due to the iterative movement. This is because the mobile sensors do not know the largest hole beforehand. This problem can be solved by using so-called virtual movement or logical movement strategy. Instead of moving physically in each round, the mobile sensors perform virtual movements, i.e., logical movements, to the logical locations. After several rounds of bidding, they determine and move to their final destinations. This approach reduces the overall movement distance, but increases the message overheads between the sensors. Thus, a proxy-based bidding protocol proposed in \cite{wang2007bidding} adopted in the virtual movement strategy to improve the performance in terms of energy efficiency. Each mobile sensor considers a static sensor as its proxy when the mobile sensor accepts the bid from this static sensor. After that, the proxy sensor emulates the virtual movement for this mobile sensor and bids for new destinations in the next round. The proxy may delegate its proxy role to another static sensor with a higher bid in the next round. Therefore, the virtual movement is actually performed by delegating the role of proxies between the stationary sensors. Simulation results in \cite{wang2004proxy} showed that the proxy-based approach can save up to 50\% of moving distance and thus reduce significantly the energy consumption while achieving the same coverage as that proposed in \cite{wang2003bidding}. However, in both  \cite{wang2003bidding} and \cite{wang2007bidding}, the authors did not consider the energy consumption balances of all of the sensors, so the lifetime of the network may be short.

\subsection{Target Coverage}
\label{sec:Target_coverage}
\begin{figure}[ht]
 \centering
\includegraphics[width=4cm, height=4cm]{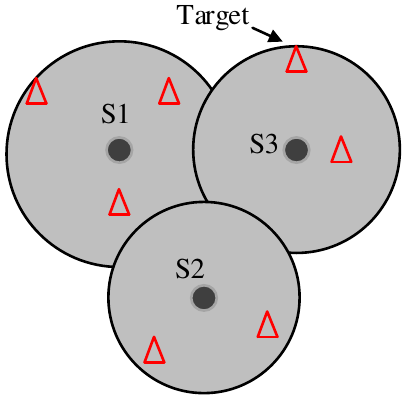}
 \caption{Target coverage.}
 \label{target_coverage}
\end{figure}
Assume that a sensing target is static. The target coverage problem aims to cover all targets by adjusting the sensing range of each sensor as shown in Fig.~\ref{target_coverage}. However, when sensors increase their sensing ranges, there are constraints such as energy consumption and sensing overlap. If we assume that each sensor has its utility which is a function of its sensing range, an appropriate model to determine sensors' sensing ranges is the NUM framework as described in Section \ref{sec:Utility_resource}. The NUM framework is suitable since it formulates the problem by maximizing the aggregate utilities of sensors subject to the constraints. In \cite{naderan2011distributed}, the NUM framework for determining the sensing range of sensors was proposed to address the target coverage in WSNs. Each utility function in this approach is logarithmically concave function of the sensing range. Therefore, the NUM problem has a strictly concave objective function. However, this is solvable only if the network has complete knowledge of all sensors' utility functions. In \cite{naderan2011distributed}, the problem was formulated with the the Lagrange multipliers with their interpretations as resource prices and overlapping prices. In particular, the resource prices are the energy prices which sensors are willing to pay. Using the gradient ascent algorithm, it was then proved that the two price variables converge to a unique solution. Simulation results indicated that when the number of sensors increases, the total objective function increases while the number of iterations needed for convergence does not change. However, to enable faster convergence for the convex problem, other methods such as the fast proximal gradient method should be used instead of the gradient ascent algorithm.

\subsection{Barrier Coverage}
\label{sec:Barrier_coverage}
\begin{figure}[ht]
 \centering
\includegraphics[width=\linewidth]{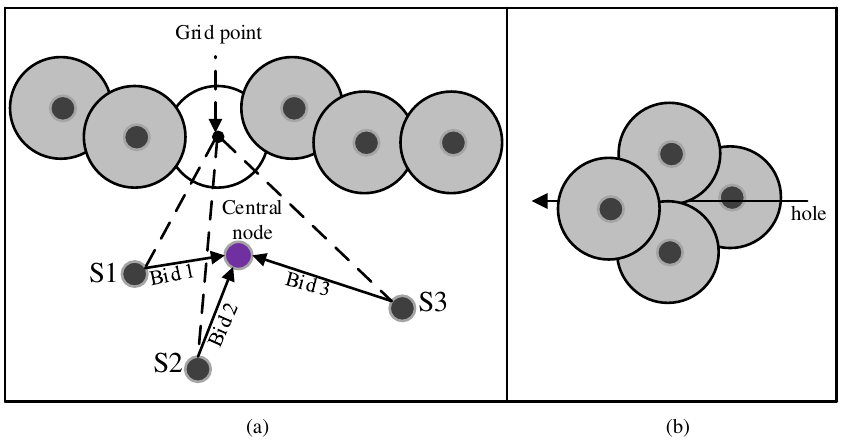}
 \caption{Market based mechanism for a) 2D coverage and b) 3D coverage.}
 \label{barrier_coverage}
\end{figure}

For the barrier coverage, sensing areas of adjacent sensors are required to overlap with each other to form a chain of sensors called a barrier \cite{saipulla2009barrier} to detect any intruders that attempt to cross the network. The authors in \cite{barr2009underwater} addressed mapping mobile sensors to grid points of 2-Dimensional (2D) barrier for underwater sensor networks using a first-price sealed-bid auction scheme. Through bidding strategy, the auction allows determining and assigning the mobile sensor closest to a grid point and thus minimizing energy consumption of mobile sensors. As shown in Fig.~\ref{barrier_coverage}(a), mobile sensors which act as bidders, i.e., buyers, will bid on the particular grid point via a central sensor, i.e., an auctioneer. Each bid includes a price which is inversely proportional to the sensor's distance to the grid point. The sensor with the highest price, i.e., the shortest distance, is assigned to the grid point. Simulation results showed that the maximum movement that one sensor must travel to its assigned location is shorter than that of a classic optimal solution, e.g., Hungarian method \cite{kuhn1955hungarian}. Although the proposed approach results in reducing energy consumption for the barrier construction, the energy balance has not achieved since only the distance factor is considered. In addition, remaining energy of each sensor can be included for the bid evaluation to obtain better energy balance, and thus maximize the lifetime of network. 

Similar to \cite{barr2009underwater}, the authors in \cite{barr2009constructing_pro} also considered the assignment of sensors to the grid points using the first-price sealed-bid auction scheme. However, this approach is applied to 3-Dimensional (3D) barrier which is more realistic in underwater sensor networks. As shown in Fig.~\ref{barrier_coverage}(a), after the chain of sensors in the 2D barrier is formed, no moving intruders
can cross the chain undetected. However, in 3D spaces (see Fig.~\ref{barrier_coverage}(b)), there may exist a hole which the intruders can pass through. Therefore, compared to the 2D barrier, the number of required grid points in 3D spaces is larger which requires more bidding rounds.

\subsection{Target Tracking}
\label{sec:Tracking_tracking}
The object tracking is also an important part of
WSNs in monitoring and surveillance applications. The core
object classification and detection process can be efficiently
performed by supervised machine learning algorithms \cite{abu2014machine}. To improve the probability of object
detection, pricing models are employed to predict the future locations
of the targeted object and to schedule the sensor nodes in an efficient energy
manner while guaranteeing the tracking quality. 

The authors in \cite{chen2006auction} adopted a first-price sealed-bid auction for the target tracking task allocation in WSNs. Specifically, when a sensor finds a target within its vicinity, and there is no leader in the network, it promotes itself as a leader candidate which acts as an auctioneer, i.e., a seller. Then, the leader sends a broadcast message to recruit sensors, i.e., buyers, for the target tracking task. Upon receiving the message, the sensors evaluate the task and respond to the leader by bids. The leader evaluates and ranks these bids to select a predefined number of buyers who have the highest bid. The chosen buyers perform the target tracking task and send the target data to the seller. Compared with the traditional task allocation protocols for target tracking, e.g., the case-based reasoning dynamic coalition scheme \cite{soh2001reflective}, the proposed approach does not need the knowledge of the neighbors in advance. Thus, it saves more storage and computing resources of the sensors. In addition, since less communication between the buyer and the sellers is required, the more energy can be saved. The simulation results indicated that the energy that the auction based scheme can save is more than $65\%$ compared to that in \cite{soh2001reflective}. However, the proposed approach did not consider the target data quality. Taking this attribute, the authors in \cite{liang2011auction} proposed an energy efficient target tracking algorithm with high accuracy, called Auction-based Adaptive Sensor Activation (AASA). First, a cluster head, i.e., an auctioneer (a seller), predicts the next location of the target via a linear prediction method and broadcasts a message to activate the sensors, i.e., buyers, in the predicted region. Once receiving this message, each buyer evaluates the task and responds to the auctioneer by a bid. The bid is a function of the residual energy of the sensor and the distance between itself and the predicted target location. The seller chooses the buyers with the highest bids for performing the target tracking task. The target data sent from the chosen buyers is used by the seller to estimate the location of the target through an improved trilateration algorithm \cite{zhou2009efficient}. Compared with the general tracking algorithm based on a cluster and prediction in which all the one-hop neighbor nodes of cluster head take part in tracking, the simulation results showed that the AASA algorithm greatly decreases the energy consumption
and increases the network lifetime while achieving high tracking quality. The same results were obtained when the AASA algorithm is compared with the prediction-based clustering algorithm \cite{deldar2011designing} as shown in \cite{zheng2014auction}. 

The above approaches only consider the scenario with a single target. When multiple targets need to be tracked, the combinatorial auction can be employed to improve resource efficiency. The authors in \cite{wu2014airborne} proposed a multi-target tracking mechanism based on the combinatorial auction. Specifically, a sensor needs to perform multiple tasks to track the targets in a battlefield. Due to the resource constraint, some close targets can be combined into a cluster, i.e., a bundle, and the sensor only needs to track a midpoint of the cluster. However, the problem is how to assign the targets to clusters such that the tracking performance in terms of the resource efficiency is maximized. The authors considered this problem as the combinatorial auction in which a sensor management agent acts as a buyer who purchases the resources such as sensing time, from the sensor, i.e., the seller, to accomplish the multi-target tracking tasks. First, the targets in a battlefield are assigned into clusters, i.e., bundles, which form a target set. Since there can be many target assignments, we may get several available target sets. Then, the sensor management agent calculates a performance-price (also known as benefit-cost) ratio for each target set. The price is the sum of resource costs for completing the tasks, and the performance is the total value of targets being destroyed.  The target set that has the highest performance-price ratio is considered as the winner of the auction. When the battlefield environment changes, e.g., a new target arrives, the auction can be conducted again. However, the computation complexity of the algorithm is high and requires much processing time.

\begin{table*}[h]
\caption{Applications of pricing models for sensing coverage and target tracking}
\label{table_seccoverage_sum}
\scriptsize 
\begin{centering}
\begin{tabular}{|>{\centering\arraybackslash}m{0.9cm}|>{\centering\arraybackslash}m{0.4cm}|>{\centering\arraybackslash}m{1.6cm}|>{\centering\arraybackslash}m{1.2cm}|>{\centering\arraybackslash}m{1.3cm}|>{\centering\arraybackslash}m{1.3cm}|>{\centering\arraybackslash}m{4.2cm}|>{\centering\arraybackslash}m{2.2cm}|>{\centering\arraybackslash}m{1.2cm}|}
\hline
\multirow{2}{*}  {\textbf{}} & \multirow{2}{*}  {\textbf{Ref.}} &  \multirow{2}{*}  {\textbf{Pricing model}}  & \multicolumn{3}{c|} {\textbf{Market structure}} & \multirow{2}{*}  {\textbf{Mechanism}} & \multirow{2}{*}  {\textbf{Objective}} & \multirow{2}{*} {\textbf{Solution}} \tabularnewline 
\cline{4-6}
 & & & \textbf{Seller} & \textbf{Buyer} & \textbf{Item}  & & &\tabularnewline
\hline
\hline
\vspace{0.5cm} Area coverage& \cite{wang2003bidding} &First-price sealed-bid auction & Mobile sensor & Static sensors & Hole coverage& Buyers submit bid prices, and the seller accepts to heal a hole with the highest price&Area coverage maximization& Competitive equilibrium\tabularnewline \cline{2-9}
&\cite{wang2007bidding} \cite{wang2004proxy} &First-price sealed-bid auction & Mobile sensor & Static sensors & Hole coverage& Same as \cite{wang2003bidding} but a virtual movement strategy is considered&Area coverage maximization, and energy efficiency& Competitive equilibrium\tabularnewline \cline{2-9}
\hline
Target coverage&\cite{naderan2011distributed}  &Network utility maximization &Sensor & Resource provider& Energy& Maximizing the aggregate utilities of sensors subject to the constraints of energy and overlapping & Sensing range maximization &Optimization solution\tabularnewline \cline{2-9}
\hline
\vspace{0.5cm} Barrier coverage& \cite{saipulla2009barrier} \cite{barr2009underwater}&First-price sealed-bid auction & Central sensor & Sensors & A grid point of 2D barrier& Buyers submit bid prices, and the seller selects a buyer with the highest price&Energy consumption minimization& Competitive equilibrium\tabularnewline \cline{2-9}
& \cite{barr2009constructing_pro}&First-price sealed-bid auction & Central sensor & Sensors & A grid point of 3D barrier& Same as \cite{saipulla2009barrier}&Energy consumption minimization& Competitive equilibrium\tabularnewline \cline{2-9}
\hline
\vspace{0.5cm} Target tracking& \cite{chen2006auction}&First-price sealed-bid auction & Leader sensor & Sensors & Target tracking task& Buyers submit bidding price, and the seller selects a predefined number of buyers with the highest prices to perform the task&Resource efficiency& Competitive equilibrium\tabularnewline \cline{2-9}
& \cite{liang2011auction} \cite{zheng2014auction}&First-price sealed-bid auction & Cluster head & Sensors & Target tracking task& Same as \cite{chen2006auction}&Resource efficiency, network lifetime maximization, and high quality of tracking& Nash equilibrium\tabularnewline \cline{2-9}
&\cite{wu2014airborne}&Combinatorial auction & Sensor & Sensor management agent & Resources& Seller assigns targets to bundles for tracking such that the performance-price ratio is maximized&Resource efficiency, and high tracking accuracy& Optimal solution\tabularnewline \cline{2-9}
\hline
\end{tabular}
\par\end{centering}
\end{table*}

\textbf{Summary:} The reviewed papers in this section have shown that the pricing models are useful for solving problems concerning sensing coverage and target tracking. The issues along with references are summarized in Table \ref{table_seccoverage_sum}. From the table, we observe that auctions are used efficiently to address issues of area and coverage as well as target tracking while the NUM framework can be applied for optimizing the target coverage. The next section reviews the use of pricing models for security and intrusion detection. The security component of a WSN guarantees the confidentiality of the collected sensors' data.

\section{Applications of pricing models for DoS attack prevention and privacy concerns}
\label{sec:Security}
WSNs are vulnerable to security threats. Various WSN security protocols were classified and evaluated in \cite{wang2006survey}. However, the traditional protocols lack interactions
among strategies of rational decision makers (i.e., the attacker and
the WSN owners) which strengthen the network security \cite{shen2011survey}. Due to this fact, pricing models have been employed to address secrurity and privacy issues in WSNs with the following approaches. 

\begin{itemize}
\item{\textit{DoS attack prevention}:} A Denial-of-Service (DoS) attack is a deliberate action of one or many sensor nodes with the aim of interrupting or suspending services of WSNs. In this case, pricing strategies can be used to detect and isolate the attackers. 
\item{\textit{Privacy concerns}:} Privacy involves the issues related to the sensitive private data of phone users in the data aggregation. Indeed, the collected data may include the location information which can reveal sensitive information, e.g., home address of users. Thus, using pricing models may provide incentive mechanisms for users' privacy protection. 
\end{itemize}

\subsection{DoS Attack Prevention}
\label{sec:Dos_prevention}

\begin{figure}[ht]
 \centering
\includegraphics[width=\linewidth, height=4.5cm]{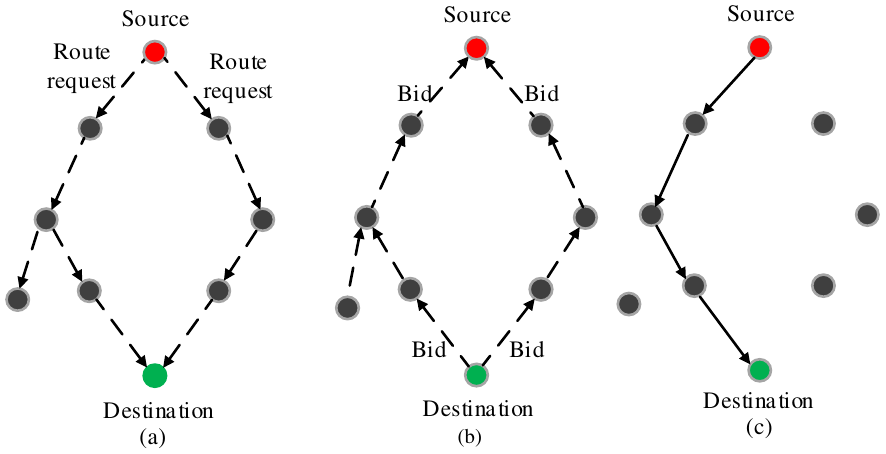}
 \caption{(a) \textit{Route request} is sent, (b) \textit{route reply} includes the bid price, and (c) path is established (represented by the solid line).}
 \label{Routing_security}
\end{figure}

The authors in \cite{agah2006security} proposed a secure sensor  
network routing protocol based on a first-price sealed-bid auction to
isolate malicious sensors. A source sensor (seller) selects a secure route which is formed by other sensors (buyers) in the network to forward packets to its destination. The secure route does not include any sensor that acts maliciously by dropping incoming packets, which is also known as the passive DoS
attack. First, the source sensor broadcasts a \textit{route request} to its neighbors as shown in Fig.~\ref{Routing_security}(a). Then, its neighbors forward this request to the others. When the destination receives this request, it sends back a \textit{reply message} including the full source route and the bid price. Each bid price represents a route and includes the sum of utilities of all sensors on the route. The utility of a sensor is calculated using the theory of equity, reciprocity, and competition \cite{bolton2000erc}. The utility depends on the sensor's battery power and reputation. Among available routes, the source sensor selects the best one, which has the highest bid, stores it and sends messages along that route. The sensors on the winning route will get good reputation as their rewards, and they will spend a percentage of the initial power as their payments. Since the sensors on a route desire to get the good reputation from the source, they are forced to cooperate to forward incoming packets. 

Although \cite{agah2006security} can mitigate malicious sensors causing the disruption to routing, there are some malicious sensors which could agree to the auction but will subvert the route later. A watch-list \cite{agah2005enforcing} which records the sensors' misbehavior can be used to recognize such malicious sensors and then let all sensors not communicate with them. It also was proved that for the given strategic game, the utility, i.e., the amount of bid that each node offers and a set of available bidding strategies of sensor $i$, there is always a Nash equilibrium at which a sensor $i$ bids the amount $b^*(v_i)=(1-\frac{1}{N})v_i$, where $N$ is the number of sensors, and $v_i$ is the total amount of bid that each node is willing
to pay. For example, for the case of 100 sensors, the
average number of dropped packets remains steady and the total
number of dropped packets is less than a half of that of the CONFIDANT protocol \cite{buchegger2002performance}. The reason is that sensors with bad reputation in the proposed protocol will
be ignored by the majority of sensors. Extending these approaches, the QoS of the network can also be improved in terms of security and delay by incorporating the information of hops from the source to the destination into each bid price. 

\subsection{Privacy Concerns}
\label{sec:Privacy_concerns} 
One of challenges in participatory sensing and crowdsensing is that the deployment of these systems can
reveal users' private information such as location and identity. Therefore, pricing mechanisms may be useful to control users' privacy. Specifically, if the payment is lower than the privacy price, the users may not have enough motivation to perform sensing tasks. 

\subsubsection{Privacy evaluation}
\label{sec:Privacy_Evaluations} 
It is natural to determine the value of the privacy. The authors in \cite{christin2013s} conducted a survey with 200 anonymous participants. The results showed that the participants' average rating about the importance of privacy is 5.82 on a scale from 1 (not important) to 7 (very
important). This confirms that the privacy is globally important to the users. For the incentives, 41\% of participants would provide their information for free, 27\% would get a monetary reward, 22\% would like to get free data access, and 14\% of them expected additional application features. Furthermore, 18\% would receive coupons while 6\% would like to be given a higher reputation. This indicates that incentive mechanisms are very important in privacy-aware participatory
sensing systems. 

The authors in \cite{danezis2005much} aimed at determining how much price must be offered to persuade someone to expose their location information for a certain time period through using schemes from experimental economics and psychology. Specifically, the authors employed the sealed-bid second-price reverse auction and invited volunteers, i.e., sellers, to participate in a fictitious study. The study involves the collection and the processing of the location information from their mobile phones. The auction mechanism was chosen to guarantee that sellers are motivated to report the true value of their location privacy. The application, i.e., the buyer (auctioneer), asks the volunteers to submit asking prices for their location privacy. The buyer then invites a certain number of participants with the lowest asking prices and pays an amount of money which is equivalent to the lowest price of the seller who is lost. The study is performed among computer science students at the University of Cambridge. The results indicated that a median asking price of 10 pounds is needed for the location privacy. However, when
the possibility of commercial interests was mentioned, the median asking price increases to 20 pounds. Since this study is based on a small group of students who are likely to have lower privacy preferences than the general population, the result is only considered as a lower bound on the location privacy preference. Moreover, only the winners get reward while the users who lose in the asking process get no compensation at all for their privacy revelation. Consider more users' incentive, the authors in \cite{gisdakis2014sppear} and \cite{li2014providing} focused on the rewarding process. Accordingly, all anonymous users receive rewards which are either credits \cite{gisdakis2014sppear} or tokens \cite{li2014providing}, as long as they submit at least $n$ reports. The reward may be monetary return or promotion for using location-based services afterwards as mentioned in \cite{duan2014motivating}.

\subsubsection{Privacy-preserving mechanisms}
\label{sec:Privacy_preserving} 
Privacy-preserving mechanisms aim to guarantee that user's data is not associated with its identity \cite{kapadia2009opportunistic}. In \cite{danezis2005much}, \cite{gisdakis2014sppear} and \cite{li2014providing}, a privacy protection mechanism for participants is ignored. Considering this problem, the authors in \cite{singla2013incentives} proposed a seal-bid second-price reverse auction that requires users only to submit their asking prices with obscure locations. After the asking process, the winning users reveal their true locations when submitting the sensed data. However, in this case, it can be seen that the buyer's utility is sacrificed. The buyer does not have the knowledge of users' accurate locations and cannot evaluate the actual utility of each user. The buyer then may overpay or underpay for the received data. 

Another method to preserve the user privacy is to use the pseudonym
to replace a user's real identity. The authors in \cite{sun2013privacy} adopted the sealed-bid second-price reverse auction in which the asks are encrypted before being submitted using a cryptographic
method, called Time Lapse Cryptography (TLC)\cite{rabin2006time}. Furthermore, before being forwarded to a server, i.e., an auctioneer, the asks are signed by the users by employing the Nyberg-Rueppel
signature scheme \cite{camenisch1995blind} to keep the confidentiality of the asks. Finally, a set of users who have the encrypted asking prices lower than a payment threshold is selected as winners to provide sensing data. The payment threshold influences the outcome of the auction. However, how to calculate the threshold was not given in the proposed approach. 

In addition to the ask privacy, the seller anonymity and fairness problems were considered through using a sealed-bid multi-attribute reverse auction in \cite{shi2013sealed}. The attributes of an ask which are the price, the location accuracy, and the sampling frequency are converted to the ask utility score through the linear utility function in \cite{bichler2005configurable}. The asks are encrypted by using the Paillier's encryption \cite{paillier1999public} before being submitted to the server. Then, the private set intersection algorithm in \cite{freedman2004efficient} was adopted for the winner determination. The proposed scheme does not need to open asks to perform the winner determination and the public verifiability which allows anybody to check the identities
of bidders and confirm whether their bids are
valid or not. Thus, the strong ask privacy is maintained in the auction execution. According to the security analysis, the proposed scheme satisfies the security requirements of an electronic auction (e-auction), such as anonymity, ask privacy, and public verifiability. The same approaches can be found in \cite{dimitriou2015privacy} and \cite{krontiris2015platform}. However, in \cite{dimitriou2015privacy}, instead of using the Paillier's encryption, each seller masks its ask with a hash value which is computed from a random number. The one-wayness of the hash function guarantees that ask values remain hidden, thus eliminating
the possibility that a seller or the server can leak information about the others before the asking process ends. As a result, the confidentiality of asks is better maintained compared to \cite{shi2013sealed}. Finally, the server collects all submitted asks and determines the winners who have the highest utility scores. However, to improve the winning chances for the losers in the next rounds, the number of previously lost auction
rounds should also be included as one of the ask attributes as proposed in \cite{krontiris2015platform}.

\subsubsection{Faked sensing attacks prevention}
\label{sec:Faked_prevention} 
In addition to the above security issues, crowdsensing networks are also vulnerable to
faked sensing attacks by users to save sensing costs
and avoid privacy leakage \cite{fatemieh2010secure}. Traditional secure mechanisms, e.g., \cite{feng2014trac}, were proposed to evaluate the received sensing reports. However, it is challenging for the server to calculate the actual costs. Therefore, price based approaches were proposed to stimulate the users to submit their actual reports and then pay the users according to their sensing accuracies. The authors in \cite{xiao2015secure} formulated the problem as a secure mobile crowdsensing game among selfish users and a server through a sealed-bid first-price reverse auction scheme. The users choose their strategies for their ask prices while the server determines the payment strategy to maximize the individual utility. The set of optimal strategies yields a unique Nash equilibrium for the asking strategy of the users and pricing strategy of the server. However, the actual costs, which form the utility functions, are unknown by the server and users. Therefore, when the game is repeated, e.g., the server requires the sensing data continuously, reinforcement learning methods such as Q-learning \cite{watkins1992q} can be employed by the server to determine its payment strategy according to the utility history for the previous trials. The proposed learning strategy can motivate the users to send higher-quality sensing reports and significantly decrease the number of faked sensing reports. This was confirmed by the simulation results that the probability of a user sending an actual sensing report is greater than $97\%$ after 300 time slots. 

The same approach was also adopted in target tracking applications \cite{cao2015target} in which a fusion center, i.e., a buyer, needs to collect sensing data from different users, i.e., sellers, to determine a state of the target. However, unlike \cite{xiao2015secure}, the buyer's uncertainty about the value of a user's ask is described by a continuous probability distribution which allows calculating the expected utility functions of users directly instead of using the Q-learning algorithm. The solution of the auction is obtained from solving
the multiple-choice knapsack problem to achieve the maximum utility for the fusion center. The optimality of the solution is guaranteed while the rationality and the truthfulness properties are achieved. 

\begin{table*}[h]
\caption{Applications of pricing models for strengthening and privacy security}
\label{table_secprivacy_sum}
\scriptsize 
\begin{centering}
\begin{tabular}{|>{\centering\arraybackslash}m{1.1cm}|>{\centering\arraybackslash}m{0.4cm}|>{\centering\arraybackslash}m{1.6cm}|>{\centering\arraybackslash}m{1cm}|>{\centering\arraybackslash}m{1.1cm}|>{\centering\arraybackslash}m{1.4cm}|>{\centering\arraybackslash}m{4.3cm}|>{\centering\arraybackslash}m{2.2cm}|>{\centering\arraybackslash}m{1.2cm}|}
\hline
\multirow{2}{*}  {\textbf{}} & \multirow{2}{*}  {\textbf{Ref.}} &  \multirow{2}{*}  {\textbf{Pricing model}}  & \multicolumn{3}{c|} {\textbf{Market structure}} & \multirow{2}{*}  {\textbf{Mechanism}} & \multirow{2}{*}  {\textbf{Objective}} & \multirow{2}{*} {\textbf{Solution}} \tabularnewline 
\cline{4-6}
 & & & \textbf{Seller} & \textbf{Buyer} & \textbf{Item}  & & &\tabularnewline
\hline
\hline
DoS attack prevention& \cite{agah2006security} \cite{agah2005enforcing} &First-price sealed-bid auction & Source sensor & Forwarding sensors & Reputation& Buyers submit bid prices, and the seller selects buyers with the highest prices&Secure routing improvement& Nash equilibrium\tabularnewline \cline{2-9}
\hline
\parbox[t]{2mm}{\multirow{9}{*}{\rotatebox[origin=c]{90}{\hspace{-6cm} Privacy concerns}}}  
&\cite{danezis2005much}&Seal-bid second-price reverse auction &Phone users &Platform & Location privacy& Sellers submit their asking prices and location information. The buyer selects the winners owning the lowest asking prices &Incentive
compatibility, and buyer's revenue maximization  &Nash equilibrium\tabularnewline \cline{2-9}
&\cite{gisdakis2014sppear} \cite{li2014providing}&Seal-bid second-price reverse auction &Phone users &Platform & Location privacy& Same as \cite{danezis2005much} &Incentive improvement &Nash equilibrium\tabularnewline \cline{2-9}
& \cite{singla2013incentives}  &Seal-bid second-price reverse auction &Phone users &Platform & Location privacy&Same as \cite{danezis2005much} but only winners submit their location information  &Incentive
compatibility, buyer's revenue maximization, and privacy security &Nash equilibrium\tabularnewline \cline{2-9}
& \cite{sun2013privacy}  &Seal-bid second-price reverse auction &Phone users &Server & Real identity& Sellers submit their encrypted asks. The buyer selects the winners owning the asking prices lower than a threshold &Incentive
compatibility, buyer's revenue maximization, and privacy security improvement &Nash equilibrium\tabularnewline \cline{2-9}
& \cite{shi2013sealed} &Multi-attribute reverse auction &Phone users &Server & Location accuracy and sampling frequency& Sellers submit their multi-attribute asks corresponding to utility scores which are encrypted by the Paillier's encryption. The buyer determines the winners with the highest utility scores &Incentive
compatibility, privacy security, and well fairness &Bayes-Nash equilibrium \tabularnewline \cline{2-9}
& \cite{dimitriou2015privacy} \cite{krontiris2015platform}  &Multi-attribute reverse auction &Phone users &Server & Location accuracy and sampling frequency& Sellers submit their multi-attribute asks which are masked with hash value, and the buyer determines the winners with the highest utility scores &Incentive
compatibility, privacy security improvement, well fairness, and confidentiality maintenance of asks &Bayes-Nash equilibrium \tabularnewline \cline{2-9}
& \cite{xiao2015secure}&Seal-bid first-price reverse auction &Phone users &Server &Sensing data& Sellers submit their asking prices, and the buyer selects the winners owning the lowest asking prices &Faked sensing report minimization, and individual utility maximization of sellers and buyer&Nash equilibrium\tabularnewline \cline{2-9}
&\cite{cao2015target}&Seal-bid first-price reverse auction &Phone users &Fusion center &Sensing data& Sellers submit their asks, and the buyer selects users to buy data to achieve maximum utility& Sellers' rationality and truthfulness, and buyer's utility maximization&Nash equilibrium\tabularnewline \cline{2-9}
\hline
\end{tabular}
\par\end{centering}
\end{table*}

\textbf{Summary:} This section discusses the existing security
approaches which are entirely based on auction schemes as summarized in Table \ref{table_secprivacy_sum}. Although these schemes have tackled some issues related to the DoS attack prevention and the privacy, future researches are required to address other security related issues, e.g., intrusion detection and coexistence with malicious sensors. The following section reviews the application of pricing models to address some other issues in IoT apart from those discussed above.

\section{Miscellaneous issues}
\label{sec:Miscellaneous}
In this section, we review some other issues related to the use of pricing models for faulty sensor detection, surveillance, and profit maximization of IoT service providers. Moreover, the pricing models for IoT services are reviewed. 
\subsection{Pricing Models for Faulty Sensor Detection}
\label{sec:Faul_prevention} 
In addition to the security, another risk which affects the reliability is the presence of faulty sensors. Sensor faults of WSNs can be divided into two categories, i.e., hard and soft \cite{jiang2009new}. A sensor is hard fault if the sensor cannot communicate with other sensors due to the failure of a certain module, e.g., the failure of the communication module and energy depletion of the sensor. A sensor is soft fault meaning that the failed sensor can continue to work and communicate with other sensors, but the sensed or transmitted data is not correct. The authors in \cite{mccausland2013auction} proposed a risk-aware robotic sensor network applied to critical infrastructure protection with the aim of detecting and mitigating the faulty sensors using an auction. The proposed approach is a task allocation mechanism which recruits the sensors with the best fitness to risk-mitigation plan. Specifically, a sensor which is first aware of a high risk network event becomes an auctioneer. Then, the auctioneer broadcasts a risk-mitigation task announcement message to all available sensors. This message contains the event location and the necessary risk details so that each receiving sensor can appropriately bid on the task. Upon receiving the message, by using the Sugeno fuzzy model \cite{sugeno1988structure}, each sensor calculates its bid based on three primary data sources including the available battery level, the distance from the sensor to the event, and the coverage redundancy. The sensors with the highest bid are selected as the winners to perform the risk-mitigation task. Since the bid is proportional to the available energy of sensors, the proposed mechanism avoids overwhelming the sensors with low energy. However, this approach did not consider how to determine the auctioneer in the situation where there are multiple sensors detecting a high risk event simultaneously.

\subsection{Pricing Models in Pervasive Monitoring Applications}
\label{sec:Persavive_Moni}
\begin{figure}[ht]
 \centering
\includegraphics[width=\linewidth, height=2.8cm]{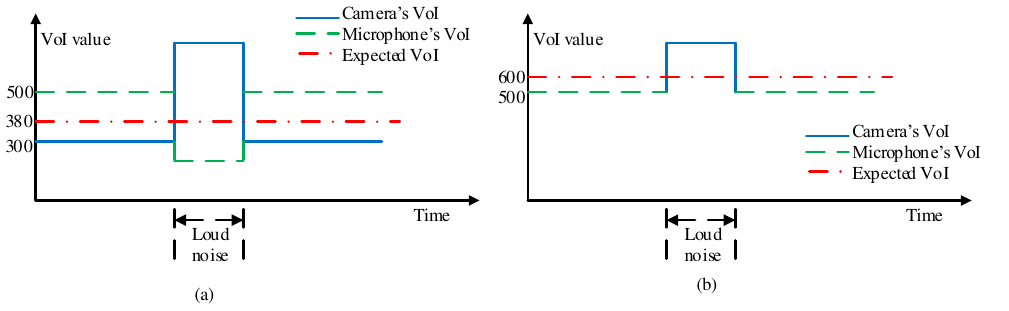}
 \caption{(a) Expected VoI value when two devices work simultaneously and (b) expected VoI value when the switch option model is used.}
 \label{Pervasive_monitoring_switch_option}
\end{figure}

This section considers using pricing models for pervasive monitoring applications such as secure house and secure gallery. In such applications, there are several security devices installed. To reduce resource costs while still maximizing the expected Value of Information (VoI), a switch option model was adopted. Switch option (see Chapter 10 in \cite{schwartz2004real} for more details) is a special type of the real options which has been widely used in economics to select the mode of production for factories or switching between investment options to maximize the profit or to minimize loss. The authors in \cite{geyik2011sensor} adopted the switch option model to maximize the VoI in a parking garage monitoring network. The system consists of two services: a microphone service of reading from an acoustic sensor to monitor the sound volume and a camera service that provides views of the area covered by the microphone monitors. The VoI from a device depends on the information volume obtained from that device and the resources consumed by the device \cite{szymanski2011market}. As shown in Fig.~\ref{Pervasive_monitoring_switch_option}(a), when two devices are simultaneously active, the expected VoI obtained from the devices is low since typically the camera requires a large amount of resources. Therefore, in a normal mode, i.e., no event, the microphone service is switched on while the camera service is switched off. Only when there is a loud noise which can signify a pervasive event, the microphone service is switched off since its VoI becomes lower due to the limited information provided. At this time, the camera service is activated alternatively to provide the information. Therefore, as indicated in Fig.~\ref{Pervasive_monitoring_switch_option}(b), the expected VoI is always higher than that of the situation activating both the devices. However, how to determine a threshold to switch the two devices was not given. 

\subsection{Pricing Models for the Platform's and IoT Service Provider's Utility Maximization}
\label{sec:Pricing_Platform_Provider}
Maximizing the utility of a platform can be achieved by either maximizing the collected data gain or minimizing overall costs of recruiting participants to provide sensing service. The authors in \cite{yang2012crowdsourcing} modeled an incentive negotiation procedure as a two-stage Stackelberg game with the aim of minimizing the platform's incentive budget. In the first stage, the platform which acts as a leader decides the payment to each participant. In the second stage, the participants, i.e., followers, strategize sensing time to maximize their own utilities. The second stage can be considered as a non-cooperative game, and it was proved that for a given payment, there exists a unique Nash equilibrium involving a set of optimal strategies of the participants. The Nash equilibrium together with the optimal payment strategy of the platform constitutes a solution called Stackelberg equilibrium. However, this approach requires complex computational procedures on participants' smart devices. Moreover, it also requires that the platform needs to know the actual sensing cost of all participants. 

A more simple mechanism was proposed in \cite{subramanian2013offline} by using reverse auctions. Accordingly, the platform, i.e., the buyer, has a utility function associated with its set of tasks. After receiving the asks from the participants, the platform shortlists the potential winners, i.e., sellers, using a greedy linear time algorithm that allows determining the user with the best marginal utility. Then, each user in the shortlist is kept, dropped or replaced depending on the effective utility that it contributes to the platform. Finally, the platform makes the payment strategy which is similar to the VCG mechanism. Specifically, each user is paid according to its marginal utility that it brings to the platform, and according to the increase in utility by replacing any user by the next best user plus the bid of the next best user. Theoretical analysis indicated that the proposed strategy maximizes the platform's utility while guaranteeing four properties of an efficient auction that are computational efficiency, individual nationality, profitability and truthfulness. This also was demonstrated by simulation results in which the platform's utility of the proposed algorithm was improved by roughly 15\% comparing with the scheme in \cite{yang2012crowdsourcing} since it has to pay the selected users less. However, from the users' side, the more-frugal payment can reduce their incentives in providing data.

After data is collected from the platform, the useful information will be extracted to sell to the service providers. The service providers' utilities are often the profits that they earn from re-selling the information to the IoT customers. The information can be treated as items traded in the market, and the main problem is how to set prices for the sensing information so that the service providers' profits are maximized. Consider a competitive market involving IoT service providers, the authors in \cite{niyatoeconomics} formulated the competition in setting the price as a non-cooperative game. The players are the service providers, i.e., sellers, and the strategy of each player is to set the price of the sensing information to maximize his own payoff, i.e., profit. The set of the best strategies of the players yield the highest payoffs, i.e., the Nash equilibrium. Based on the proposed sensing service pricing model, the marginal cost to reproduce the information is likely to be zero. It means that some service providers can buy information and resell to the others, thereby hierarchical games, e.g., Stackelberg game, are required. 

Consider such a hierarchical information market, the authors in \cite{mei2013pricing} investigated the information service pricing by formulating the Stackelberg game, among service providers, brokers, and customers. The brokers acquire information from the providers, i.e., leaders, and sell the
services to the customers, i.e., followers. Two strategies, the bundling strategy and the pure components strategy, are adopted to calculate information selling prices which maximize profits for the service providers and the brokers. The theoretical analysis was also given to specify which strategy (i.e., bundling or pure components) should be used based on the relationship between the information collection cost and the willingness-to-pay value of the customers. However, the method to let service providers form bundling has to be devised.

\subsection{Pricing Models for Evaluating the IoT Deployment}
\label{sec:Pricing_Evaluating_IoT}

As discussed earlier, mobile sensors were found useful to address issues in WSNs, e.g., sensing coverage and data collection. In addition to the technical advantages, the authors in \cite{mathurcost} adopted the cost-benefit analysis to evaluate the deployment utility of mobile sensors compared with static sensors from the economic perspective. Here, the benefit of a mobile node is to replace a certain number of static nodes while still guaranteeing the functionality of the network, e.g., area coverage, mobile relay, target coverage, and barrier coverage. Consider the $k$-coverage with $k=1$ meaning that every point in the target area is covered by at least 
$1$ sensor, it was indicated that one mobile sensor is equal to five static sensors. The Micaz mote \cite{johnson2009comparative} was adopted for the cost model. Micaz is a widely used sensor, and thus its cost is used for the evaluation. The cost of a static sensor is $\$100$ while that of a mobile sensor is $\$150$ including extra cost of peripherals (e.g., motor and wheels). Therefore, the overall economic benefit of using a mobile node against static sensors is $\$100 \times 5 - \$150=\$350$.

Different from \cite{johnson2009comparative}, the authors in \cite{decker2008cost} adopted the cost-benefit model to analyze the impact of deploying the different IoT devices based on the profits of stakeholders in supply chain applications. A supply chain is a system to move goods from a producer to customers via shippers. To monitor and improve the condition of goods transportation in the supply chains, IoT devices, e.g., barcode, RFID, and sensors, are embedded with the goods. The RFID transponders can provide sensor information of the goods, e.g. temperature or pressure or humidity, during the transport and enable the monitoring of goods. By calculating the cost and the benefit of each stakeholder, i.e., the producer, the shipper, and the customer, the authors showed that although the initial cost for deploying smart IoT devices, e.g., wireless sensors, is higher, the profits of the stakeholders are higher than that of only using barcodes or RFID. 

In practice, the RFID devices can be integrated with the sensors to improve the operational and functional capacities of the IoT systems \cite{zhang2006integration}. For example, the RFID readers and the sensors can combine with
multi-functional devices, such as PDAs and cell phones so that consumers are able to read any RFID sensor tag in many applications \cite{liu2008taxonomy}. In terms of reader-relay integrated architectures, an integration of WSNs and RFIDs was proposed in \cite{al2012novel}. This architecture has two layers in which the upper layer consists of integrated nodes of RFID readers and wireless relays, called reader-relay nodes. The lower layer is light nodes which are either sensors or tags, called porter nodes. The two-layered hierarchical architecture aims at distributing the sensing and relaying tasks over the components of integrated networks optimally. However, this also imposes several
challenges and requirements. The most important issue refers to the additional costs related to designing and deploying integrated hardware components. The authors in \cite{al2012pricing} analyzed the factors which can reduce costs when the integrated system is deployed. There are fours major factors affecting the cost reduction, i.e., the porter nodes density, porter nodes traffic load, the transmission distance between the porter nodes and the reader-relay nodes, and the delivery guarantee. For example, when the number of porter nodes is increased, each node in the network can reduce their power consumption to forward the packet from a source to a sink, thereby decreasing the cost. However, the authors did not specify which pricing model used to evaluate these factors. 

\textbf{Summary:} This section reviews the use of pricing models in some practical IoT applications, e.g., pervasive monitoring systems and supply chains. Through the pricing models, both economic and technical evaluations are provided. Moreover, some pricing models for service markets are also introduced to determine the selling prices and the profits of the stakeholders. The following section reviews the application of pricing models to address issues in machine-to-machine communications.

\section{Pricing Models in M2M Communication}
\label{sec:M2M_pricing}
\begin{figure}[ht]
 \centering
\includegraphics[width=\linewidth, height=5.5cm]{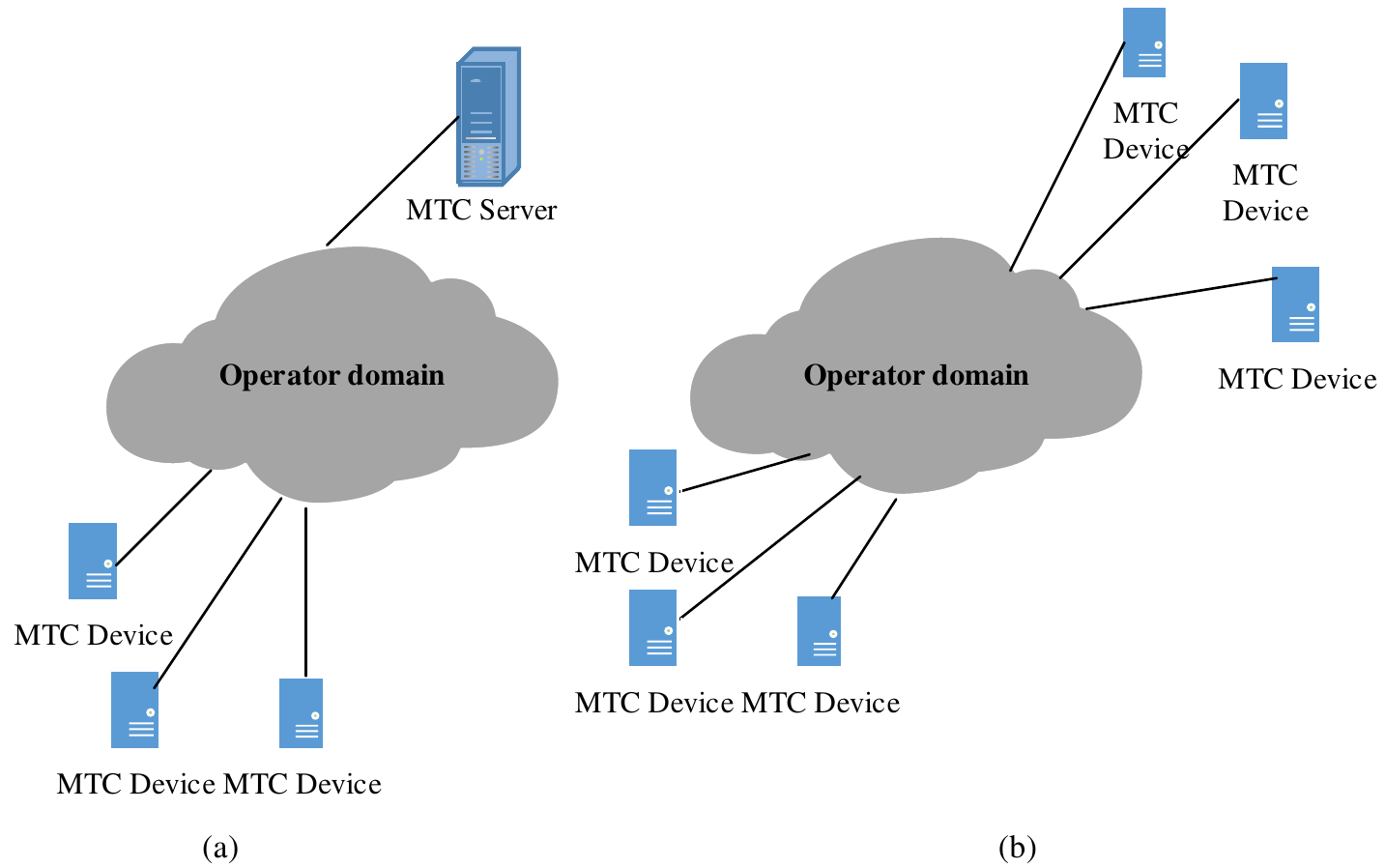}
 \caption{(a) MTC devices communicating with MTC server, and (b) MTC devices communicating directly with each other without intermediate MTC server.}
 \label{M2M_communication}
\end{figure}

Machine-to-Machine (M2M) communication or Machine-
Type Communication (MTC) is a term used to describe
technologies that enable devices, e.g, computers, embedded
processors, smart sensors, actuators, and mobile devices to
communicate with each other via wired or wireless communication networks without human intervention \cite{3GPPMTC2011}. Essentially, M2M belongs to the IoT, since it has the four components mentioned in Section II. B \cite{chen122012machine}. There are two typical communication scenarios for the MTC communication: MTC devices communicating with one or some MTC servers (see Fig.~\ref{M2M_communication}(a)) and MTC devices communicating directly with each other without intermediate MTC server (see Fig.~\ref{M2M_communication}(b)). Several approaches have been proposed to address the M2M issues such as the access control, the security, and the QoS guarantee. There are also several surveys of M2M \cite{kim2014m2m}, but they did not mention economic approaches. The idea of applying economic models for M2M communication is to enable the M2M entities to be interconnected, networked, and controlled remotely, with low-cost, scalable, and reliable architecture. 

Such an approach can be found in \cite{lee2012optimal} in which the authors aimed at  guaranteeing the required throughput in the successive reserved time slots by combining the second-price sealed-bid auction with the option pricing, so-called auction-based option pricing model. The model consists of $n$ devices, i.e., buyers, which request the radio resource from a service provider, i.e., a seller. Based on the second-price sealed-bid auction rule, the service provider can select $m$ among $n$ buyers as winners and the price that each winner must pay is the $(m+1)$th highest bid price. In addition, the proposed approach provides a resource reservation scheme for the buyers in the future. Accordingly, buyers can reserve the requested resources at a pre-specified price and a pre-defined time via a call option contract. The contract will protect the buyer against a higher resource price in the future and also provide for the buyer the right to purchase the requested resource quantity even when the buyer is dropped from the auction. Moreover, a convex cost function was introduced which allows the buyers to choose the strike price, i.e., the price at which the call option can be exercised, to minimize their reservation costs while obtaining the guaranteed QoS in terms of throughput. Under the proposed
scheme, the transmission resources could be allocated with the maximum utilization. However, the assumption that the different M2M devices require the same throughput in every time slot is not realistic. 

\begin{figure}[ht]
 \centering
\includegraphics[width=\linewidth, height=3.5cm]{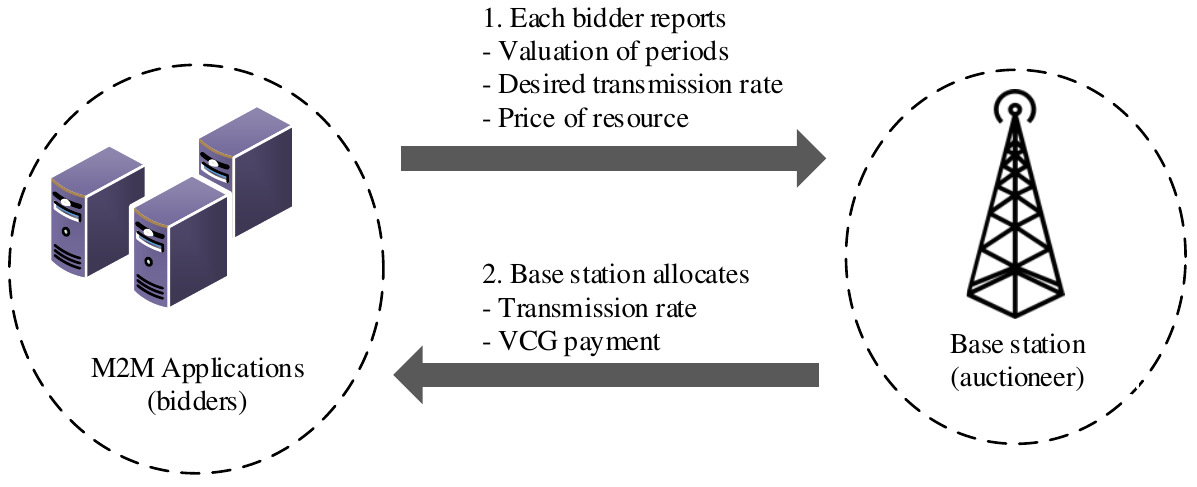}
 \caption{Rate allocation for M2M applications based on auction.}
 \label{M2M_communication_auction}
\end{figure}

Given the framework of the sealed-bid auction, the authors in \cite{lin2014multi} adopted the VCG auction mechanism to allocate the packet arrival rate to M2M applications. The M2M applications are assumed to have different valuations on the same period. For example, at the traffic peak hours, the valuation of an M2M application in a vehicular network may be much higher than those of others, e.g., an M2M application to monitor environment. The auction-based model is illustrated in Fig.~\ref{M2M_communication_auction} in which each application, i.e., a buyer, submits its bid to a base station, i.e., an auctioneer, for the packet arrival rate. Similar to the multi-attribute auction scheme discussed earlier, each bid includes multiple parameters such as the application's valuation, the packet arrival rate requirement, and the resource price which the buyer is willing to pay. The base station selects the winners that maximize the social welfare and then charges each winner according to the VCG payment strategy for a truth-telling valuation report. The simulation results showed that the transmission success probability can reach up to 97\%. However, the assumption that the packet arrival rate of each M2M application is constant in each time slot is not realistic. Also, the proposed solution did not provide any incentive mechanism for the M2M applications to report their truthful valuation. Finally, the M2M application server might not have complete information or a perfect prediction about the packet arrival rate. 

To predict the packet arrival rate of the M2M applications, the authors in \cite{lin2015auction} introduced a packet attempt estimation stage through the maximum likelihood method before conducting the auction. The estimation stage will predict the number of applications which are trying to send packets and thus allow the base station to assign time slots more efficiently. The simulation results showed that the transmission success probability can reach up to 99\%, i.e., improved by 2\% compared to that in \cite{lin2014multi}.

\begin{figure}[ht]
 \centering
\includegraphics[width=\linewidth, height=5.5cm]{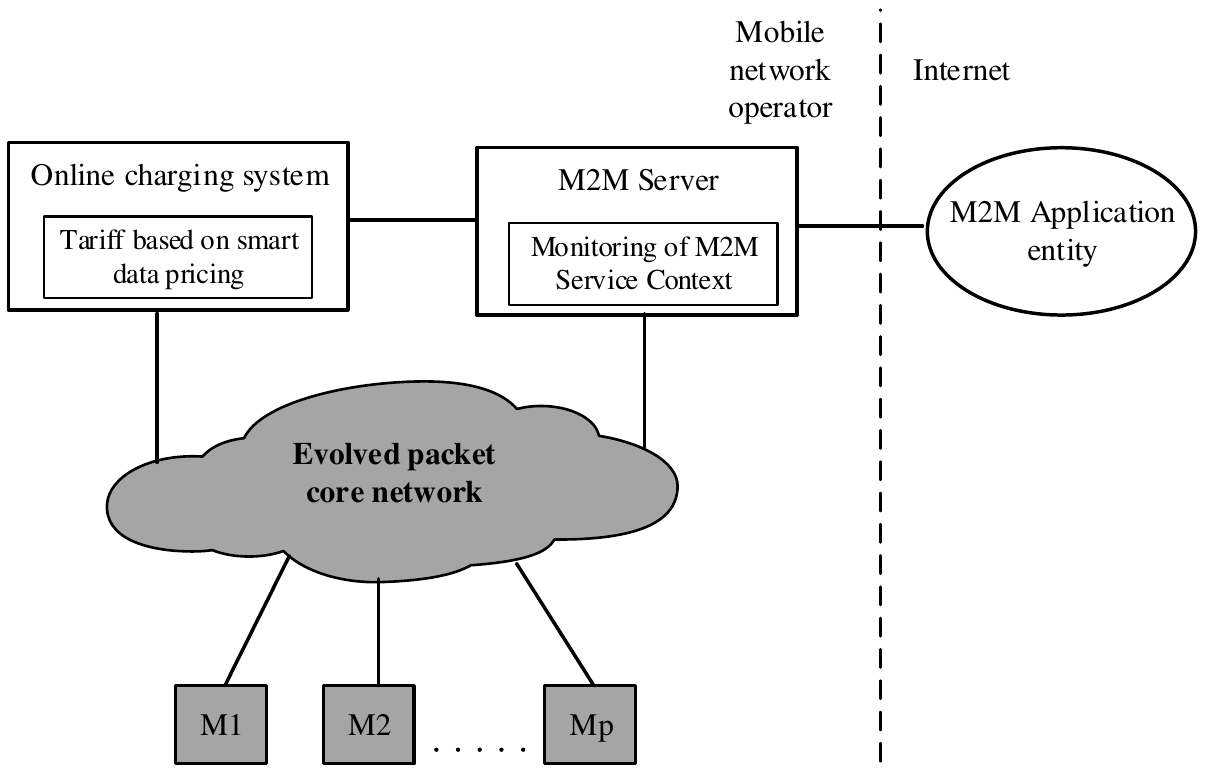}
 \caption{Architecture for M2M service context model.}
 \label{M2M_communication_congestion}
\end{figure}

Consider an M2M service context model in which a group of machines are used jointly for a certain purpose (i.e., M2M service), the authors in \cite{maric2015online} addressed the charge minimization for M2M communication based on the dynamic smart data pricing. The model is shown in Fig.~\ref{M2M_communication_congestion} in which a group of machines $M1, M2,\ldots,M_p$ are used to perform monitor functions, e.g., gas meter, camera, and thermometer. Accordingly, when a machine establishes an M2M communication with the application, the machine will be charged by an online charging system. The charging process is based on the smart data pricing, meaning that the network traffic which is sent through the mobile network in the congestion duration is charged at a higher price. The M2M server allows to grant or postpone the M2M communication, depending on the current communication requirements, e.g, emergency or no emergency, and communication charge, e.g., congestion or normal. For example, if there is an emergency situation, e.g., a fire in a house, a requirement for priority is marked in packets that allows them to be communicated with the lowest delay. However, in a normal situation, if the M2M communication is required while the network is congested, it can be postponed until the congestion subsides. Therefore, the mechanism avoids the high charge for the service providers and also improves the efficiency of the network resource management. The same approach can be found in \cite{ali2012study} in which the tariff based on the smart data pricing from electricity provider is used to control operations of home appliances, i.e., in the smart grid environment. 

The authors in \cite{maric2015online} did not consider the situations in which there are more than one emergency communication required simultaneously. Consider these situations, the authors in \cite{huang2013utility} proposed a dynamic rate allocation scheme for M2M services through the utility function. The model consists of a large number of different M2M services, i.e., buyers, which require to simultaneously access the Radio Access Network (RAN), i.e., the seller. The objective is to optimize the rate allocation to the M2M services under some constraints, e.g., total rate capacity of the RAN. Typically, the optimization problem can be solved by using the utility function as discussed in the previous approaches, e.g., \cite{eswaran2007distributed}. However, the utility of the M2M services in \cite{huang2013utility} is calculated proportionally to the delay and the price that the M2M services must pay for the allocated rate. Therefore, the optimization problem is to minimize the utility function under the constraints. It was proved that the objective utility function is concave, and there are algorithms that can converge to a unique solution for the rate allocation and the price. The simulation results showed that the utility in terms of delay and price are lower than that of the commonly used load balancing scheme \cite{sun2010traffic}. However, using the Newton's method to obtain the converged optimal rate in the proposed approach can result in slow convergence. 

\textbf{Summary:} This section discusses the application of pricing models in M2M communication. The existing approaches addressed the resource allocations as summarized in Table \ref{sec:table_secM2M_sum}. However, with the diverse characteristics of M2M traffic in conjunction with the massive number of devices, mobility, latency, reliability, security and power consumption still need more investigations.

\begin{table*}[h]
\caption{Applications of pricing models in M2M communication}
\label{sec:table_secM2M_sum}
\scriptsize 
\begin{centering}
\begin{tabular}{|>{\centering\arraybackslash}m{0.2cm}|>{\centering\arraybackslash}m{0.4cm}|>{\centering\arraybackslash}m{1.6cm}|>{\centering\arraybackslash}m{1.3cm}|>{\centering\arraybackslash}m{1.3cm}|>{\centering\arraybackslash}m{1.1cm}|>{\centering\arraybackslash}m{5cm}|>{\centering\arraybackslash}m{2.2cm}|>{\centering\arraybackslash}m{1.2cm}|}
\hline
\multirow{2}{*}  {\textbf{}} & \multirow{2}{*}  {\textbf{Ref.}} &  \multirow{2}{*}  {\textbf{Pricing model}}  & \multicolumn{3}{c|} {\textbf{Market structure}} & \multirow{2}{*}  {\textbf{Mechanism}} & \multirow{2}{*}  {\textbf{Objective}} & \multirow{2}{*} {\textbf{Solution}} \tabularnewline 
\cline{4-6}
 & & & \textbf{Seller} & \textbf{Buyer} & \textbf{Item}  & & &\tabularnewline
\hline
\hline
\parbox[t]{2mm}{\multirow{9}{*}{\rotatebox[origin=c]{90}{\hspace{-2cm} Resource allocation}}}
&\cite{lee2012optimal} &Second-price sealed-bid auction and option pricing& Service provider &Devices & Time slots& Buyers submit bid prices, and the seller selects buyers with the highest prices. The buyers can reserve the time slots in the future&Resource efficiency, and throughput, i.e., QoS, guarantee& Nash equilibrium\tabularnewline \cline{2-9}
&\cite{lin2014multi} \cite{lin2015auction} & Vickrey-Clarke-Groves auction& Base station &M2M applications&Packet arrival rate& Buyers submit bids, each of which includes application's valuation, packet arrival rate, and resource price. The seller selects winners by solving the system evaluation maximization problem and charges each winner with VCG payment &Resource efficiency, and welfare social maximization& Nash equilibrium\tabularnewline \cline{2-9}  
& \cite{maric2015online}&Smart data pricing& Service provider &Machines&Time slots&When receiving the communication requirements from the buyers, the seller can grant or postpone them based on the current type of the requirement and its tariff &Resource efficiency, charge minimization, and QoS guarantee&Market equilibrium\tabularnewline \cline{2-9} 
&\cite{ali2012study} &Smart data pricing& Electricity provider &Home appliances&Time slots&Same as \cite{maric2015online}&Resource efficiency, charge minimization, and QoS guarantee&Market equilibrium\tabularnewline \cline{2-9} 
&\cite{huang2013utility} &Utility maximization& Radio access network &M2M services& Transmission rate&Buyers submit requests to the seller for the throughput. The seller finds optimal rate and charge for the buyers through the utility function& Maximized throughput, load balance, and minimized cost&Optimization solution\tabularnewline \cline{2-9} 
\hline
\end{tabular}
\par\end{centering}
\end{table*}

\section{SUMMARY, OPEN ISSUES AND FUTURE RESEARCH DIRECTIONS}
\label{sec:Open_issues}

\subsection{Summary}
\label{sec:Open_issues_summary}
The previous sections show that economic aspects are important not only for business development, but also for system design and implementation. Pricing introduces another dimension of the problem that the traditional methods of system optimization are not applicable. This is evident as a lot of researchers are studying economic issues of IoT in which many works reviewed in our survey have proposed using different approaches to address various challenges. The survey are organized according to the issues of IoT. Then, different economic and pricing approaches are introduced and analyzed to solve the issues. This helps network managers to select desirable pricing schemes which satisfy their needs, e.g., QoS improvement and payment minimization. Besides the existing approaches, there are still
some challenges as well as research directions as discussed in the following subsection.

\subsection{Challenges, Open Issues, and Future Researches of Applying Pricing Models In the Context of IoT}
\label{sec:Open_issues_Challenges_Future_research}

\subsubsection{Combination of auction and other approaches}
\label{sec:Open_issues_Auction_combined}
Most of the aforementioned auction approaches consider only an
auction alone to solve the issues in sensing networks. In practice, the advantages of auctions might be
combined with other non-auction methods to achieve better
performance or more flexible schemes. Such an example is a combination of the reverse Vickrey auction and the posted price mechanism. In the reverse Vickrey auction, the seller with the lowest price is the winner and is paid the price of the second lowest ask. Since the winner will get the price more than its expectation, sellers are always stimulated to reveal their true cost when submitting asks. Therefore, the payment policy of the reverse Vickrey auction can guarantee the truthfulness of the mechanism. Whereas, as discussed in Section \ref{subsec:Posted_price} (and see Fig.~\ref{data_aggregation_posted_pricing}), the posted price mechanism does not require a soliciting process as the reverse Vickrey auction. Therefore, the payment policy of the Vicrkey auction can be employed in the posted price mechanism to achieve both the truthfulness properties and improve the speed of the mechanism.

\subsubsection{Mobility of phone users}
\label{sec:Open_issues_Phone_Users}
 In the reverse auction-based data collection approaches, e.g., \cite{lee2010sell} and \cite{lee2010dynamic}, phone users are assumed to be stationary, and they submit their offer prices before sending their data to the server. The server selects the phone users with the lowest prices as winners to provide data. There may be a situation in which the phone users are selected as winners but then move out of the area of interest. This may degrade the utility of the platform. Therefore, the server needs to keep track of the mobility pattern of each phone user. The lightweight triangulation method \cite{kim2006extracting} can be used to estimate the next location of each phone user. The estimation value is then considered as one of attributes to select the winners.

\subsubsection{Confidentiality of bids}
\label{sec:Open_issues_Confidentiality}
 In auction approaches, bids of bidders are assumed to be submitted simultaneously. However, in practice bidders may not send their bids precisely at the same time. If the auctioneer or any other party can recover some of the bids before the bid opening, it can inform a colluding bidder to cheat and win the auction. Security protocols, e.g., the lightweight security protocol \cite{sugeno1988structure}, which guarantees the bid secrecy up until the opening phase is required.

\subsubsection{Payment security}
\label{sec:Open_issues_Payment_security}
Payment process is crucial to stimulate users to provide their sensing data. However, the messages including privacy information, e.g., payment and user identity, can be received by other irrelevant network users which can be malicious users. Therefore, security problems for the payment process should be analyzed, and the techniques for the payment protection, e.g., using a bank along with an e-cash scheme, need to be developed.

\subsubsection{Complex resource market}
\label{sec:Open_issues_Complex_market}
Traditional auctions usually contains three entities: buyers, sellers, and auctioneers. Buyers purchase resources from sellers via the assistance of auctioneers. However, in a more complex market, the trades can be made between buyers and sellers directly, or among buyers via an auctioneer, or among auctioneers. Also, resources can be bought back, forfeited or lent. Such diversified market for resources auctions need to be considered as the future work.

\subsubsection{Collusion and Price of Anarchy}
\label{sec:Anarchy}
 In a competitive market, service providers set selling prices to maximize their own profits given the strategies of other providers. A Nash equilibrium can be an appropriate solution as proposed in \cite{niyatoeconomics}. However, due to selfish behaviors of the service providers, they can collude with each other to optimize their profits collectively, and thus the Nash equilibrium become inefficient \cite{dubey1986inefficiency}. The collusion among service providers and the price of anarchy \cite{goemans2005sink} can be investigated in the market.

\subsubsection{Contract theory}
\label{sec:Contract_theory}
Relay selection based on the Dutch auction as proposed in \cite{lima2008game} may require several iterations to select the best neighboring sensor for the data forwarding, and thus increasing the packet delay. Alternatively, we can employ a contract theory for the relay selection to reduce the number of iterations. A contract theory is to analyze how sellers and buyers construct contractual arrangements, in the presence of asymmetric information \cite{hart1986theory}. The asymmetric information means that one part in the contract has more information about the contract than the rest. Using contract theory in the relay selection process in WSNs can be described as follows. First, the source sensor, i.e., the buyer, designs incentive offers involving items, e.g., the payment and desired signal-to-noise-ratio at the destination. The source sensor then broadcasts these offers to its neighboring sensors, i.e., sellers. If the neighboring sensors are willing to relay data, they notify the source with the contracts that they agree to accept. The source with a limited budget can solve the relay selection problem as a knapsack problem. In general, the contract approach is simple
to implement which requires minimal signaling overhead while still guaranteeing QoS, i.e., data timeliness. Therefore, contract theory can also be employed for the data collection among data sellers, e.g., phone users, and a buyer, e.g, service provider or data collector.


\section{Conclusions}
\label{sec:Conclusion}
This paper has provided a comprehensive survey of the economic and pricing theory as well as their applications in data collection and communication of IoT. Firstly, we have presented a general architecture of an IoT system including its components and services. To understand the motivations of using the economic and pricing theory in IoT, we have introduced fundamentals of various pricing models with their general objectives. Then, we have provided detailed reviews, analysis, and comparisons of approaches using the economic and pricing theories for solving a variety of issues in sensing networks, i.e., WSNs, participatory sensing, crowdsensing networks, and M2M communication. Finally, we have discussed some open issues and important research directions. A survey of the applications of the economic and pricing theory for issues in other layers in IoT(e.g., cloud platform) is considered as the future work.
\section*{Acknowledgements}
 
This work was supported in part by the National Research Foundation of Korea (NRF) grant funded by the Korean government (MSIP) (2014R1A5A1011478), Singapore MOE Tier 1 (RG18/13 and RG33/12) and MOE Tier 2 (MOE2014-T2-2-015 ARC 4/15), and the U.S. National Science Foundation under Grants US NSF ECCS-1547201, CCF-1456921, CNS-1443917, ECCS-1405121, and NSFC61428101

\bibliographystyle{IEEEtran}
\bibliography{PricingIoTBib}
\end{document}